\documentclass[seceq,preprint]{ptptex}
\usepackage{amsmath,amssymb,mathtools}
\usepackage{wrapft}

\notypesetlogo                       
\preprintnumber[3cm]{
YITP-14-106}

\title{
Symmetries and Feynman Rules for Ramond Sector\\ in Open Superstring Field Theory
}

\author{
Hiroshi \textsc{Kunitomo}\footnote{%
E-mail:\  {\tt kunitomo@yukawa.kyoto-u.ac.jp}}
}

\inst{
Yukawa Institute for Theoretical Physics, Kyoto University, \\
Kyoto 606-8502, Japan
}

\abst{
We examine the symmetries of the action suppelemented by the constraint 
in the WZW-type open superstring field theory. 
It is found that this pseudo-action has additional symmetries 
provided we impose the constraint after the transformation.
Respecting these additional symmetries, we propose a prescription for 
the new Feynman rules for the Ramond sector. It is shown that the new rules reproduce
the well-known on-shell four- and five-point amplitudes with external fermions.
}

\begin{document}

\maketitle

\section{Introduction}

One of the first criteria to determine whether a string field theory is 
acceptable or not is that its on-shell physical amplitudes are equivalent 
to the well-known results in the first-quantized formulation. 
This is well studied for bosonic string field theories.
It has been proved that an arbitrary amplitude at any loop order
is correctly reproduced in the cubic open string 
field theory\cite{Witten:1985cc,Giddings:1986wp,Zwiebach:1990az} 
and the nonpolynomial closed string field 
theory.\cite{Saadi:1989tb,Kugo:1989aa,Zwiebach:1992ie} 
In contrast, such considerations are still not sufficient for 
superstring field theories. 

The most promising open superstring field theory at present is
the Wess-Zumino-Witten (WZW)-type formulation proposed by Berkovits utilizing 
the large Hilbert space.\cite{Berkovits:1995ab,Berkovits:2001im}
Its Neveu-Schwarz (NS) sector is compactly described
by the WZW-type gauge-invariant action with the help of the pure-gauge 
string field.\cite{Berkovits:1995ab}
It does not require explicit insertions 
of the picture-changing operators and thus does not suffer from the divergence 
coming from their collisions. 
In return for this advantage, the action becomes nonpolynomial, 
which obscures whether it gives the correct amplitudes or not.
It has been confirmed so far that only
four-\cite{Berkovits:1999bs} and five-boson\cite{Michishita:Riken, Michishita:2012ku}  
amplitudes at the tree level are correctly reproduced.

On the other hand, the Ramond (R) sector of the formulation is 
less well understood. 
While the equations of motion can be given,\cite{Berkovits:2001im} 
it is difficult to construct the covariant action.
Then, as an alternative to an action in the usual sense, Michishita constructed 
an action supplemented by an appropriate constraint by introducing an auxiliary 
field.\cite{Michishita:2004by} The variation of this pseudo-action leads 
to the equations of motion that reduce to the desired ones after eliminating 
the auxiliary field by imposing the constraint. 
Although this is not a usual action from which we can uniquely derive the Feynman rules,
it could be used as a clue to propose the Feynman rules\footnote{
We call them the self-dual rules in this paper since they allow 
only the self-dual part (the part satisfying the linearized constraint)
of the R string fields to propagate.} 
that reproduce the correct on-shell four-point amplitudes
with external fermions.\cite{Michishita:2004by}
After a while, however, it was found that these self-dual rules 
do not lead to the correct five-point amplitudes with two external
fermions.\cite{Michishita:Riken, Michishita:2012ku}

In order to determine the reason why the self-dual rules do not reproduce
the correct amplitudes, 
we will examine the gauge symmetries of the theory. 
It has been known that the pseudo-action of the R sector has fewer gauge 
symmetries than those at the linearized level.\cite{Michishita:2004by} 
We will find that the missing symmetries exist provided we impose
the constraint after transforming it.
While these are not symmetries in the usual sense, we will assume that 
they have to be respected in the calculation and propose a prescription for 
new Feynman rules.
Then, by using these new rules, we will calculate the four- and five-point amplitudes 
with external fermions. It will be shown that they are in fact equivalent to
the amplitudes in the first quantized formulation.

This paper is organized as follows.
In \S~\ref{Feynman}, we will first summarize the basic properties of the WZW-type
open superstring field theory.
The symmetries of the pseudo-action will then be studied. It will be found 
that it is invariant under the additional gauge symmetries 
if we suppose it to be subject to the constraint after the transformation.
Respecting these symmetries, we will propose a prescription for the new Feynman rules,
in which the R propagator has an off-diagonal form.
By using these new rules, we will explicitly calculate the on-shell four- and 
five-point amplitudes with external fermions in \S \ref{amplitudes}. 
During the process to confirm that the four-point amplitudes are
reproduced as in the self-dual rules, it will be clarified what kind 
of diagrams produce differences between two results by the two sets of the Feynman rules. 
For the five-point amplitudes, such diagrams certainly appear in the calculation of 
two-fermion-three-boson amplitudes that cannot be correctly reproduced 
by the self-dual rules. 
We will show that the results are actually improved,
and in consequence all the five-point amplitudes come to be equivalent to
those in the first quantized formulation.
The final section, \S\ref{conclusion}, is devoted
to the conclusion and the discussion.

\section{New Feynman rules in WZW-type open superstring field theory}\label{Feynman}

After summarizing the known basic properties of 
the WZW-type open superstring field theory,\cite{Berkovits:1995ab, Michishita:2004by}
we will recall the self-dual Feynman rules proposed in Ref.~\citen{Michishita:2004by}
in this section. 
Then, examining the gauge symmetries of the pseudo-action, we will propose a prescription
for the new Feynman rules, 
which are more natural in the viewpoint of the symmetry.

\subsection{WZW-type open superstring field theory}

The conventional (small) Hilbert space $\mathcal{H}_{small}$ of the first-quantized RNS superstring is 
in general described by a tensor product of those of an $N=1$ superconformal matter with
central charge $c=15$, the reparametrization ghosts $(b(z), c(z))$
with $c=-26$, and the superconformal ghosts $(\beta(z), \gamma(z))$
with $c=11$. The superconformal ghost system is known to be represented also by
a chiral boson $\phi(z)$ $(c=13)$ and a pair of fermions $(\eta(z), \xi(z))$
$(c=-2)$ through the bosonization formula,\cite{Friedan:1985ge}
\begin{equation}
 \beta(z) = e^{-\phi(z)}\partial\xi(z),\qquad \gamma(z) = \eta(z) e^{\phi(z)}.
\end{equation}
The large Hilbert space $\mathcal{H}_{large}$ can be introduced by replacing 
the Hilbert space of the superconformal ghosts in $\mathcal{H}_{small}$ 
by that of the bosonized fields $(\phi(z), \eta(z), \xi(z))$, which is twice
as large as $\mathcal{H}_{small}$ due to the zero-mode $\xi_0$.
The correlation function in $\mathcal{H}_{large}$ is normalized as
\begin{equation}
 \langle \xi e^{-2\phi}c\partial c\partial c^2\rangle = 1,
\end{equation}
and can be nonzero if and only if the ghost and picture numbers $(G,P)=(2,-1)$ in total.

The NS (R) string field,  $\Phi$ ($\Psi$), is
defined using $\mathcal{H}_{large}$.
Their free equations of motion are given by\footnote{
In this paper the zero-mode $\eta_0$ of $\eta(z)$ is denoted as $\eta$, for simplicity.}
\begin{subequations} \label{free eom}
 \begin{align}
 Q\eta  \Phi =& 0,\label{free NS eom}\\
 Q\eta  \Psi =& 0,\label{free R eom}
\end{align}
\end{subequations}
which are invariant under the gauge transformations
\begin{subequations} \label{free gauge}
\begin{align}
\delta\Phi =& Q\Lambda_0+\eta \Lambda_1,\label{free NS gauge}\\
\delta\Psi =& Q\Lambda_{\frac{1}{2}}+\eta \Lambda_{\frac{3}{2}}.\label{free R gauge}
\end{align} 
\end{subequations}
Here the $Q$ is the first-quantized BRST operator.
The string fields $\Phi$ and $\Psi$ are Grassmann even and have $(G,P)=(0,0)$ 
and $(0,1/2)$, respectively. The gauge parameter string fields
$\Lambda_{n/2}$ $(n=0,\cdots,3)$ are Grassmann odd and have $(G,P)=(-1,n/2)$.
The on-shell states satisfying (\ref{free eom}) up to the gauge transformation 
(\ref{free gauge}) are equivalent to the conventional physical spectrum defined
by the BRST cohomology in $\mathcal{H}_{small}$.

For the NS sector, the free equation of motion (\ref{free NS eom})
was ingeniously extended 
by Berkovits\cite{Berkovits:1995ab} to the nonlinear equation of motion
\begin{equation}
 \eta(e^{-\Phi}(Qe^\Phi))=0,
\end{equation}
derived from the variation of the the WZW-type action,
\begin{equation}
 S_{NS} = \frac{1}{2}\langle (e^{-\Phi}Qe^\Phi)(e^{-\Phi}\eta e^\Phi)
-\int^1_0dt (e^{-t\Phi}\partial_t e^{t\Phi})\{(e^{-t\Phi}Qe^{t\Phi}),
(e^{-t\Phi}\eta e^{t\Phi})\}\rangle.\label{NS action}
\end{equation}
This action (\ref{NS action}) is invariant under 
the nonlinear extension of (\ref{free NS gauge}),
\begin{equation}
 e^{-\Phi}(\delta e^\Phi) 
= Q'\Lambda_0'+\eta \Lambda_1,
\end{equation}
where $\Lambda_0'=e^{-\Phi}\Lambda_0e^\Phi$.
The shifted BRST operator $Q'$ is
defined as an operator that acts on a general string field $A$ as
\begin{equation}
 Q'A=QA+(e^{-\Phi}Qe^{\Phi})A-(-1)^AA(e^{-\Phi}Qe^\Phi).
\end{equation}
We can show that this is also nilpotent, $Q'(Q'A)=0$, due to the identity, 
\begin{equation}
Q(e^{-\Phi}Qe^\Phi)+(e^{-\Phi}Qe^\Phi)^2\equiv 0. 
\end{equation}

In contrast, 
only the equations of motion (\ref{free eom}) can be extended to the nonlinear form,
\begin{subequations}\label{full eom} 
\begin{align}
 \eta(e^{-\Phi}(Qe^\Phi))+(\eta \Psi)^2 =& 0,\label{full NS eom}\\
 Q'\eta \Psi =& 0,\label{full R eom}
\end{align} 
for the full theory including the interaction with the R sector.
\end{subequations}
The gauge transformations (\ref{free gauge}) can also be extended to the nonlinear form,
\begin{subequations} \label{nonlinear gauge}
 \begin{align}
  e^{-\Phi}(\delta e^\Phi)=& Q'\Lambda_0+\eta \Lambda_1-\{\eta \Psi, \Lambda_{\frac{1}{2}}\},
  \label{nonlinear gauge NS}\\
 \delta\Psi =& Q'\Lambda_{\frac{1}{2}}+\eta \Lambda_{\frac{3}{2}}+[\Psi, \eta \Lambda_1],
 \label{nonlinear gauge R}
 \end{align}
\end{subequations}
so as to keep the equations of motion (\ref{full eom}) invariant.
However, we cannot construct an action, or even its quadratic term,
so as to have $(G,P)=(2,-1)$ unless we introduce the (inverse) picture-changing 
operator or an additional string field.

Then, as an alternative formalism, an action for the R sector, 
\begin{equation}
S_R = -\frac{1}{2}\langle (Q\Xi) e^\Phi(\eta \Psi)e^{-\Phi}\rangle,
\label{R action}
\end{equation}
was proposed\cite{Michishita:2004by} 
by introducing an auxiliary Grassmann even R string field $\Xi$ with $(G,P)=(0,-1/2)$.
The equations of motion derived from the variation of $S=S_{NS}+S_R$ are
\begin{subequations}\label{eom michi}
\begin{align}
 \eta(e^{-\Phi}(Qe^\Phi))+\frac{1}{2}\{\eta \Psi, Q'\Xi'\} =& 0,\label{eom michi phi}\\
 Q'\eta \Psi =& 0,\label{eom michi psi}\\
 \eta Q'\Xi' =& 0,\label{eom michi xi}
\end{align} 
\end{subequations}
where $\Xi'=e^{-\Phi}\Xi e^\Phi$, which reduce to (\ref{full eom}) 
if we eliminate the $\Xi$ by imposing the constraint,
\begin{equation}
 Q'\Xi'=\eta \Psi.\label{const}
\end{equation}
In this sense, the action (\ref{R action}) is not an action in the usual sense 
but a pseudo-action supplemented by the constraint (\ref{const}).\cite{Bergshoeff:2001pv}

\subsection{Gauge fixing and the self-dual Feynman rules}

We next review how tree-level amplitudes are calculated in this formulation.
Let us first derive the Feynman rules for the NS sector from
the action (\ref{NS action}).
Since its quadratic part,
\begin{equation}
 S_{NS}^{(2)}= \frac{1}{2}\langle(Q\Phi)(\eta \Phi)\rangle,
\end{equation}
is invariant under
\begin{equation}
\delta \Phi = Q\Lambda_0+\eta \Lambda_1,\label{linear gauge NS} 
\end{equation}
we have to fix these symmetries to obtain the propagator.
We take here the simplest gauge conditions:
\begin{equation}
 b_0\Phi=\xi_0\Phi=0.\label{NS conditions}
\end{equation}
The NS propagator in this gauge becomes
\begin{align}
\overbracket[0.5pt]{\!\!\Phi\Phi\!\!}\ =\
\Pi_{NS} =& \frac{\xi_0 b_0}{L_0}\nonumber\\
=& \int_0^\infty d\tau (\xi_0 b_0)e^{-\tau L_0}.
\label{NS propagator}
\end{align}
The interaction vertices can be read by expanding 
(\ref{NS action}) in the power of $\Phi$. The three and four string vertices are
\begin{subequations} 
 \begin{align} 
S_{NS}^{(3)} =& -\frac{1}{3!}\Big(\langle
  \Phi(Q\Phi)(\eta \Phi)\rangle+\langle\Phi(\eta \Phi)(Q\Phi)\rangle\Big),
\label{NS 3 vertex}\\
 S_{NS}^{(4)} =& \frac{1}{4!}\Big(
\langle\Phi^2(Q\Phi)(\eta \Phi)\rangle-\langle\Phi^2(\eta \Phi)(Q\Phi)\rangle 
-2\langle \Phi (Q\Phi) \Phi (\eta \Phi)\rangle\Big),
\label{NS 4 vertex}
\end{align}
\end{subequations}
respectively, which are necessary for the calculation in the next section.
It was confirmed that these Feynman rules reproduce the same on-shell physical 
amplitudes with four\cite{Berkovits:1999bs,Fuji:2006me} and 
five\cite{Michishita:Riken, Michishita:2012ku} external bosons
as those in the first quantized formulation.

For the R sector, however, the Feynman rules are not logically derived from 
the pseudo-action (\ref{R action}) since it is not an action in the usual 
sense. We summarize here the Feynman rules proposed in \citen{Michishita:2004by}. 
We first suppose that the $\Xi$ and $\Psi$ are independent string fields. 
Then the propagator can be easily read from the quadratic term,
\begin{equation}
 S_R^{(2)}= -\frac{1}{2}\langle(Q\Xi)(\eta \Psi)\rangle,
\end{equation}
as in the case of the NS sector. Fixing the gauge symmetries
\begin{subequations}\label{linear gauge R}
 \begin{align}
 \delta \Psi =& Q\Lambda_{\frac{1}{2}}+\eta \Lambda_{\frac{3}{2}},\\
 \delta \Xi =& Q\Lambda_{-\frac{1}{2}}+\eta \tilde{\Lambda}_{\frac{1}{2}},
\end{align}
\end{subequations}
by the same conditions as (\ref{NS conditions}),
\begin{equation}
b_0\Psi=\xi_0\Psi=0,\qquad
 b_0\Xi=\xi_0\Xi=0,\label{R conditions}
\end{equation}
we can obtain the (off-diagonal) R propagator in this gauge as
\begin{align}
\overbracket[0.5pt]{\!\!\Psi\Xi\!\!}\ =\
\overbracket[0.5pt]{\!\!\Xi\Psi\!\!}\ \equiv 
\Pi_R=&-2\frac{\xi_0 b_0}{L_0}\nonumber\\
=&-2\int_0^\infty d\tau (\xi_0 b_0)e^{-\tau L_0}.\label{R propagator}
\end{align}
The auxiliary field $\Xi$ is eliminated from the external on-shell states
by the linearized constraint $Q\Xi=\eta\Psi$.\footnote{
Note that the linearized constraint is sufficient to impose on
the external (asymptotic) on-shell states.}
The rule not uniquely determined is how we take 
into account the constraint at the off-shell.
A prescription for the self-dual Feynman rules is 
to replace $Q\Xi$ and $\eta\Psi$ 
in the vertices with their self-dual part $\omega=(Q\Xi+\eta\Psi)/2$,
by which the part that vanishes under the (linearized) constraint is decoupled.
From the cubic, quartic and quintic terms of the action (\ref{R action}),
\begin{subequations}\label{R vertices}
 \begin{align}
 S_R^{(3)} =&
  \frac{1}{2}\Big(\langle\Phi(Q\Xi)(\eta \Psi)\rangle
  +\langle\Phi(\eta \Psi)(Q\Xi)\rangle\Big),
\label{R 3 vertex}\\
 S_R^{(4)} =& -\frac{1}{4}\Big(\langle(\Phi^2(Q\Xi)(\eta \Psi)\rangle
 -\langle\Phi^2(\eta \Psi)(Q\Xi)\rangle\Big)
\nonumber\\
&\hspace{30mm}
+\frac{1}{4} \Big(\langle\Phi(Q\Xi)\Phi(\eta \Psi)\rangle
-\langle\Phi(\eta\Psi)\Phi(Q \Xi)\rangle\Big),
\label{R 4 vertex}\\
  S_R^{(5)} =&
  \frac{1}{12}\Big(\langle\Phi^3(Q\Xi)(\eta \Psi)\rangle
  +\langle\Phi^3(\eta \Psi)(Q\Xi)\rangle\Big)
\nonumber\\
&\hspace{30mm}
-\frac{1}{4}\Big(\langle\Phi^2(Q\Xi)\Phi(\eta \Phi)\rangle
+\langle\Phi^2(\eta \Phi)\Phi(Q\Xi)\rangle\Big),
\label{R 5 vertex}
\end{align}
\end{subequations}
the three, four and five string vertices in this prescription, 
needed to calculate the five-point amplitudes later, are obtained as
\begin{subequations}\label{sd vertices}
\begin{align}
\tilde{S}_R^{(3)} =& \langle \Phi \omega^2 \rangle,\label{sd 3 vertex}\\
\tilde{S}_R^{(4)} =& 0,\label{sd 4 vertex}\\
\tilde{S}_R^{(5)} =& \frac{1}{6} \langle \Phi^3 \omega^2 \rangle
-\frac{1}{2}\langle \Phi^2 \omega \Phi \omega \rangle,\label{sd 5 vertex}
\end{align}
\end{subequations}
respectively. In particular, the two-fermion-two-boson vertex
(or generally two-fermion-even-boson vertices) vanishes 
in this prescription.\cite{Michishita:2004by} 
The propagator has the form
\begin{equation}
 \overbracket[0.5pt]{\!\!\omega\omega\!\!}\ =\
\frac{1}{4}(Q\Pi_R\eta+\eta\Pi_RQ),\label{sd R propagator}
\end{equation}
from (\ref{R propagator}).
It was shown that these self-dual Feynman rules reproduce 
the well-known four-point amplitudes,\cite{Michishita:2004by} 
but unfortunately do not do 
the five-point amplitudes with two external 
fermions.\cite{Michishita:Riken, Michishita:2012ku}
The extra contributions including no propagator
are not completely cancelled by those from
the five string interaction (\ref{sd 5 vertex}), and remain nonzero.

\subsection{Gauge symmetries and the new Feynman rules}\label{gauge symm}

In order to find out the reason why the self-dual Feynman rules 
do not work well,
let us examine the gauge symmetries in detail.
The total (pseudo-) action, $S=S_{NS}+S_R$,
is invariant under the gauge transformations,
\begin{subequations}\label{gtf}
\begin{align}
& e^{-\Phi}(\delta e^\Phi) =
Q'\Lambda_0'+\eta \Lambda_1,\\
& \delta\Psi = \eta \Lambda_{\frac{3}{2}}+[\Psi, \eta \Lambda_1],\qquad
 \delta\Xi = Q\Lambda_{-\frac{1}{2}}+[Q\Lambda_0,\Xi].
\end{align}
\end{subequations}
Since these symmetries are compatible with the self-dual anti-self-dual 
decomposition of the R strings,
\begin{equation}
 \delta (Q'\Xi'\pm\eta\Psi) = [(Q'\Xi'\pm\eta\Psi), \eta\Lambda_1],
\end{equation}
they are also the symmetries of the constraint (\ref{const}),
and so respected by the self-dual Feynman rules.\footnote{
It is not clear whether it is sufficient to take into account 
the linearized constraint to define the self-dual part $\omega$ or not.}
Nevertheless, these symmetries do not include all the symmetries of
the linearized level, (\ref{linear gauge R}).
The missing transformations
extended to the nonlinear form
\begin{subequations}\label{nonlinear gtf Michi}
 \begin{align}
 e^{-\Phi}(\delta e^\Phi) =& 
-\frac{1}{2}\{Q'\Xi',\Lambda_{\frac{1}{2}}\}
+\frac{1}{2}\{\eta \Psi,\tilde{\Lambda}_{\frac{1}{2}}\},\\
 \delta\Psi =& Q'\Lambda_{\frac{1}{2}},\qquad 
\delta\Xi=e^\Phi(\eta \tilde{\Lambda}_{\frac{1}{2}})e^{-\Phi},
\end{align}
\end{subequations}
transform the action to the form proportional to the constraint:
\begin{equation}
 \delta S = \frac{1}{4}\langle \Lambda_{\frac{1}{2}}[(Q'\Xi')^2,(Q'\Xi'-\eta \Psi)]\rangle
+\frac{1}{4}\langle \tilde{\Lambda}_{\frac{1}{2}}[(\eta\Psi)^2,(Q'\Xi'-\eta \Psi)]\rangle.
\label{variation}
\end{equation}
In other words, the action is invariant under (\ref{nonlinear gtf Michi})
provided we impose the constraint after the transformation.
Their consistent part with the constraint,
obtained by putting $\tilde{\Lambda}_{1/2}=-\Lambda_{1/2}$, reduce to
the symmetries (\ref{nonlinear gauge}) of the equations of motion (\ref{full eom})
if we eliminate the $\Xi$ by the constraint.
These are not the symmetries in the usual sense, but have to be important properties
to characterize 
the action. Therefore, it is natural to consider that a reason why the self-dual 
rules do not work is because the replacement to the self-dual part $\omega$ of $Q\Xi$ 
and $\eta\Psi$ breaks these symmetries.
This leads us to propose the following alternative prescription for 
the (tree-level) Feynman rules:
\begin{itemize}
 \item Use the off-diagonal propagator (\ref{R propagator}) for the R string.
 \item Use the vertices (\ref{R vertices}) as they are without
restricting both of $Q\Xi$ and $\eta\Psi$ to their self-dual part.
 \item Add two possibilities, $\Xi$ and $\Psi$, of each 
external fermion and impose the linearized constraint $Q\Xi=\eta\Psi$
on the on-shell external states.
\end{itemize}
We claim this prescription respecting all the gauge symmetries, including 
those in the above sense, is more appropriate for the Feynman rules read from 
the pseudo-action (\ref{R action}).

\section{Amplitudes with external fermions}\label{amplitudes}

In this section, we will explicitly calculate the on-shell four- 
and five-point amplitudes with external fermions
using the new Feynman rules. It will be shown
that the equivalent amplitudes to those in the first quantized formulation 
are correctly reproduced.

\subsection{Four-point amplitudes}\label{four point}

The on-shell four-point amplitudes with external fermions were already 
calculated using the self-dual Feynman rules, and shown to be equivalent 
to those obtained in the first quantized formulation.\cite{Michishita:2004by}
We first show that the new Feynman rules also reproduce the same results.

\begin{figure}[htbp]
\begin{minipage}{0.15\hsize}
\mbox{}
\end{minipage}
\begin{minipage}{0.35\hsize}
 \begin{center}
 \includegraphics[width=4cm]{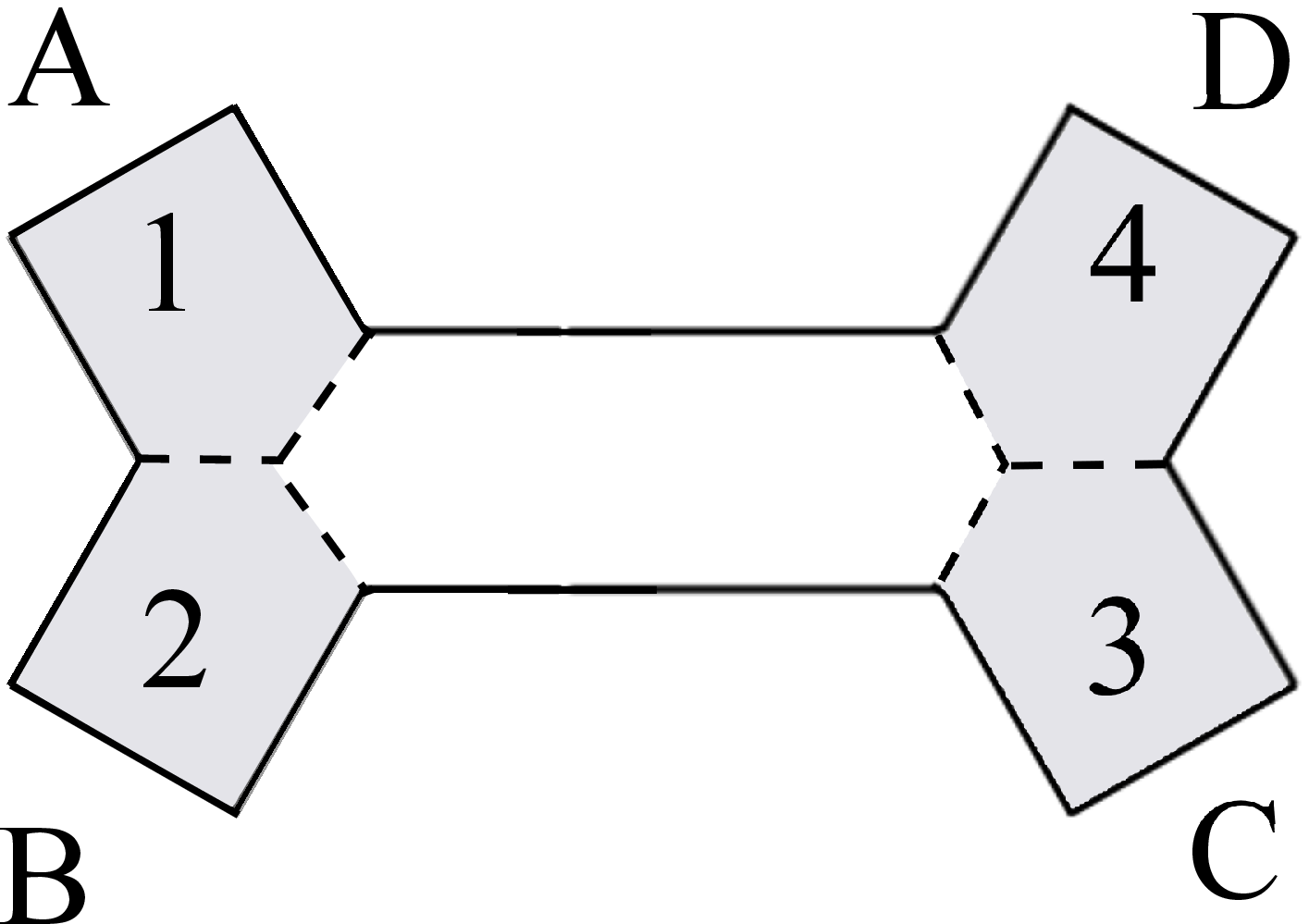}\\
(a)
 \end{center}
\end{minipage}
\begin{minipage}{0.35\hsize}
 \begin{center}
 \includegraphics[width=4cm]{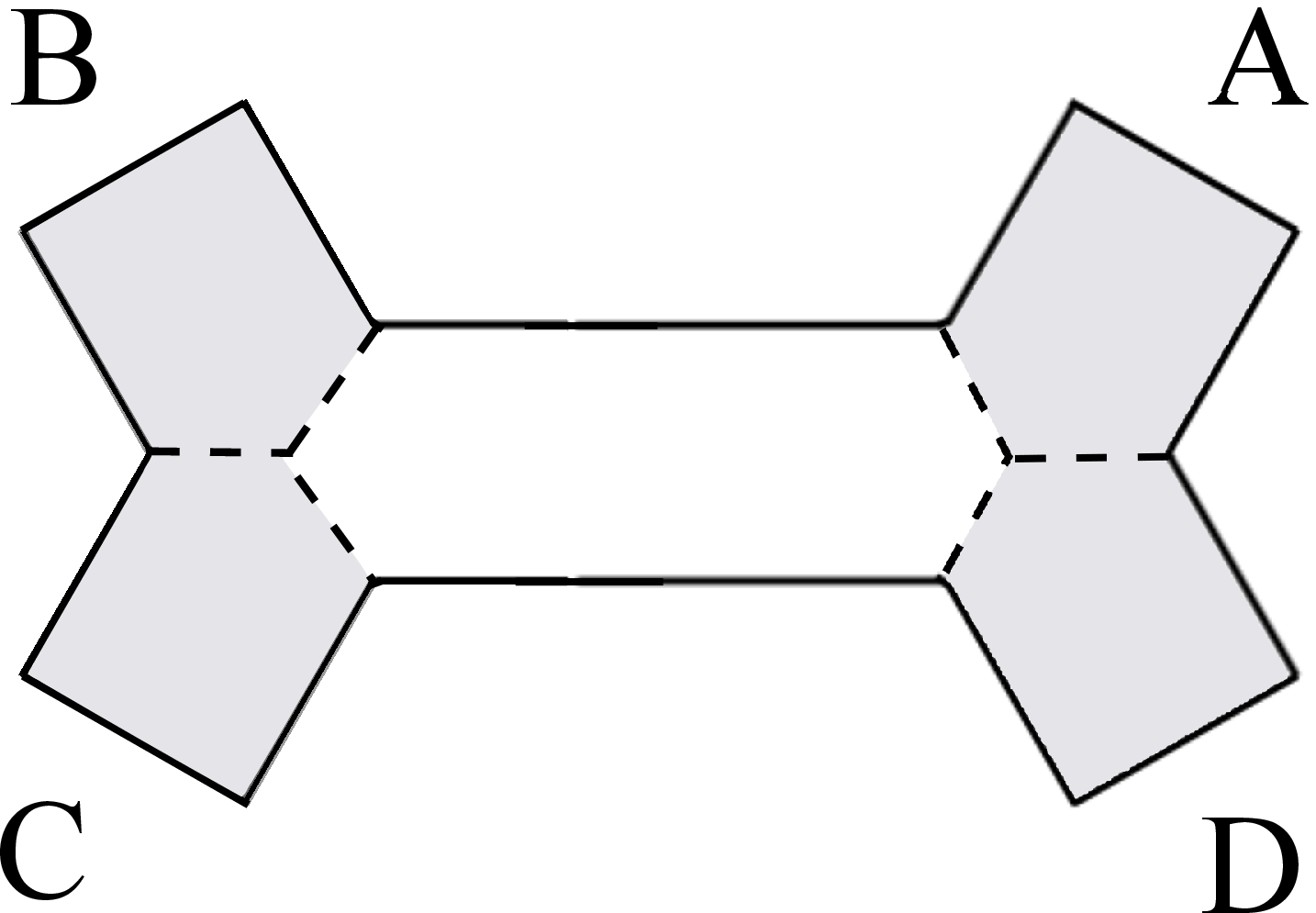}\\
(b)
 \end{center}
\end{minipage}
\vspace{2mm}
\caption{Two Feynman diagrams
for four-fermion amplitude: (a) $s$-channel and (b) $t$-channel. 
Each leg is numbered from 1 to 
4 as depicted in (a). The fermion legs are colored with gray.} 
\label{FFFF}
\end{figure}

\begin{figure}[htbp]
 \begin{center}
  \includegraphics[width=7.5cm]{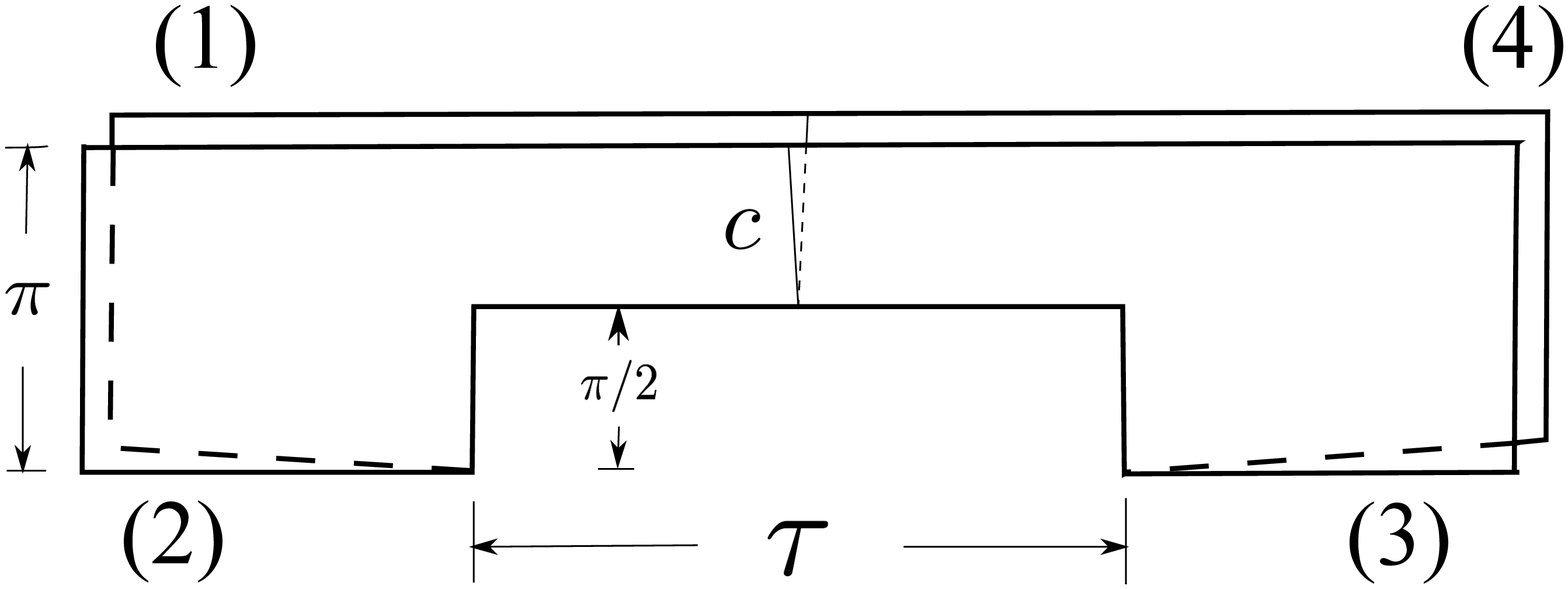}
 \end{center}
\caption{The Witten diagram for four-point amplitudes. Each of the four legs is numbered from
$(1)$ to $(4)$, corresponding to those in Fig.~\ref{FFFF}. They should be read 
as semi-infinite strips.}
\label{Witten4}
\end{figure}

Let us start from the calculation of the four-fermion amplitude $\mathcal{A}_{FFFF}$
with fixed color ordering. 
Since there is no four-fermion vertex in (\ref{R 4 vertex}), 
the contributions only come from the $s$- and $t$-channel diagrams
in Fig.~\ref{FFFF}. We take a convention that the fermion legs 
and propagators in the Feynman diagrams are colored with gray.
If we label each four external states $A,$ $B$, $C$, and $D$, the $s$-channel contribution
is calculated as
\begin{align}
 \mathcal{A}_{FFFF}^{(s)} =& \left(\frac{1}{2}\right)^2
\int_0^\infty d\tau\
\langle \Big(Q\Xi_A(1)\ \eta \Psi_B(2)+\eta \Psi_A(1)\ Q\Xi_B(2)\Big)\nonumber\\
&\hspace{40mm} \times
(\xi_c b_c)\ \Big(Q\Xi_C(3)\ \eta \Psi_D(4)+\eta \Psi_C(3)\ Q\Xi_D(4)\Big)\rangle_W,
\label{FFFF s}
\end{align}
where the correlation is evaluated as the conformal field theory on the Witten diagram
given in Fig.~\ref{Witten4}. 
The $\xi_c$ and $b_c$ denote the corresponding fields integrated along the path $c$
depicted on the diagram.
Each leg is numbered from 1 to 4,
but this is redundant if we always arrange the external 
states in order of the numbers from the left as in (\ref{FFFF s}). 
Taking this convention, we omit hereafter to indicate them.
Then the $t$-channel contribution can similarly be calculated as
\begin{align}
 \mathcal{A}_{FFFF}^{(t)} =& \left(\frac{1}{2}\right)^2\int_0^\infty d\tau\
\langle \Big(Q\Xi_B\ \eta \Psi_C+\eta \Psi_B\ Q\Xi_C\Big)
(\xi_c b_c)\Big(Q\Xi_D\ \eta \Psi_A+\eta \Psi_D\ Q\Xi_A\Big)\rangle_W.
\end{align}
These two contributions are essentially the same as those obtained using the self-dual 
rules\cite{Michishita:2004by} and are combined into the conventional four-point 
amplitude as
\begin{align}
 \mathcal{A}_{FFFF} =& \mathcal{A}_{FFFF}^{(s)}+\mathcal{A}_{FFFF}^{(t)}
\nonumber\\
=&\frac{1}{4}\int_0^\infty d\tau\Bigg( 
\langle \Big(Q\Xi_A\ \eta \Psi_B+\eta \Psi_A\ Q\Xi_B\Big)
(\xi_c b_c)\ \Big(Q\Xi_C\ \eta \Psi_D+\eta \Psi_C\ Q\Xi_D\Big)\rangle_W
\nonumber\\
&\hspace{1.5cm}
+\langle \Big(Q\Xi_B\ \eta \Psi_C+\eta \Psi_B\ Q\Xi_C\Big)
(\xi_c b_c)\Big(Q\Xi_D\ \eta \Psi_A+\eta \Psi_D\ Q\Xi_A\Big)\rangle_W\Bigg),
\nonumber\\
=&\int_0^\infty d\tau\Big(\
\langle\eta \Psi_A\ \eta \Psi_B\ (\xi_c b_c)\ \eta \Psi_C\ \eta \Psi_D\rangle_W
+\langle\eta \Psi_B\ \eta \Psi_C\ (\xi_c b_c)\ \eta \Psi_D\ \eta \Psi_A\rangle_W
\Big),\label{fourR}
\end{align}
where, in the last equality, we eliminate the auxiliary field $\Xi$ from
the on-shell external states by imposing the linearized constraint $Q\Xi=\eta\Psi$.
Recalling that the BRST-invariant fermion vertex operator in the $-1/2$ picture 
is included in $\Psi$ in the form $\Psi=\xi_0V^{(-1/2)}$,
we can explicitly map the last expression in (\ref{fourR}) to 
that evaluated on the upper half-plane:\cite{Giddings:1986iy} 
\begin{equation}
\mathcal{A}_{FFFF}= \int_0^1 d\alpha \langle\ \xi_0\ \left(\int d^2z\mu_\alpha(z,\bar{z})b(z)\right) 
\ V_A^{(-\frac{1}{2})}(-\alpha^{-1})\ V_B^{(-\frac{1}{2})}(-\alpha)
V_C^{(-\frac{1}{2})}(\alpha)\ V_D^{(-\frac{1}{2})}(\alpha^{-1})\rangle_{UHP}.
\label{FFFF UHP}
\end{equation}
Here $\mu_\alpha(z,\bar{z})$ is the appropriate Beltrami 
differential for an $\alpha$-dependent parametrization of the 
modulus.\cite{Berkovits:1999bs}

\begin{figure}[htbp]
\begin{minipage}{0.33\hsize}
 \begin{center}
 \includegraphics[width=4cm]{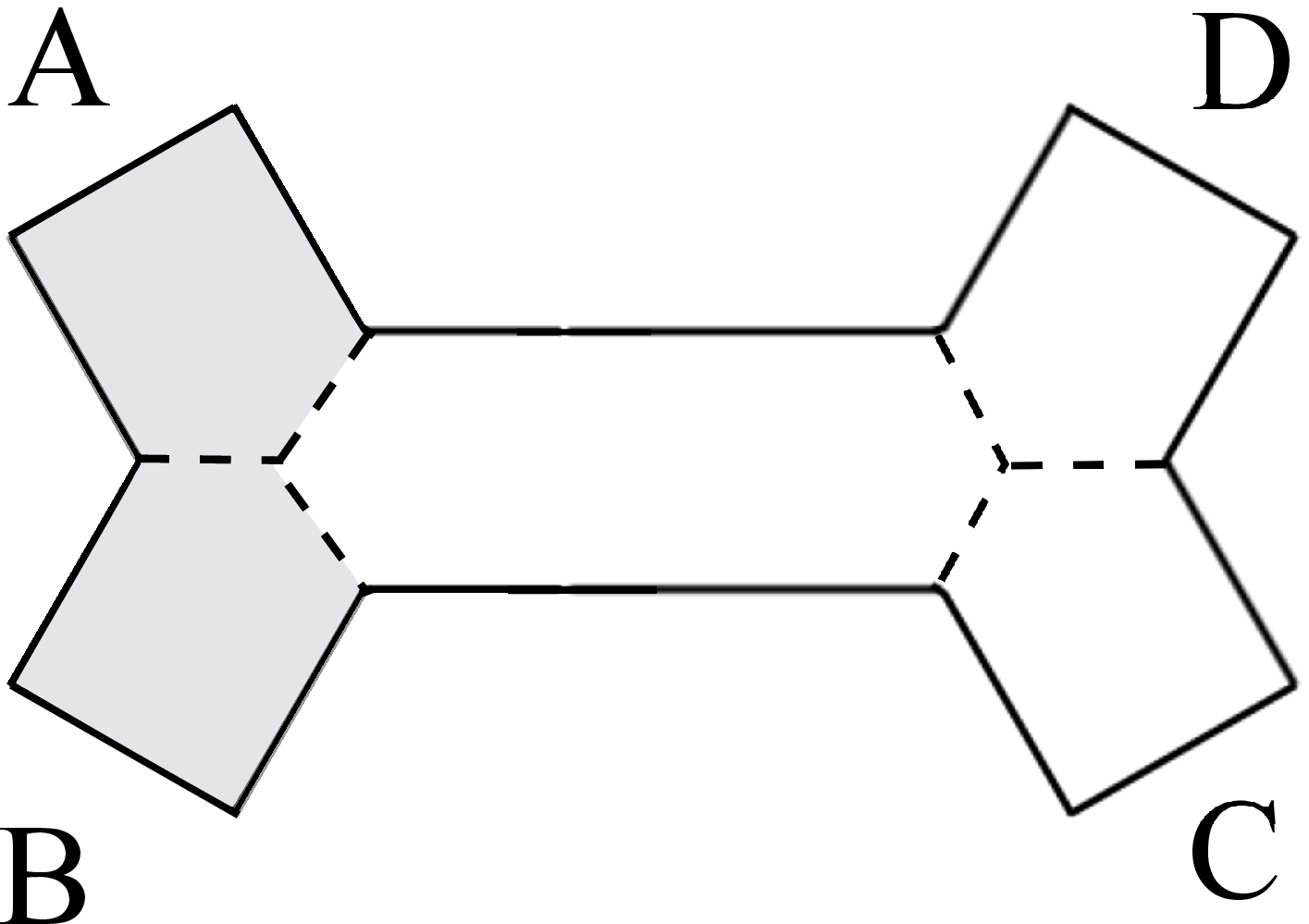}\\
(a)
 \end{center}
\end{minipage}
\begin{minipage}{0.33\hsize}
 \begin{center}
 \includegraphics[width=4cm]{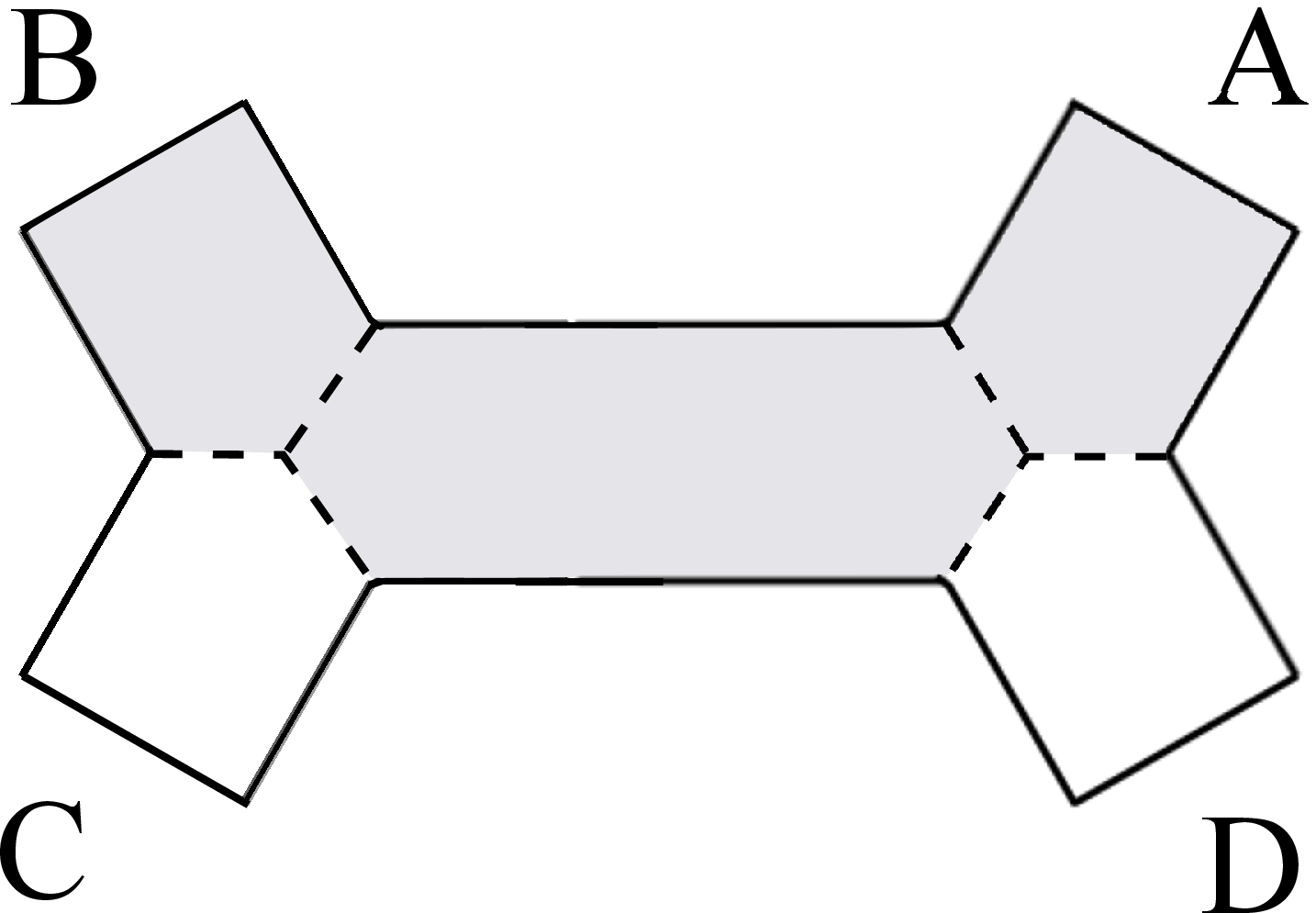}\\
(b)
 \end{center}
\end{minipage}
\begin{minipage}{0.33\hsize}
 \begin{center}
 \includegraphics[width=3cm]{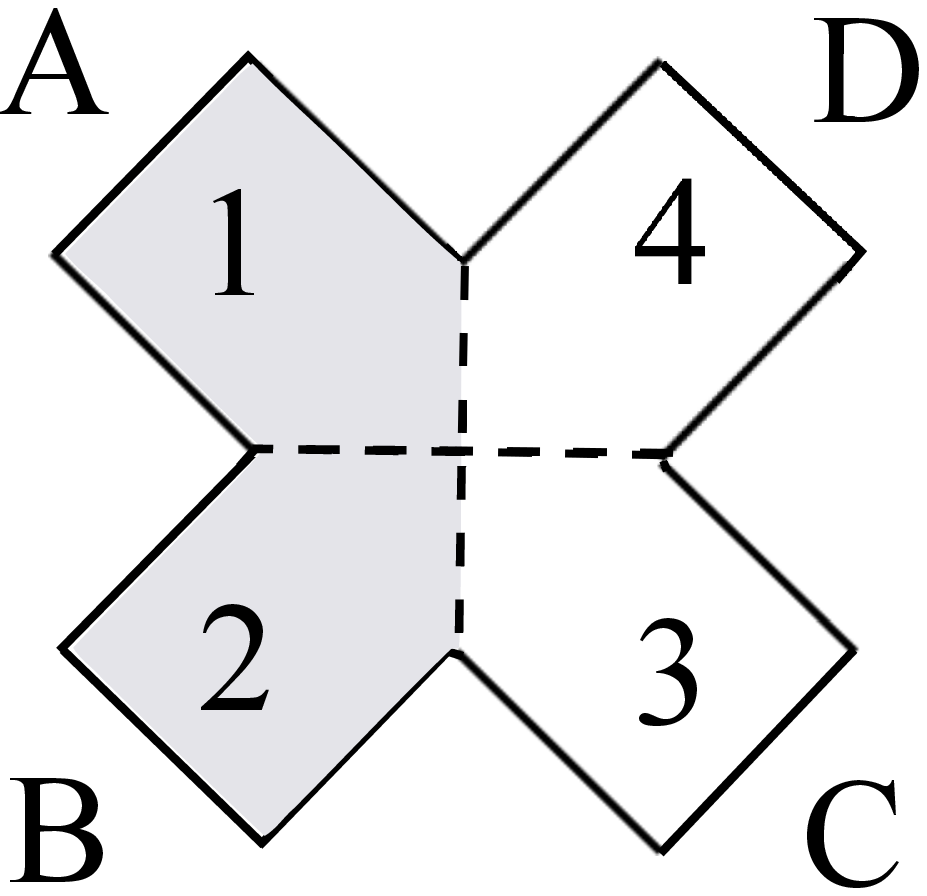}\\
(c)
 \end{center}
\end{minipage}
\vspace{2mm}
\caption{Three Feynman diagrams for two-boson-two-fermion 
amplitude with ordering $FFBB$:\break
(a) $s$-channel, (b) $t$-channel and
(c) four-string interaction.   The fermion legs 
and propagator are colored gray. } 
\label{FFBB}
\end{figure}

The two-fermion-two-boson amplitude with color-ordering $FFBB$
has three contributions from $s$-channel, $t$-channel and four-string interaction 
diagrams in Fig.~\ref{FFBB}.
The $s$-channel contribution is evaluated as
 \begin{align}
 \mathcal{A}_{FFBB}^{(s)} =& 
\frac{1}{2}\cdot\left(-\frac{1}{2}\right)\int_0^\infty d\tau\ 
\langle\Big(Q\Xi_A\ \eta \Psi_B+\eta \Psi_A\ Q\Xi_B\Big)\nonumber\\ 
&\hspace{50mm} \times
(\xi_c b_c)\ \Big(Q\Phi_C\ \eta \Phi_D+\eta \Phi_C\ Q\Phi_D\Big)\rangle_W
\nonumber\\
=&-\frac{1}{2}\int_0^\infty d\tau\
\langle\Big(Q\Xi_A\ \eta\Psi_B\ +\eta\Psi_A\ Q\Xi_B\Big)\
(\xi_c b_c)\ Q\Phi_C\ \eta\Phi_D\rangle_W\nonumber\\
&\hspace{5mm}
+\frac{1}{4}\langle\Big(
Q\Xi_A\ \eta\Psi_B\ +\eta\Psi_A\ Q\Xi_B\Big)\ \Phi_C\ \Phi_D\rangle_W,
\label{s FFBB}
\end{align}
where, in the second equality, we moved the $Q$ and $\eta$ on the 
external bosons
so that each of them 
has single forms $Q\Phi_C$ and $\eta\Phi_D$.
Consequently, the extra contribution, which does not include
the propagator (proper time integration), is produced from the boundary at
$\tau=0$ when we exchange 
the order of the $Q$ and $b_c$, due to the relation 
\begin{align}
 \int_0^\infty d\tau\ \{Q, b_0\} e^{-\tau L_0}=& \int_0^\infty d\tau\ L_0 e^{-\tau L_0}
\nonumber\\
=& -\int_0^\infty d\tau\ \frac{\partial}{\partial\tau} e^{-\tau L_0}.\label{collapse}
\end{align}

The $t$-channel contribution is similarly calculated as
\begin{align}
 \mathcal{A}_{FFBB}^{(t)} =& -\left(-\frac{1}{2}\right)^2\cdot (-2) \int_0^\infty d\tau
\nonumber\\
&\hspace{10mm} \times
\Bigg(\langle Q\Xi_B\ \Phi_C\ (\eta \xi_c b_c Q)\ \Phi_D\ \eta\Psi_A\rangle_W
+\langle \eta\Psi_B\ \Phi_C\ (Q \xi_c b_c \eta)\ \Phi_D\ Q\Xi_A\rangle_W
\Bigg)
\nonumber\\
=& -\frac{1}{2}\int_0^\infty d\tau\Bigg(
\langle Q\Xi_B\ Q\Phi_C\ (\xi_c b_c)\ \eta\Phi_D\ \eta\Psi_A\rangle_W
+\langle\eta\Psi_B\ Q\Phi_C\ (\xi_c b_c)\ \eta\Phi_D\ Q\Xi_A\rangle_W\Bigg)
\nonumber\\
&\hspace{5mm}
-\frac{1}{2}\langle\eta\Psi_A\ Q\Xi_B\ \Phi_C\ \Phi_D\rangle_W,
\label{t FFBB}
\end{align}
which was deformed again so that each external boson has 
the same forms, $Q\Phi_C$ and $\eta\Phi_D$, as in the $s$-channel contribution.
Adding the contribution from the four-string interaction, 
which we can read from (\ref{R 4 vertex}) as
\begin{equation}
 \mathcal{A}_{FFBB}^{(4)} =
-\frac{1}{4}\langle \Big(Q\Xi_A\ \eta \Psi_B-\eta \Psi_A\ Q\Xi_B\Big)\ \Phi_C\ \Phi_D\rangle_W,
\label{contact FFBB}
\end{equation}
the total amplitude becomes
\begin{align}
 \mathcal{A}_{FFBB} =& \mathcal{A}_{FFBB}^{(s)}+\mathcal{A}_{FFBB}^{(t)}+\mathcal{A}_{FFBB}^{(4)}
\nonumber\\
=&-\frac{1}{2}\int_0^\infty d\tau\Bigg(
\langle Q\Xi_A\ \eta\Psi_B\ (\xi_c b_c)\ Q\Phi_C\ \eta\Phi_D\rangle_W
+\langle \eta\Psi_A\ Q\Xi_B\ (\xi_c b_c)\ Q\Phi_C\ \eta\Phi_D\rangle_W
\nonumber\\
&\hspace{2cm}
+\langle \eta\Psi_B\ Q\Phi_C\ (\xi_c b_c)\ \eta\Phi_D\ Q\Xi_A\rangle_W
+\langle Q\Xi_B\ Q\Phi_C\ (\xi_c b_c)\ \eta\Phi_D\ \eta\Psi_A\rangle_W
\Bigg)
\nonumber\\
=& -\int_0^\infty d\tau\Big(
\langle\eta \Psi_A\ \eta \Psi_B\ (\xi_c b_c)\ 
Q\Phi_C\ \eta \Phi_D\rangle_W
+\langle \eta \Psi_B\ Q\Phi_C\ (\xi_c b_c)\ \eta \Phi_D\ \eta \Psi_A\rangle_W
\Big).\label{amp FFBB}
\end{align}
We again eliminated the $\Xi$ in the last expression by the linearized constraint.
It should be noted here that the extra contributions with no propagator 
in (\ref{s FFBB}) and (\ref{t FFBB})
are cancelled by that from the four-string interaction diagram
(\ref{contact FFBB}) without imposing the constraint.
Using the fact that the BRST-invariant NS vertex operator in the $-1$ picture 
is included in the $\Phi$ in a form such as $\eta\Phi=V^{(-1)}$, 
and so $Q\Phi=\{Q,\xi_0\}V^{(-1)}=V^{(0)}$,
we can again map the result (\ref{amp FFBB}) to the form
\begin{equation}
A_{FFBB}= -\int_0^1 d\alpha \langle\ \xi_0\ \left(\int d^2z\mu_\alpha(z,\bar{z})b(z)\right) 
\ V_A^{(-\frac{1}{2})}(-\alpha^{-1})\ V_B^{(-\frac{1}{2})}(-\alpha)
V_C^{(0)}(\alpha)\ V_D^{(-1)}(\alpha^{-1})\rangle_{UHP}.
\label{FFBB UHP}
\end{equation}
This is equivalent to that obtained in the first quantized formulation.\footnote{The overall 
minus sign can be absorbed into the phase convention for how the fermion vertex operator is 
embedded in $\Psi$. This has to be fixed by imposing the reality condition on $\Psi$.}
\begin{figure}[htbp]
\begin{minipage}{0.33\hsize}
 \begin{center}
 \includegraphics[width=4cm]{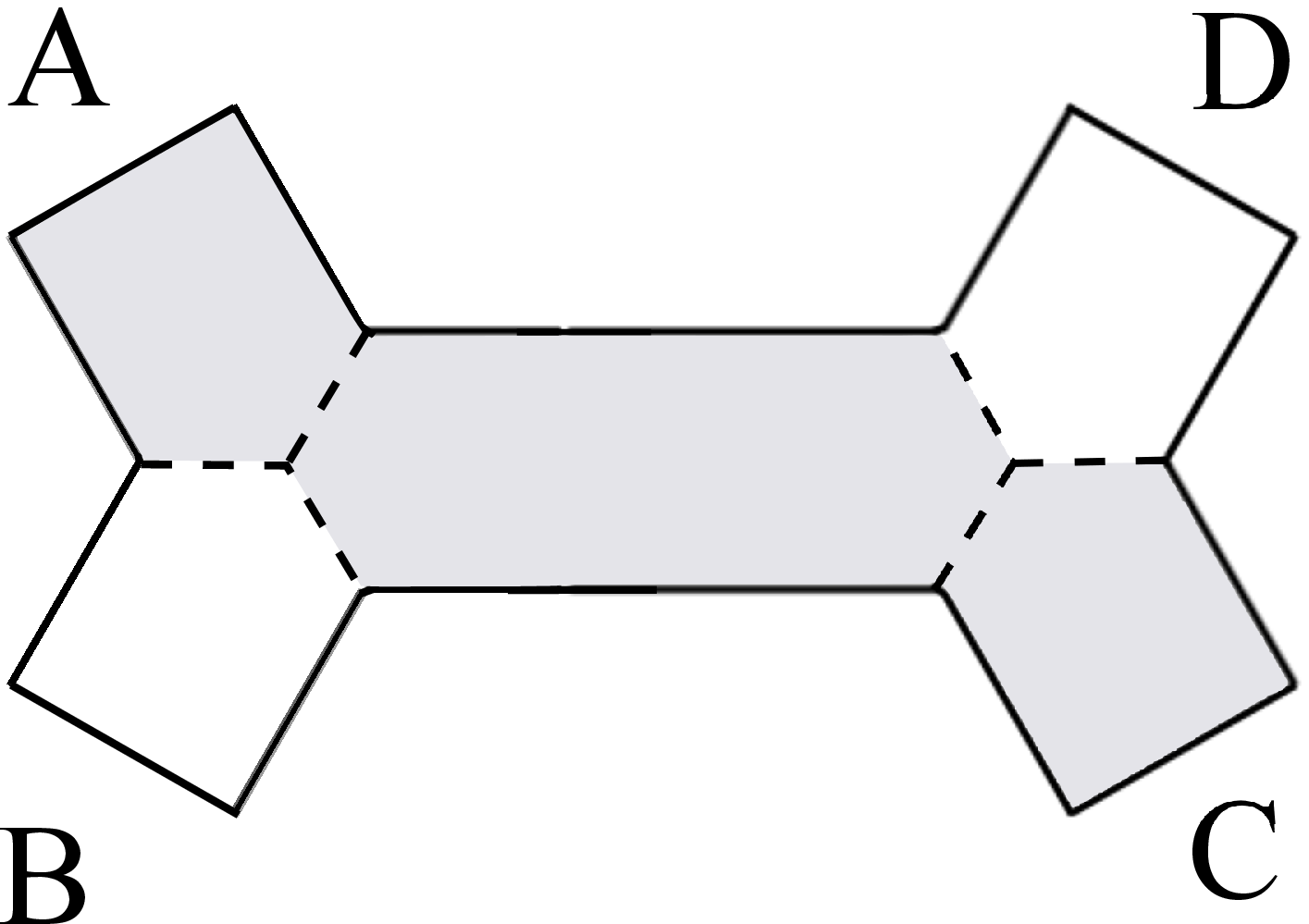}\\
(a)
 \end{center}
\end{minipage}
\begin{minipage}{0.33\hsize}
 \begin{center}
 \includegraphics[width=4cm]{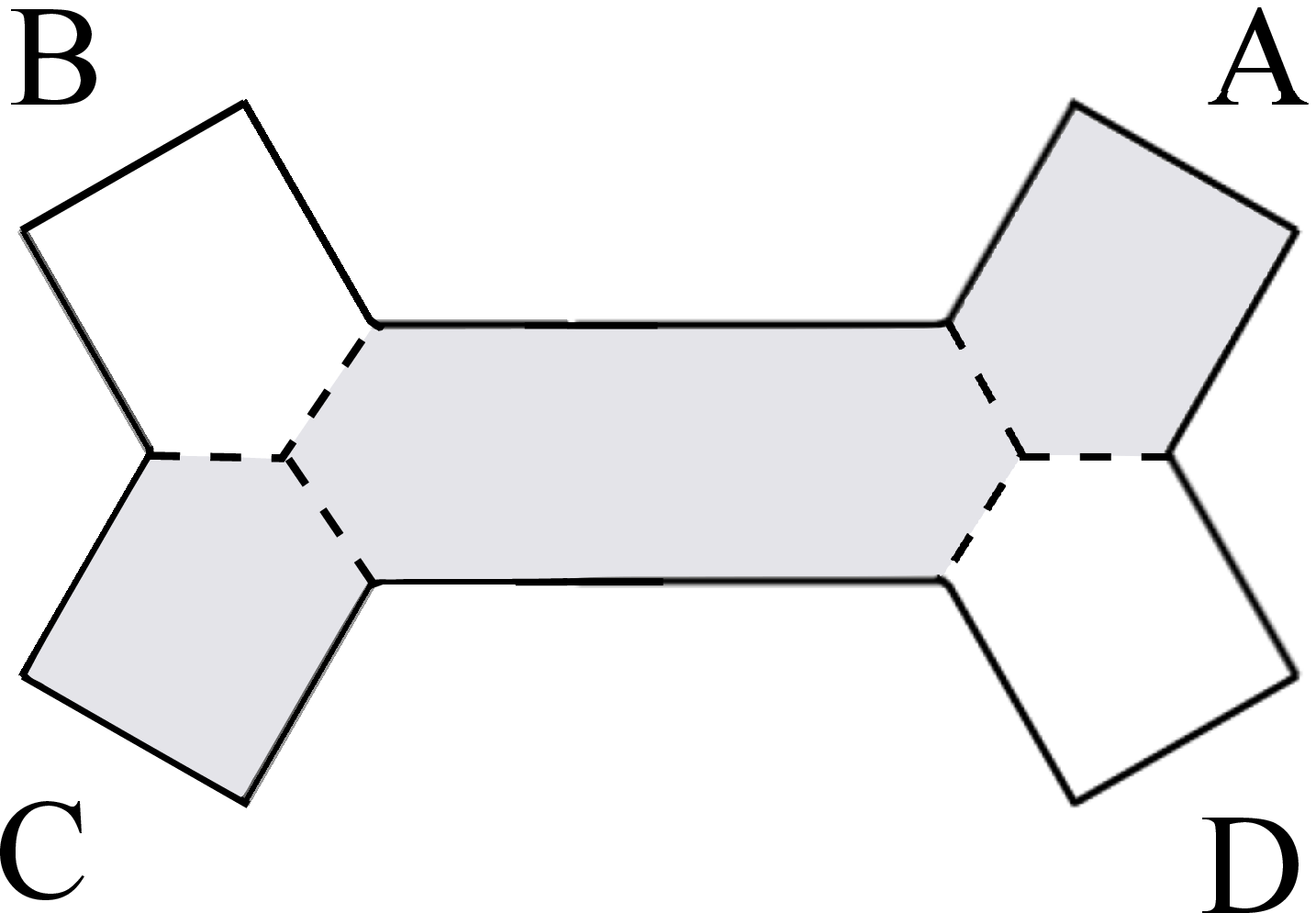}\\
(b)
 \end{center}
\end{minipage}
\begin{minipage}{0.33\hsize}
 \begin{center}
  \includegraphics[width=3cm]{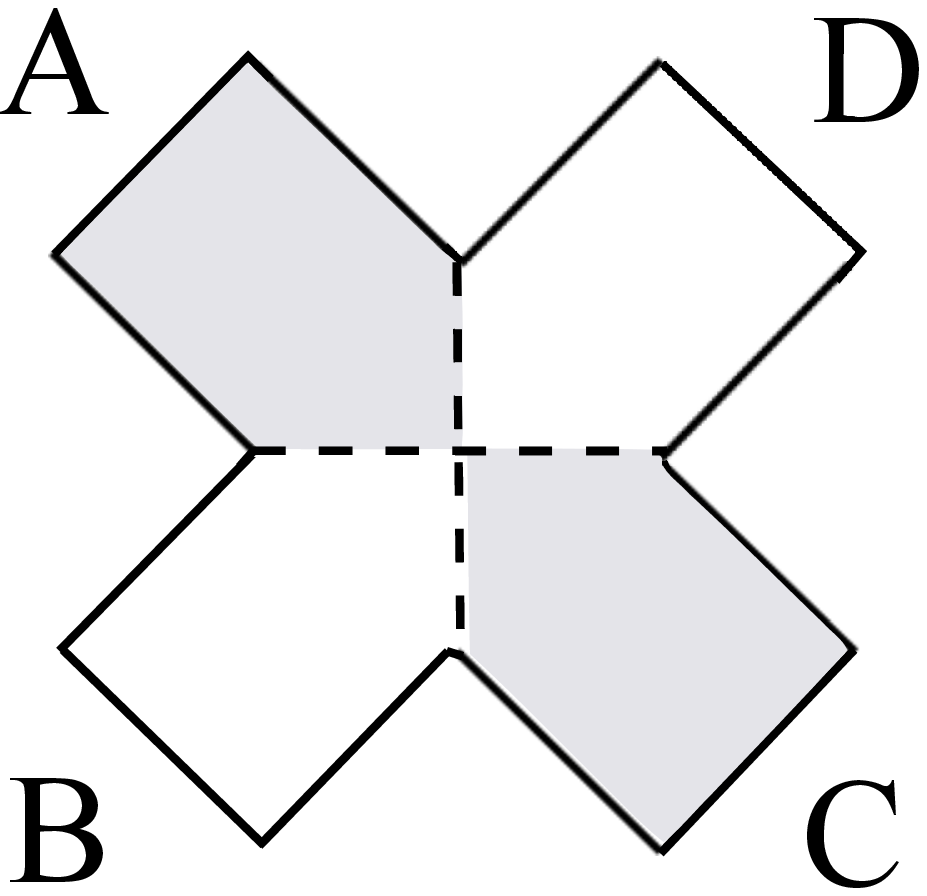}\\
(c)
 \end{center}
\end{minipage}
\vspace{2mm}
\caption{Three Feynman diagrams for two-boson-two-fermion 
amplitude with ordering $FBFB$:\break
(a) $s$-channel, (b) $t$-channel and
(c) four-string interaction.} \label{FBFB}
\end{figure}

The last four-point amplitude is that for two-boson-two-fermion scattering
with ordering $FBFB$. It also has three contributions from
the three diagrams in Fig.~\ref{FBFB}, which can be given by
\begin{subequations}
 \begin{align}
 \mathcal{A}_{FBFB}^{(s)} 
=&-\frac{1}{2}\int_0^\infty d\tau\Bigg(
\langle Q\Xi_A\ Q\Phi_B\ (\xi_c b_c)\ \eta\Psi_C\ \eta\Phi_D\rangle_W
+\langle\eta\Psi_A\ Q\Phi_B\ (\xi_c b_c)\ Q\Xi_C\ \eta\Phi_D\rangle_W\Bigg)
\nonumber\\
&\hspace{5mm}
-\frac{1}{2}\langle Q\Xi_A\ \Phi_B\ \eta\Psi_C\ \Phi_D\rangle_W,
\label{s FBFB}\\
 \mathcal{A}_{FBFB}^{(t)} 
=&-\frac{1}{2}\int_0^\infty d\tau\Bigg(
\langle Q\Phi_B\ Q\Xi_C\ (\xi_c b_c)\ \eta\Phi_D\ \eta\Psi_A\rangle_W
+\langle Q\Phi_B\ \eta\Psi_C\ (\xi_c b_c)\ \eta\Phi_D\ Q\Xi_A\rangle_W
\Bigg)
\nonumber\\
&\hspace{5mm}
+\frac{1}{2}\langle \eta\Psi_A\ \Phi_B\ Q\Xi_C\ \Phi_D\rangle_W,
\label{t FBFB}\\
 \mathcal{A}_{FBFB}^{(4)} =& \frac{1}{2}\Bigg(
\langle Q\Xi_A\ \Phi_B\ \eta \Psi_C\ \Phi_D\rangle_W
-\langle \eta \Psi_A\ \Phi_B\ Q\Xi_C\ \Phi_D\rangle_W
\Bigg),\label{four FBFB}
\end{align}
\end{subequations}
by deforming so that the external boson states have the common forms
$Q\Phi_B$ and $\eta\Phi_D$.
The extra contributions in (\ref{s FBFB}) and (\ref{t FBFB}) are again 
cancelled by that from the four-string interaction (\ref{four FBFB}) at this stage.
In consequence the amplitude becomes the well-known form:
\begin{align}
 \mathcal{A}_{FBFB} =& \mathcal{A}_{FBFB}^{(s)} + \mathcal{A}_{FBFB}^{(t)} + \mathcal{A}_{FBFB}^{(4)} 
\nonumber\\
=& -\frac{1}{2}\int_0^\infty d\tau\Bigg(
\langle Q\Xi_A\ Q\Phi_B\ (\xi_c b_c)\ \eta \Psi_C\ \eta \Phi_D \rangle_W
+\langle\eta \Psi_A\ Q\Phi_B\ (\xi_c b_c)\ Q\Xi_C\ \eta \Phi_D \rangle_W\nonumber\\
&\hspace{20mm}
+\langle Q\Phi_B\ \eta \Psi_C\ (\xi_c b_c)\ \eta \Phi_D\ Q\Xi_A\rangle_W
+\langle Q\Phi_B\ Q\Xi_C\ (\xi_c b_c)\ \eta\Phi_D\ \eta \Psi_A\rangle_W
\Bigg)\nonumber\\
=&-\int_0^\infty d\tau\Bigg(
\langle \eta\Psi_A\ Q\Phi_B\ (\xi_c b_c)\ \eta\Psi_C\ \eta\Phi_D\rangle_W
+\langle Q\Phi_B\ \eta\Psi_C\ (\xi_c b_c)\ \eta\Phi_D\ \eta\Psi_A\rangle_W
\Bigg),\nonumber\\
=& -\int_0^1 d\alpha \langle\ \xi_0\ \left(\int d^2z\mu_\alpha(z,\bar{z})b(z)\right) 
\ V_A^{(-\frac{1}{2})}(-\alpha^{-1})\ V_B^{(0)}(-\alpha)
V_C^{(-\frac{1}{2})}(\alpha)\ V_D^{(-1)}(\alpha^{-1})\rangle_{UHP}.
\end{align}
In this way the new Feynman rules also give the same on-shell four-point
amplitudes as those obtained by the self-dual rules.

In general, one can see that the two sets of rules give
different results in the contribution from the diagram with either 
(i) at least two fermion propagators or
(ii) two-fermion-even-boson interaction at least one of whose fermions 
is connected to the propagator. 
%
%
The difference in case (i) comes from the form of the fermion propagators.
If the diagram has two fermion propagators, the self-dual rule using the propagator
(\ref{sd R propagator}) gives a contribution of the form
\begin{equation}
 \mathcal{A} \sim\ \langle \cdots (Q\Pi_R\eta+\eta\Pi_RQ) \cdots 
(Q\Pi_R\eta+\eta\Pi_RQ) \cdots \rangle_W.
\end{equation}
If we follow the new Feynman rules, on the other hand, the contribution of the same
diagram becomes
\begin{equation}
 \mathcal{A} \sim\ \langle \cdots Q\Pi_R\eta \cdots Q\Pi_R\eta\cdots\rangle_W
+\langle \cdots \eta\Pi_R Q \cdots \eta\Pi_R Q\cdots\rangle_W,
\end{equation}
using the fermion interactions,
\begin{equation}
 S_R^{(n+2)} =\ -\frac{1}{2}\sum_{m=0}^n \frac{(-1)^m}{(n-m)!m!}
\langle (Q\Xi)\Phi^{n-m}(\eta\Psi)\Phi^m\rangle,\label{R interaction}
\end{equation}
and the (off-diagonal) R propagator (\ref{R propagator}).\footnote{
Similarly, it is easy to see that the two rules give the same contributions
if the diagram has only one R propagator.}
In case (ii), the difference is due to the fact that
(\ref{R interaction}) can be rewritten as
\begin{equation}
 S_R^{(n+2)} =\ -\frac{1}{4}\sum_{m=0}^n \frac{(-1)^m}{(n-m)!m!}\Big(
\langle (Q\Xi)\Phi^{n-m}(\eta\Psi)\Phi^m\rangle
-(-1)^n
\langle (\eta\Psi)\Phi^{n-m}(Q\Xi)\Phi^m\rangle
\Big).
\end{equation}
Therefore, the two-fermion-even-boson vertices for the self-dual rules vanish,
as previously mentioned. In the new Feynman rules, in contrast, 
the two-fermion-even-boson interactions can contribute 
if at least one of the two fermions is connected to the propagator.
%
We will next show that these differences 
in fact improve the discrepancy in the five-point amplitudes.

\subsection{Five-point amplitudes with external fermions}

Then we calculate the on-shell five-point amplitudes with external 
fermions. We follow the convention above; 
\textit{i.e.}, we label the five external strings by $A,\ B,\ C,\ D$, and $E$, 
and omit to explicitly indicate the numbers, depicted
in Figs.~\ref{Feynman FFFFB 333}(a), \ref{Feynman FFFFB 34}(a), 
and \ref{Feynman FFBBB 5}, by arranging the external states in order
of these numbers from the left.

\subsubsection{Four-fermion-one-boson amplitude}

\begin{figure}[htbp]
\begin{minipage}{0.33\hsize}
 \begin{center}
 \includegraphics[width=5cm]{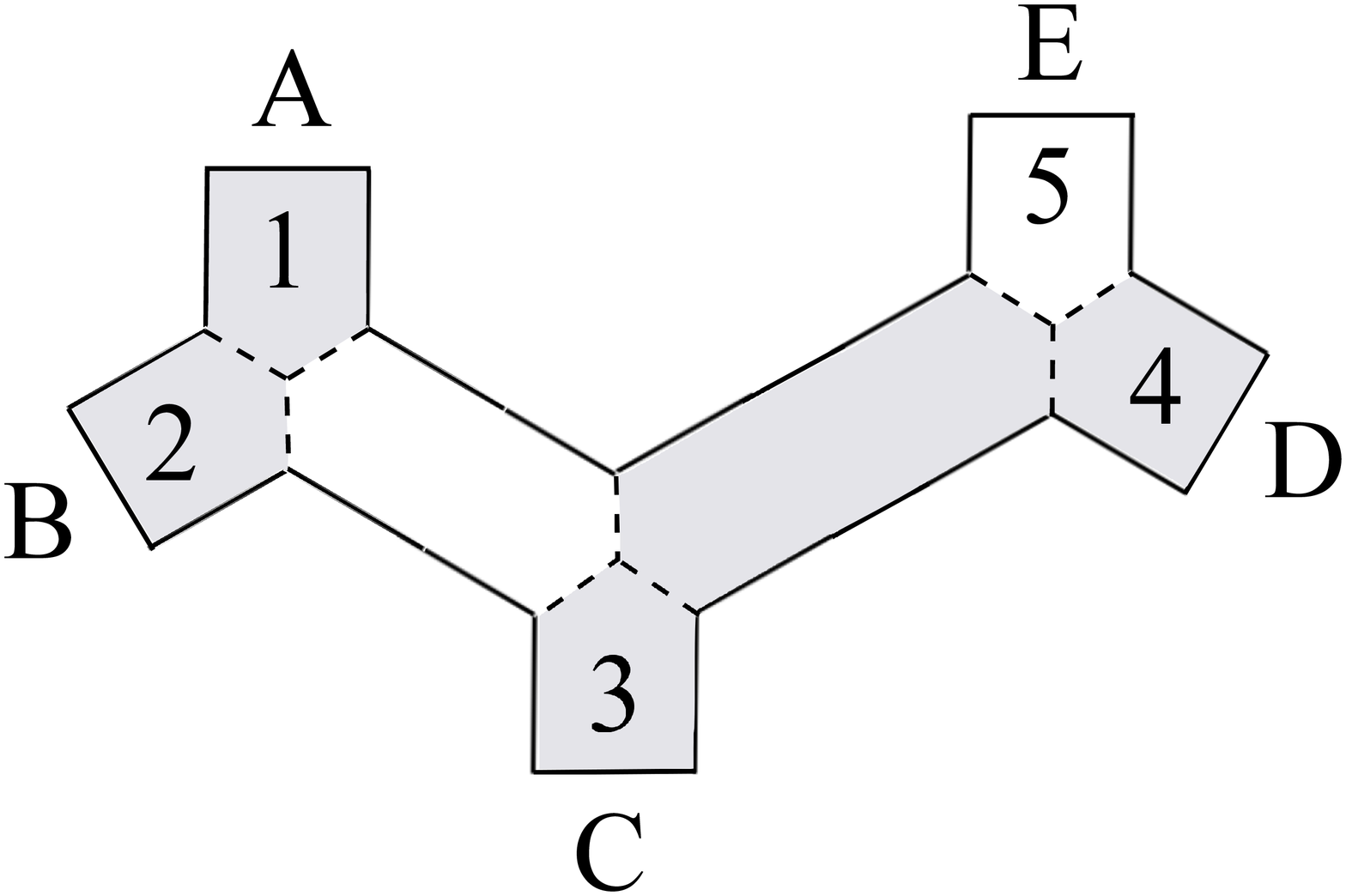}\\
(a)
 \end{center}
\end{minipage}
\begin{minipage}{0.33\hsize}
 \begin{center}
 \includegraphics[width=5cm]{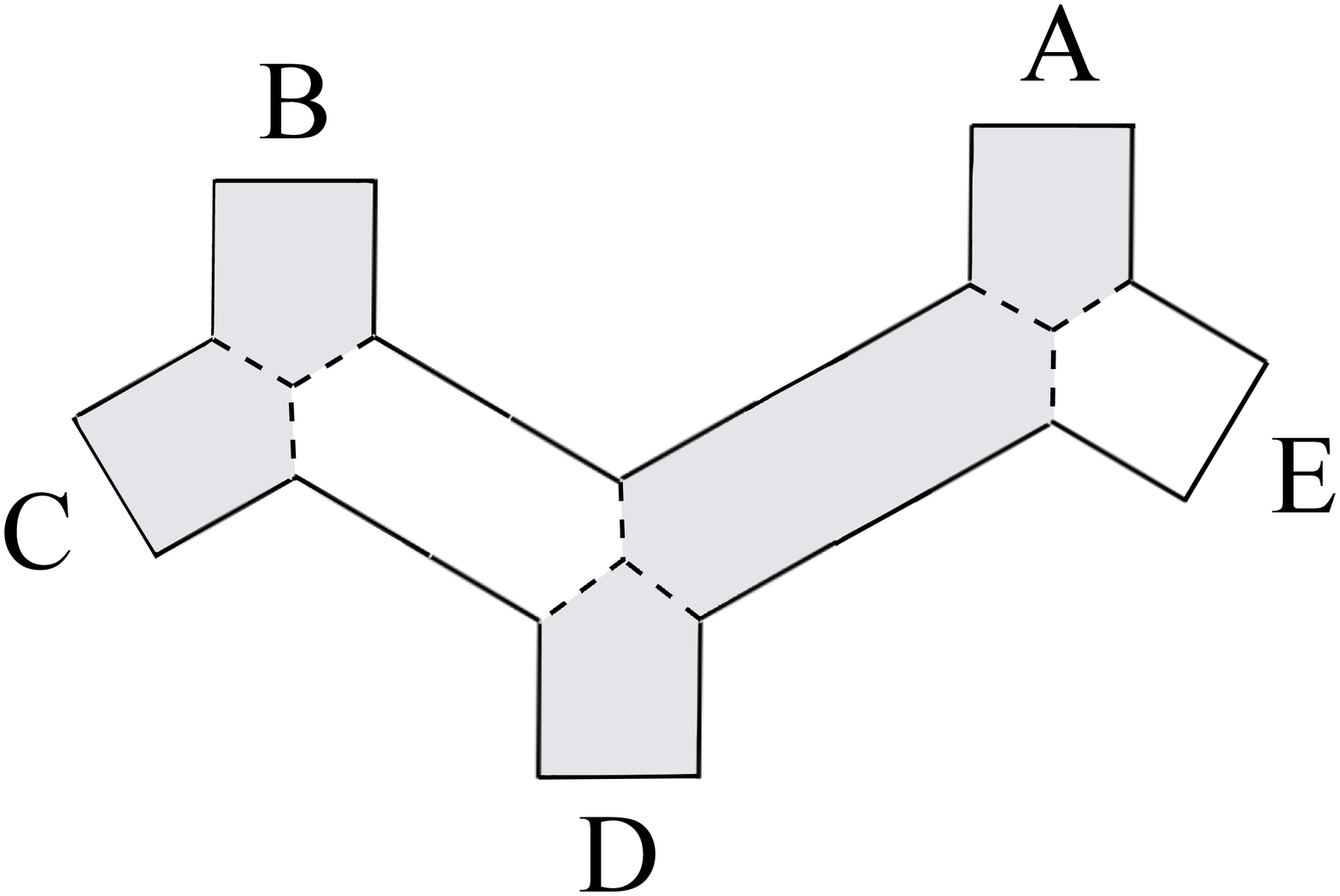}\\
(b)
 \end{center}
\end{minipage}
\begin{minipage}{0.33\hsize}
 \begin{center}
 \includegraphics[width=5cm]{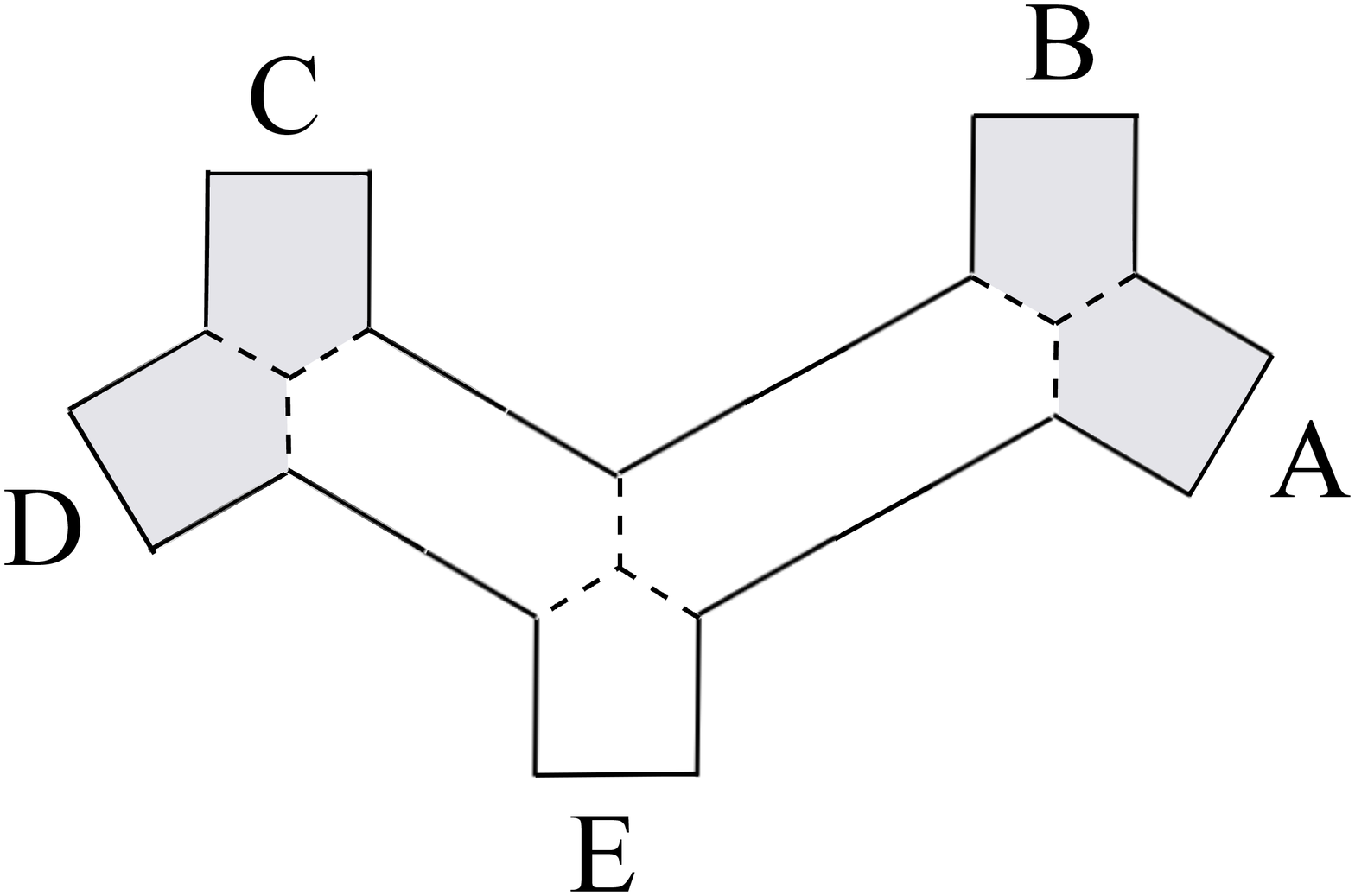} \\
(c)
 \end{center}
\end{minipage}
\begin{minipage}{0.15\hsize}
\mbox{}
\end{minipage}
\begin{minipage}{0.33\hsize}
 \begin{center}
 \includegraphics[width=5cm]{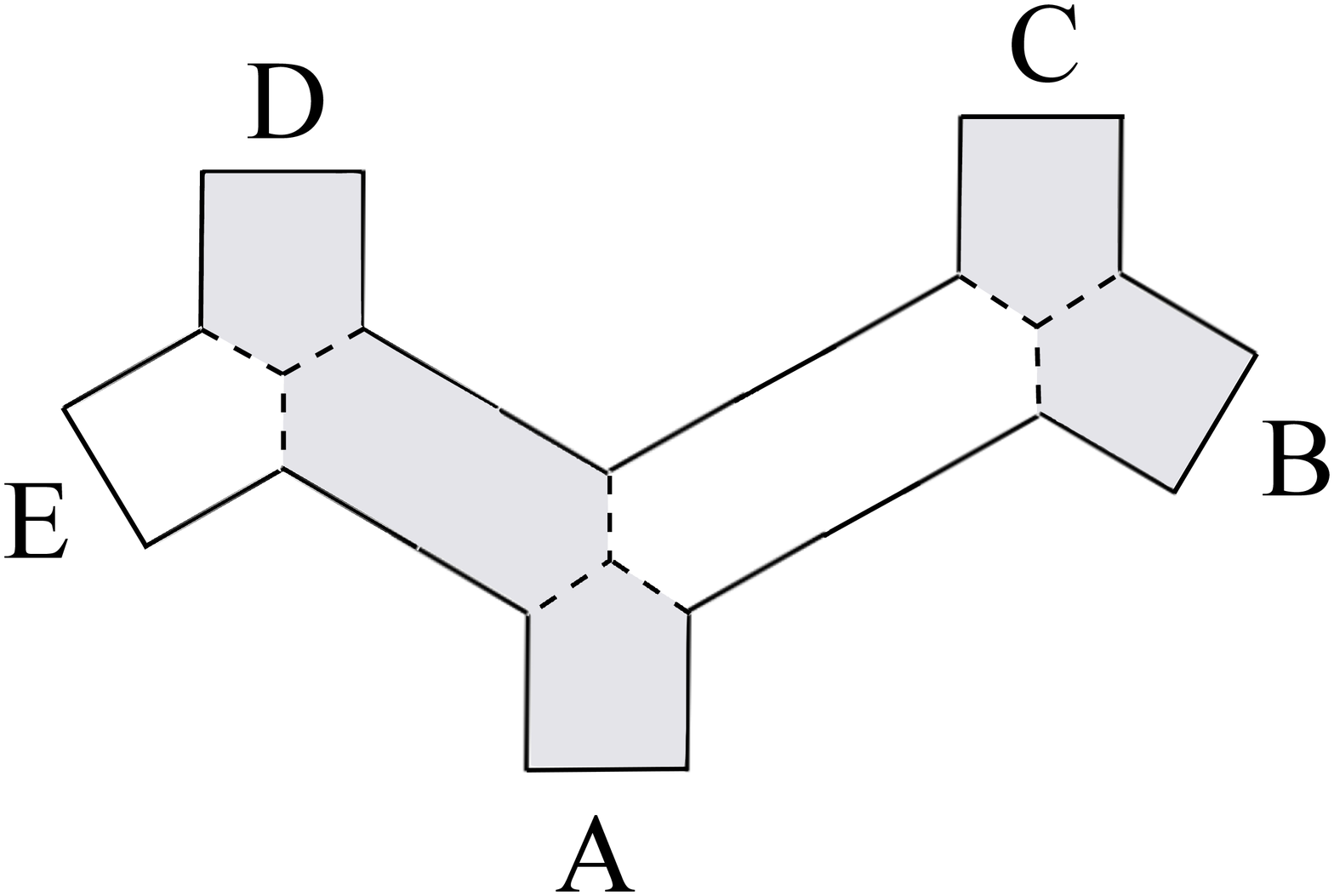} \\
(d)
 \end{center}
\end{minipage}
\begin{minipage}{0.33\hsize}
 \begin{center}
 \includegraphics[width=5cm]{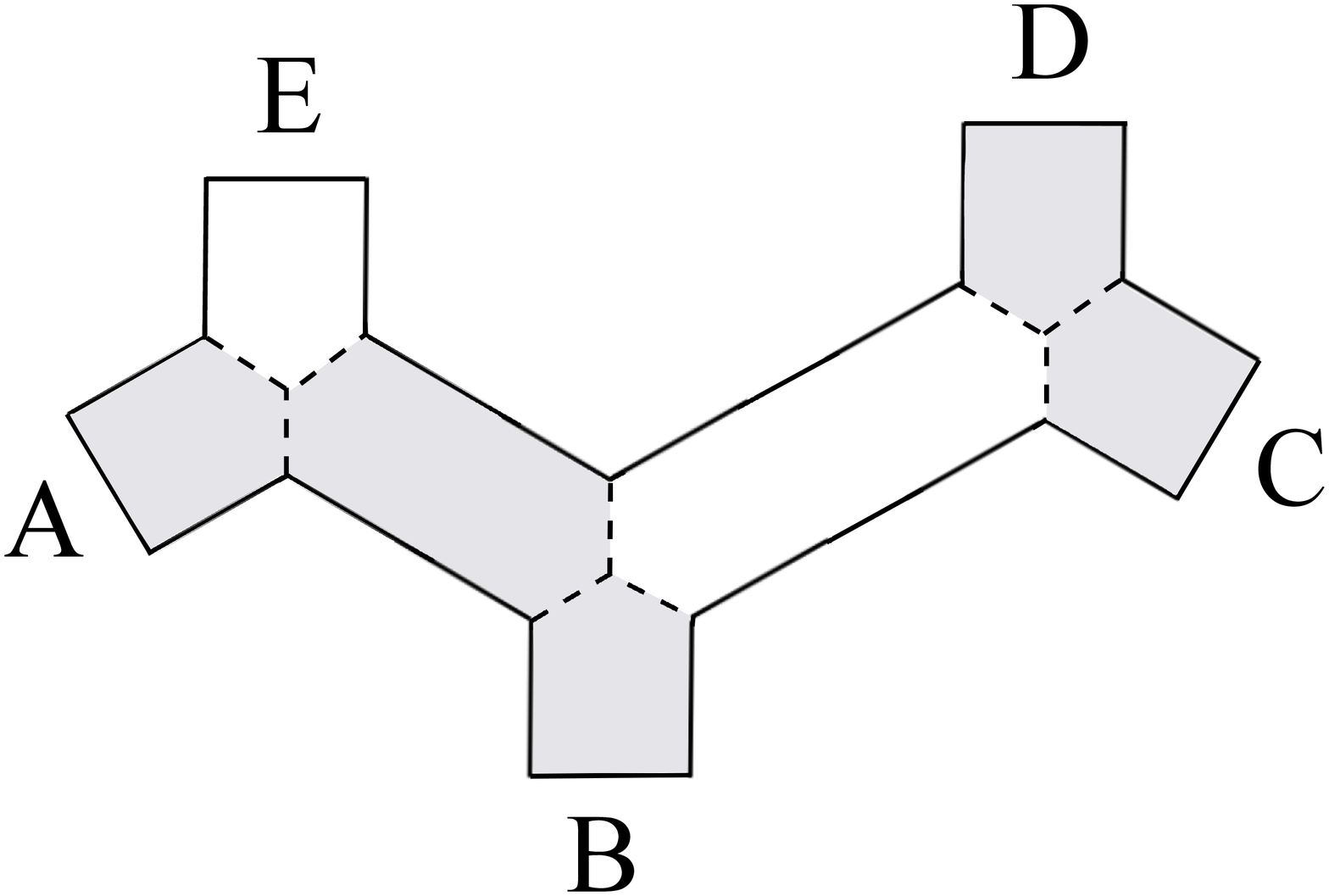} \\
(e)
 \end{center}
\end{minipage}
\vspace{2mm}
\caption{Five 2P Feynman diagrams for four-fermion-one-boson amplitude. 
}\label{Feynman FFFFB 333}
 \end{figure}

\begin{figure}[htbp]
 \begin{center}
  \includegraphics[width=7.5cm]{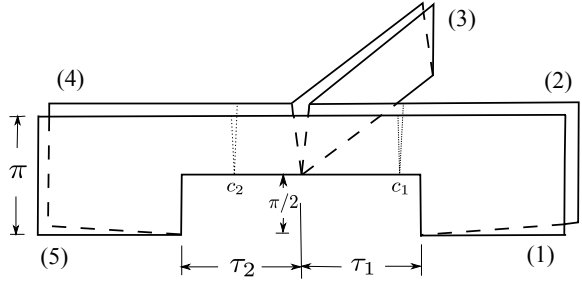}
 \end{center}
\caption{The Witten diagram for five-point amplitudes. Each of the five legs,
to which the numbers from (1) to (5) are assigned,
should be read as semi-infinite strips.}\label{Witten5}
\end{figure}

Let us begin with the calculation of the four-fermion-one-boson amplitude.
The dominant contributions come from the diagrams containing 
three three-string vertices and two propagators, which we call 
in this paper the two-propagator (2P) diagrams.
There are five different channels for color-ordered amplitudes 
as in Fig.~\ref{Feynman FFFFB 333}.
The contribution from the first diagram Fig.~\ref{Feynman FFFFB 333}(a)
is given by
\begin{align}
\mathcal{A}_{FFFFB}^{(2\textrm{P})(a)} =& -\left(\frac{1}{2}\right)^3\cdot(-2)
\int_0^\infty d\tau_1\int_0^\infty d\tau_2\
\nonumber\\
&\hspace{0mm} \times
\Bigg(\langle\Big(Q\Xi_A\ \eta \Psi_B+\eta \Psi_A\ Q\Xi_B\Big)\
(\xi_{c_1} b_{c_1})\
Q\Xi_C\ (\eta  \xi_{c_2} b_{c_2} Q)\ \eta \Psi_D\ \Phi_E\rangle_W\nonumber\\
&\hspace{5mm}
+\langle\Big(Q\Xi_A\ \eta \Psi_B+\eta \Psi_A\ Q\Xi_B\Big)\ (\xi_{c_1} b_{c_1})\
\eta \Psi_C\ (Q \xi_{c_2} b_{c_2} \eta )\ Q\Xi_D\ \Phi_E\Big)\rangle_W
\Bigg),
\end{align}
where the correlation is evaluated as the conformal field theory on the Witten 
diagram depicted in Fig.~\ref{Witten5}. 
$(\xi_{c_1}, b_{c_1})$ and $(\xi_{c_2}, b_{c_2})$ denote the corresponding fields integrated
along the paths $c_1$ and $c_2$, respectively. 
Under the on-shell conditions, all the $Q$ and $\eta$ can be moved 
so as to act on the external states 
(without exchanging the order of $\xi$ and $Q$):
\begin{subequations}\label{Amp FFFFB 333} 
\begin{align}
 \mathcal{A}_{FFFFB}^{(2\textrm{P})(a)} 
=&-\frac{1}{4}\int_0^\infty d^2\tau\
\Bigg(
\langle \Big(Q\Xi_A\ \eta \Psi_B+\eta \Psi_A\ Q\Xi_B\Big)\
(\xi_{c_1} b_{c_1})\ Q\Xi_C\ b_{c_2}\ \eta \Psi_D\ Q\Phi_E\rangle_W 
\nonumber\\
&\hspace{25mm}
+\langle \Big(Q\Xi_A\ \eta \Psi_B+\eta \Psi_A\ Q\Xi_B\Big)\
(\xi_{c_1} b_{c_1})\
\eta \Psi_C\ b_{c_2}\ Q\Xi_D\ Q\Phi_E\rangle_W\Bigg)\nonumber\\
&
-\frac{1}{4}\int_0^\infty d\tau\Bigg(
\langle \Big(Q\Xi_A\ \eta \Psi_B+\eta \Psi_A\ Q\Xi_B\Big)\
(\xi_c b_c)\ 
\eta \Psi_C\ Q\Xi_D\ \Phi_E\rangle_W\nonumber\\
&\hspace{25mm}
+\langle Q\Xi_D\ \Phi_E\ (\xi_c b_c)\
\Big(Q\Xi_A\ \eta \Psi_B+\eta \Psi_A\ Q\Xi_B\Big)\ \eta \Psi_C\ \rangle_W\Bigg),
\end{align}
where we used a shorthand notation
\begin{equation}
\int_0^\infty d^2\tau \equiv \int_0^\infty d\tau_1\int_0^\infty d\tau_2.\nonumber
\end{equation}
The extra terms with less (one) propagator were produced from the boundary
at $\tau_1=0$ or $\tau_2=0$ through the relation
(\ref{collapse}) by exchanging the order of the $b_{c_i}$ and $Q$. 
%
The contributions from the other four channels depicted in 
Figs.~\ref{Feynman FFFFB 333}(b)-(e) can similarly be evaluated as
\begin{align}
 \mathcal{A}_{FFFFB}^{(2\textrm{P})(b)} 
=&-\frac{1}{4}\int_0^\infty d^2\tau\
\langle \Big(Q\Xi_B\ \eta \Psi_C+\eta \Psi_B\ Q\Xi_C\Big)\ (\xi_{c_1} b_{c_1})\nonumber\\
&\hspace{40mm}\times  
\Big(Q\Xi_D\ b_{c_2}\ Q\Phi_E\ \eta \Psi_A
+\eta \Psi_D\ b_{c_2}\ Q\Phi_E\ Q\Xi_A\Big)
\rangle_W\nonumber\\
&+\frac{1}{4}\int_0^\infty d\tau\Bigg(
\langle \Big(Q\Xi_B\ \eta \Psi_C+\eta \Psi_B\ Q\Xi_C\Big)\ (\xi_c b_c)\ 
\eta \Psi_D\ \Phi_E\ Q\Xi_A\rangle_W\nonumber\\
&\hspace{25mm}
+\langle \Phi_E\ Q\Xi_A\ (\xi_c b_c)\
\Big(Q\Xi_B\ \eta \Psi_C+\eta \Psi_B\ Q\Xi_C\Big)\ \eta \Psi_D\  
\rangle_W\Bigg),\\
 \mathcal{A}_{FFFFB}^{(2\textrm{P})(c)} 
=&-\frac{1}{4}\int_0^\infty d^2\tau\
\langle \Big(Q\Xi_C\ \eta \Psi_D+\eta \Psi_C\ Q\Xi_D\Big)\ \nonumber\\
&\hspace{40mm}\times (\xi_{c_1} b_{c_1})\ 
Q\Phi_E\ b_{c_2}\ \Big(Q\Xi_A\ \eta \Psi_B+\eta \Psi_A\ Q\Xi_B\Big)\rangle_W\nonumber\\
&
-\frac{1}{8}\int_0^\infty d\tau\Bigg(
\langle \Big(Q\Xi_C\ \eta \Psi_D+\eta \Psi_C\ Q\Xi_D\Big)\nonumber\\ 
&\hspace{40mm}
\times (\xi_c b_c)\ \Phi_E\ 
\Big(Q\Xi_A\ \eta \Psi_B+\eta \Psi_A\ Q\Xi_B\Big)\rangle_W\nonumber\\
&\hspace{20mm}
-\langle \Big(Q\Xi_A\ \eta \Psi_B+\eta \Psi_A\ Q\Xi_B\Big)\ \nonumber\\
&\hspace{40mm}\times (\xi_c b_c)\ \Big(Q\Xi_C\ \eta \Psi_D+\eta \Psi_C\
 Q\Xi_D\Big)\ \Phi_E\rangle_W\Bigg),\\
 \mathcal{A}_{FFFFB}^{(2\textrm{P})(d)} 
=&-\frac{1}{4}\int_0^\infty d^2\tau\ \Bigg(
\langle \eta \Psi_D\ \Phi_E\ (\xi_{c_1} b_{c_1})\ Q\Xi_A\nonumber\\
&\hspace{35mm}
\times b_{c_2}\ 
\Big(Q\Xi_B\ \eta \Psi_C+\eta \Psi_B\ Q\Xi_C\Big)\rangle_W\nonumber\\
&\hspace{20mm}
+\langle Q\Xi_D\ \Phi_E\ (\xi_{c_1} b_{c_1})\ \eta \Psi_A\nonumber\\ 
&\hspace{35mm}
\times b_{c_2}\ 
\Big(Q\Xi_B\ \eta \Psi_C+\eta \Psi_B\ Q\Xi_C\Big)\rangle_W\Bigg)\nonumber\\
&+\frac{1}{4}\int_0^\infty d\tau\Bigg(
\langle Q\Xi_D\ \Phi_E\ (\xi_c b_c)\ \eta \Psi_A\  
\Big(Q\Xi_B\ \eta \Psi_C+\eta \Psi_B\ Q\Xi_C\Big)\rangle_W\nonumber\\
&\hspace{20mm}
-\langle \Big(Q\Xi_B\ \eta \Psi_C+\eta \Psi_B\ Q\Xi_C\Big)\ 
(\xi_c b_c)\ Q\Xi_D\ \Phi_E\ \eta \Psi_A\rangle_W\Bigg),\\
 \mathcal{A}_{FFFFB}^{(2\textrm{P})(e)} 
=&-\frac{1}{4}\int_0^\infty d^2\tau\
\Bigg(
\langle Q\Phi_E\ Q\Xi_A\ (\xi_{c_1} b_{c_1})\ \eta \Psi_B\
b_{c_2}\ 
\Big(Q\Xi_C\ \eta \Psi_D+\eta \Psi_C\ Q\Xi_D\Big)\rangle_W\nonumber\\
&\hspace{25mm}
+\langle Q\Phi_E\ \eta \Psi_A\ (\xi_{c_1} b_{c_1})\ Q\Xi_B\
b_{c_2}\ \Big(Q\Xi_C\ \eta \Psi_D+\eta \Psi_C\ Q\Xi_D\Big)\rangle_W
\Bigg)\nonumber\\
&
-\frac{1}{4}\int_0^\infty d\tau\Bigg(
\langle \Phi_E\ Q\Xi_A\ (\xi_c b_c)\ 
\eta \Psi_B\ \Big(Q\Xi_C\ \eta \Psi_D+\eta \Psi_C\ Q\Xi_D\Big)\rangle_W\nonumber\\
&\hspace{20mm}
-\langle \Big(Q\Xi_C\ \eta \Psi_D+\eta \Psi_C\ Q\Xi_D\Big)\ 
(\xi_c b_c)\ \Phi_E\ Q\Xi_A\ \eta \Psi_B\rangle_W\Bigg).
\end{align}
\end{subequations}
Here the sum of the extra contributions with less propagator becomes
\begin{align}
\sum_{i=a}^e\mathcal{A}_{FFFFB}^{(2\textrm{P})(i)}\Big|_\textrm{extra}
=&\frac{1}{8}\int_0^\infty d\tau \Bigg(
\langle\Big(Q\Xi_A\ \eta\Psi_B+\eta\Psi_A\ Q\Xi_B\Big)
\nonumber\\
&\hspace{40mm} \times
(\xi_c b_c)\ \Big(Q\Xi_C\ \eta\Psi_D-\eta\Psi_C\ Q\Xi_D\Big)\ \Phi_E\rangle_W
\nonumber\\
&\hspace{15mm}
-2\langle\Big(Q\Xi_B\ \eta\Psi_C+\eta\Psi_B\ Q\Xi_C\Big)
\nonumber\\
&\hspace{40mm} \times
(\xi_c b_c)\ \Big(Q\Xi_D\ \Phi_E\ \eta\Psi_A-\eta\Psi_D\ \Phi_E\ Q\Xi_A\Big)\rangle_W
\nonumber\\
&\hspace{15mm}
+\langle \Big(Q\Xi_C\ \eta\Psi_D+\eta\Psi_C\ Q\Xi_D\Big)
\nonumber\\
&\hspace{40mm} \times
(\xi_c b_c)\ \Phi_E\ \Big(Q\Xi_A\ \eta\Psi_B-\eta\Psi_A\ Q\Xi_B\Big)\rangle_W
\nonumber\\
&\hspace{15mm}
-2\langle Q\Xi_D\Phi_E\ (\xi_c b_c) 
\nonumber\\
&\hspace{30mm} \times
\Big(Q\Xi_A\ \eta\Psi_B\ \eta\Psi_C
-\eta\Psi_A\ \eta\Psi_B\ Q\Xi_C\Big)\rangle_W
\nonumber\\
&\hspace{15mm}
+2\langle \Phi_E\ Q\Xi_A\ (\xi_c b_c)
\nonumber\\
&\hspace{30mm} \times
\Big(Q\Xi_B\ \eta\Psi_C\ \eta\Psi_D
-\eta\Psi_B\ \eta\Psi_C\ Q\Xi_D\Big)\rangle_W\Bigg),
\label{FFFFB 333 extra}
\end{align}
which has the identical structure to the contributions from
the diagrams which we will consider next. This extra contribution 
(\ref{FFFFB 333 extra}) vanishes under the constraint, 
but we keep them for a while.
 \begin{figure}[htbp]
\begin{minipage}{0.125\hsize}
 \mbox{}
\end{minipage}
\begin{minipage}{0.25\hsize}
 \begin{center}
 \includegraphics[width=3cm]{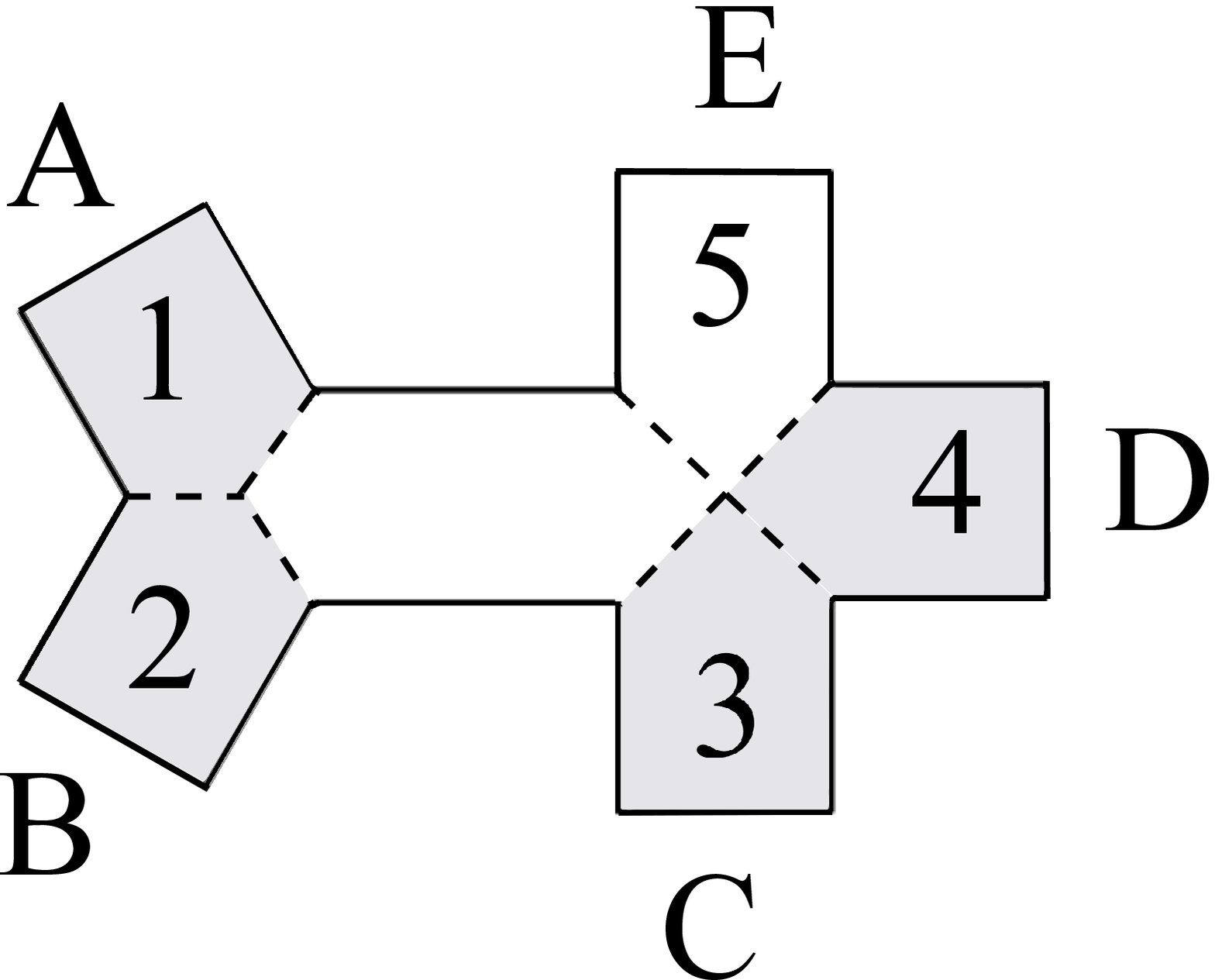}\\
(a)
 \end{center}
\end{minipage}
\begin{minipage}{0.25\hsize}
 \begin{center}
 \includegraphics[width=3cm]{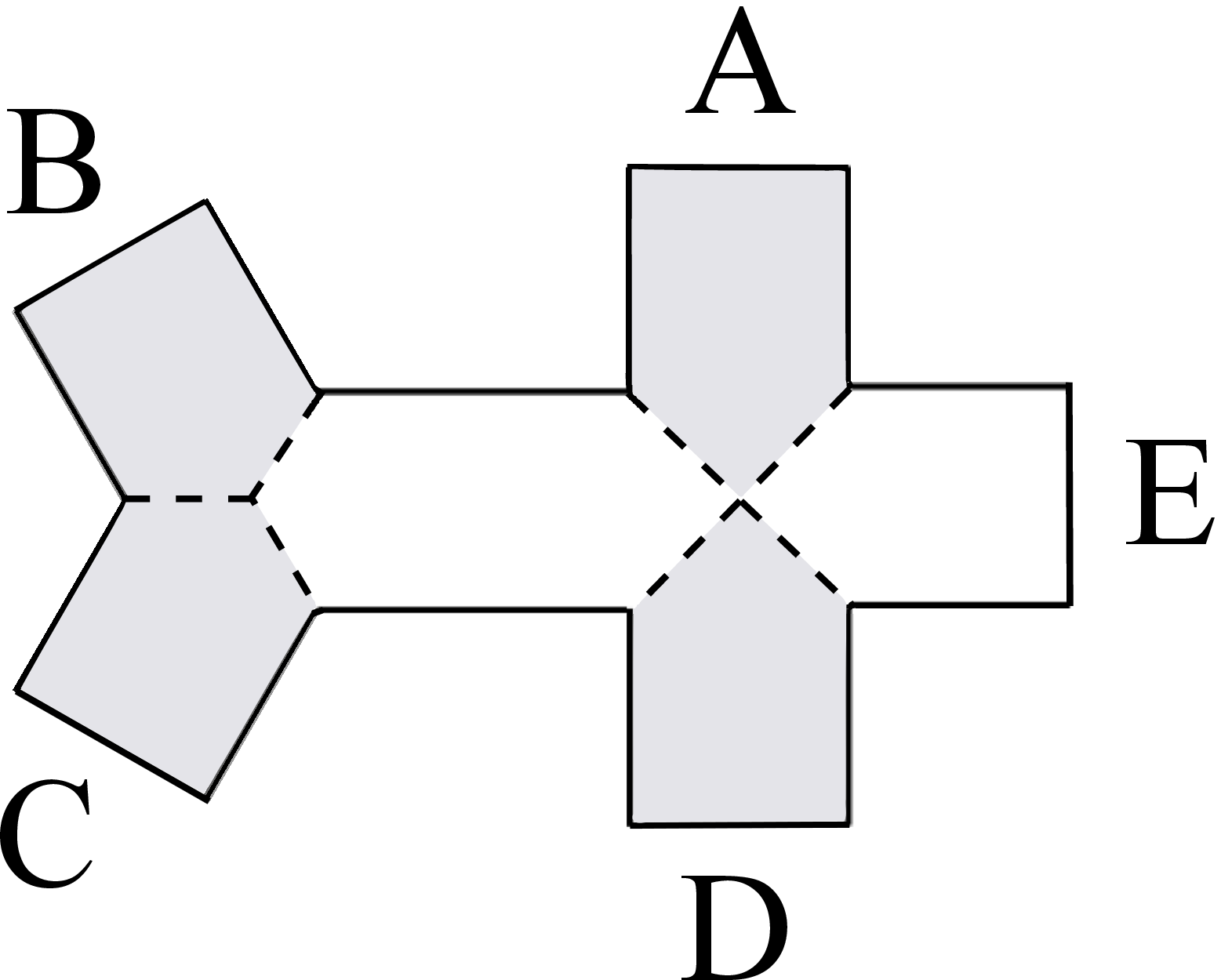}\\
(b)
 \end{center}
\end{minipage}
\begin{minipage}{0.25\hsize}
 \begin{center}
 \includegraphics[width=3cm]{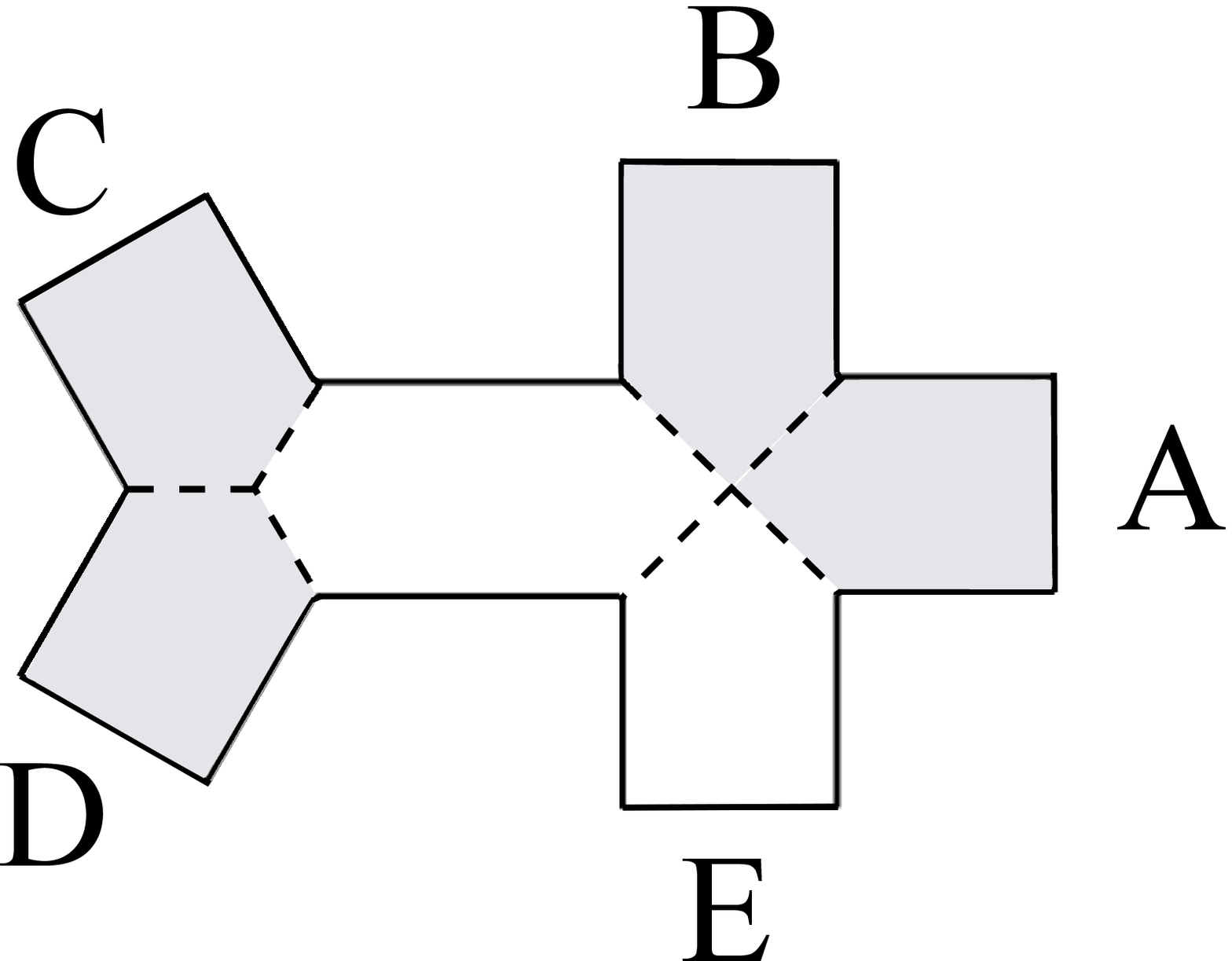} \\
(c)
 \end{center}
\end{minipage}
\vspace{2mm}
\caption{Three 1P Feynman diagrams for four-fermion-one-boson amplitude.
}\label{Feynman FFFFB 34}
\end{figure}

The second contributions come from the diagrams
called in this paper the one-propagator (1P) diagrams.
These diagrams are constructed by using one three-string vertex,
one four-string vertex and one propagator. 
In this case, we can draw the three 1P diagrams as depicted 
in Fig.~\ref{Feynman FFFFB 34}. Their contributions become
\begin{subequations} 
\begin{align}
 \mathcal{A}_{FFFFB}^{(1\textrm{P})(a)} =& -\frac{1}{8}
\int_0^\infty d\tau\
\langle\Big(Q\Xi_A\ \eta \Psi_B+\eta \Psi_A\ Q\Xi_B\Big)
\nonumber\\
&\hspace{40mm} \times 
(\xi_c b_c)\
\Big(Q\Xi_C\ \eta \Psi_D-\eta \Psi_C\ Q\Xi_D\Big)\ \Phi_E\rangle_W,\\
 \mathcal{A}_{FFFFB}^{(1\textrm{P})(b)} =& \frac{1}{4}\int_0^\infty d\tau\
\langle \Big(Q\Xi_B\ \eta \Psi_C+\eta \Psi_B\ Q\Xi_C\Big) 
\nonumber\\
&\hspace{40mm} \times 
(\xi_c b_c)\
\Big(Q\Xi_D\ \Phi_E\ \eta \Psi_A-\eta \Psi_D\ \Phi_E\ Q\Xi_A\Big)\rangle_W,\\
 \mathcal{A}_{FFFFB}^{(1\textrm{P})(c)} =& -\frac{1}{8}
\int_0^\infty d\tau\
\langle \Big(Q\Xi_C\ \eta \Psi_D+\eta \Psi_C\ Q\Xi_D\Big)
\nonumber\\
&\hspace{40mm} \times 
(\xi_c b_c)\
\Phi_E\ \Big(Q\Xi_A\ \eta \Psi_B-\eta \Psi_A\ Q\Xi_B\Big)\rangle_W,
\end{align}
\end{subequations}
each of which cancels the first three terms in (\ref{FFFFB 333 extra}),
respectively. 
Its last two terms, however, remain without being cancelled,
and vanish by imposing the constraint.
After eliminating the $\Xi$ from the external on-shell fermions
by the constraint, the total amplitude finally becomes
\begin{align}
 \mathcal{A}_{FFFFB} =&
\sum_{i=a}^e\mathcal{A}_{FFFFB}^{(2\textrm{P})(i)}
+\sum_{i=a}^c\mathcal{A}_{FFFFB}^{(1\textrm{P})(i)} 
\nonumber\\
=& -\int_0^\infty d^2\tau\
\Bigg(\langle\eta \Psi_A\ \eta \Psi_B\ (\xi_{c_1} b_{c_1})\ 
\eta \Psi_C\ b_{c_2}\ \eta \Psi_D\ Q\Phi_E\rangle_W
\nonumber\\
&\hspace{20mm}
+\langle\eta \Psi_B\ \eta \Psi_C\ (\xi_{c_1} b_{c_1})\ 
\eta \Psi_D\ b_{c_2}\ Q\Phi_E\ \eta \Psi_A\rangle_W
\nonumber\\
&\hspace{20mm}
+\langle\eta \Psi_C\ \eta \Psi_D\ (\xi_{c_1} b_{c_1})\ 
Q\Phi_E\ b_{c_2}\ \eta \Psi_A\ \eta \Psi_B\rangle_W
\nonumber\\
&\hspace{20mm}
+\langle\eta \Psi_D\ Q\Phi_E\ (\xi_{c_1} b_{c_1})\ 
\eta \Psi_A\ b_{c_2}\ \eta \Psi_B\ \eta \Psi_C\rangle_W
\nonumber\\
&\hspace{20mm}
+\langle Q\Phi_E\ \eta \Psi_A\ (\xi_{c_1} b_{c_1})\ 
\eta \Psi_B\ b_{c_2}\ \eta \Psi_C\ \eta \Psi_D\rangle_W
\Bigg).\label{amp FFFFB}
\end{align}   
This result has the same form as the five-point amplitude
in the bosonic cubic string field theory\cite{Witten:1985cc}
(CSFT) if we identify the $\eta\Psi$ and $Q\Phi$ 
with the bosonic string fields, both of which have 
the same ghost number, $G=1$.
If we recall the fact that the CSFT reproduces 
the bosonic string amplitudes,\cite{Zwiebach:1990az}
we can conclude that (\ref{amp FFFFB}) 
is equivalent to the correct superstring amplitude.
If necessary, we can map the expression to the conventional form
evaluated on the upper half-plane by the same conformal
mapping as that to be used in the CSFT.

\subsubsection{Two-fermion-three-boson amplitude with ordering $FFBBB$}

\begin{figure}[h]
\begin{minipage}{0.33\hsize}
 \begin{center}
 \includegraphics[width=5cm]{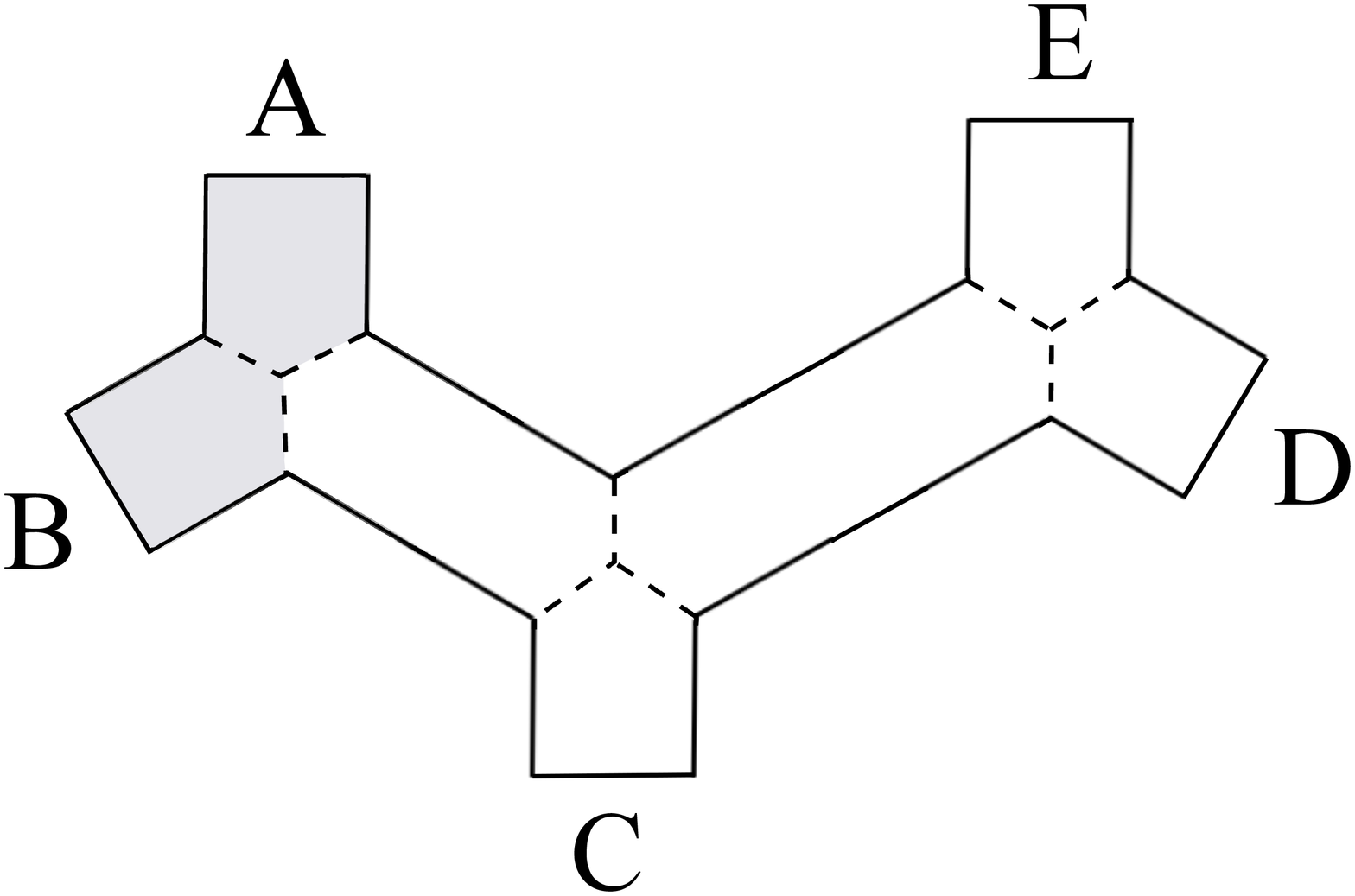}\\
(a)
 \end{center}
\end{minipage}
\begin{minipage}{0.33\hsize}
 \begin{center}
 \includegraphics[width=5cm]{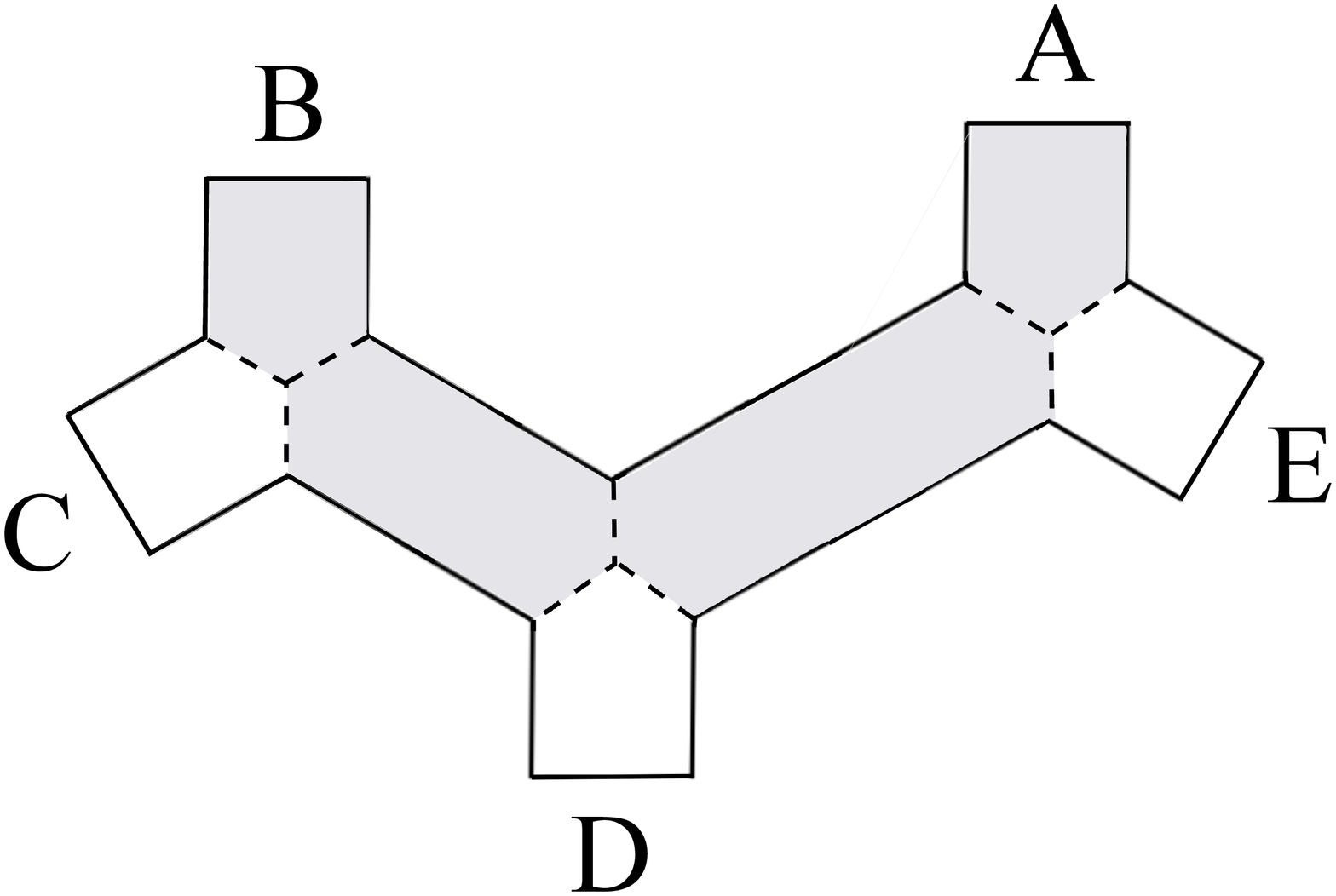}\\
(b)
 \end{center}
\end{minipage}
\begin{minipage}{0.33\hsize}
 \begin{center}
 \includegraphics[width=5cm]{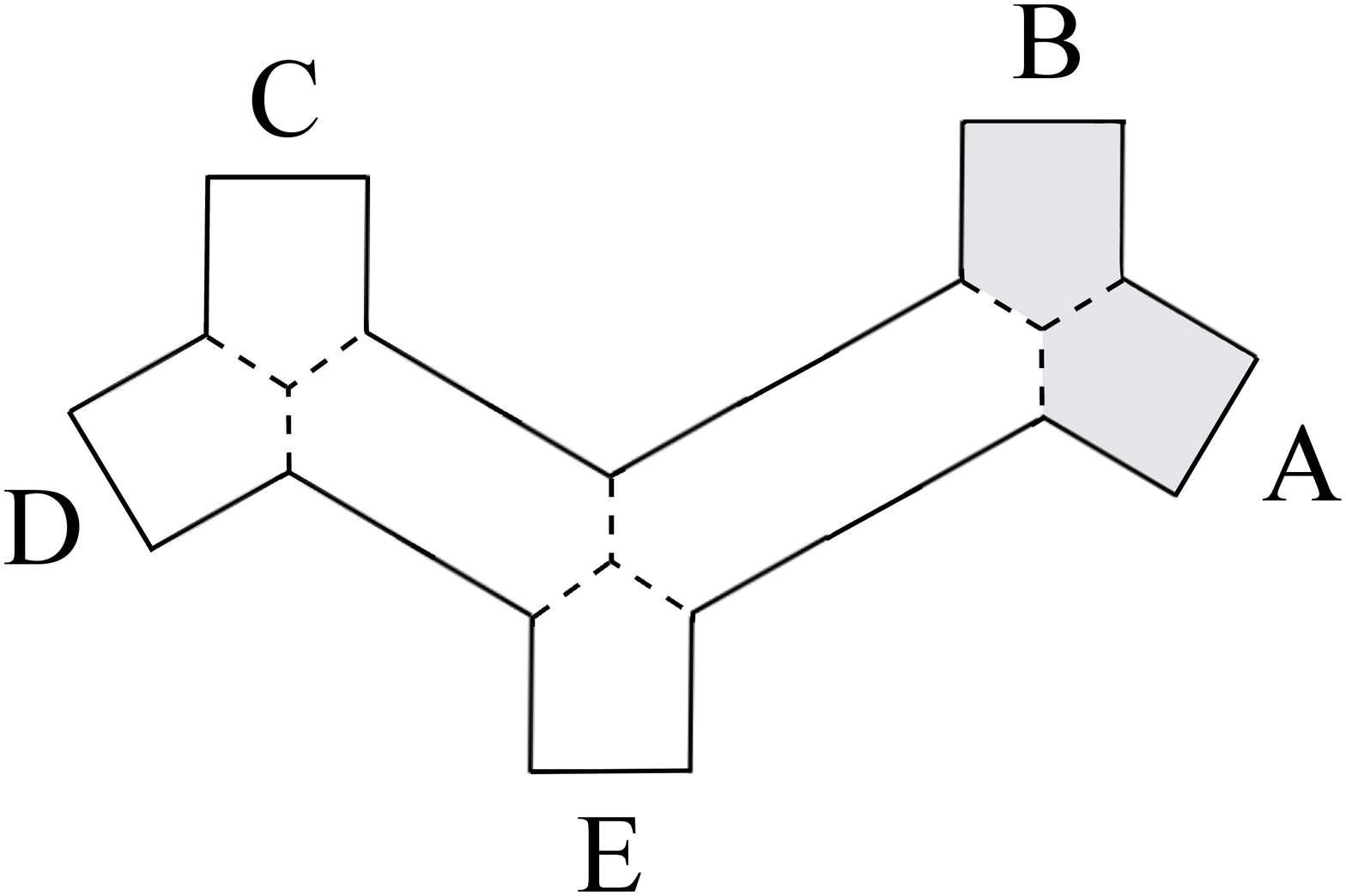} \\
(c)
 \end{center}
\end{minipage}
\begin{minipage}{0.15\hsize}
\mbox{}
\end{minipage}
\begin{minipage}{0.33\hsize}
 \begin{center}
 \includegraphics[width=5cm]{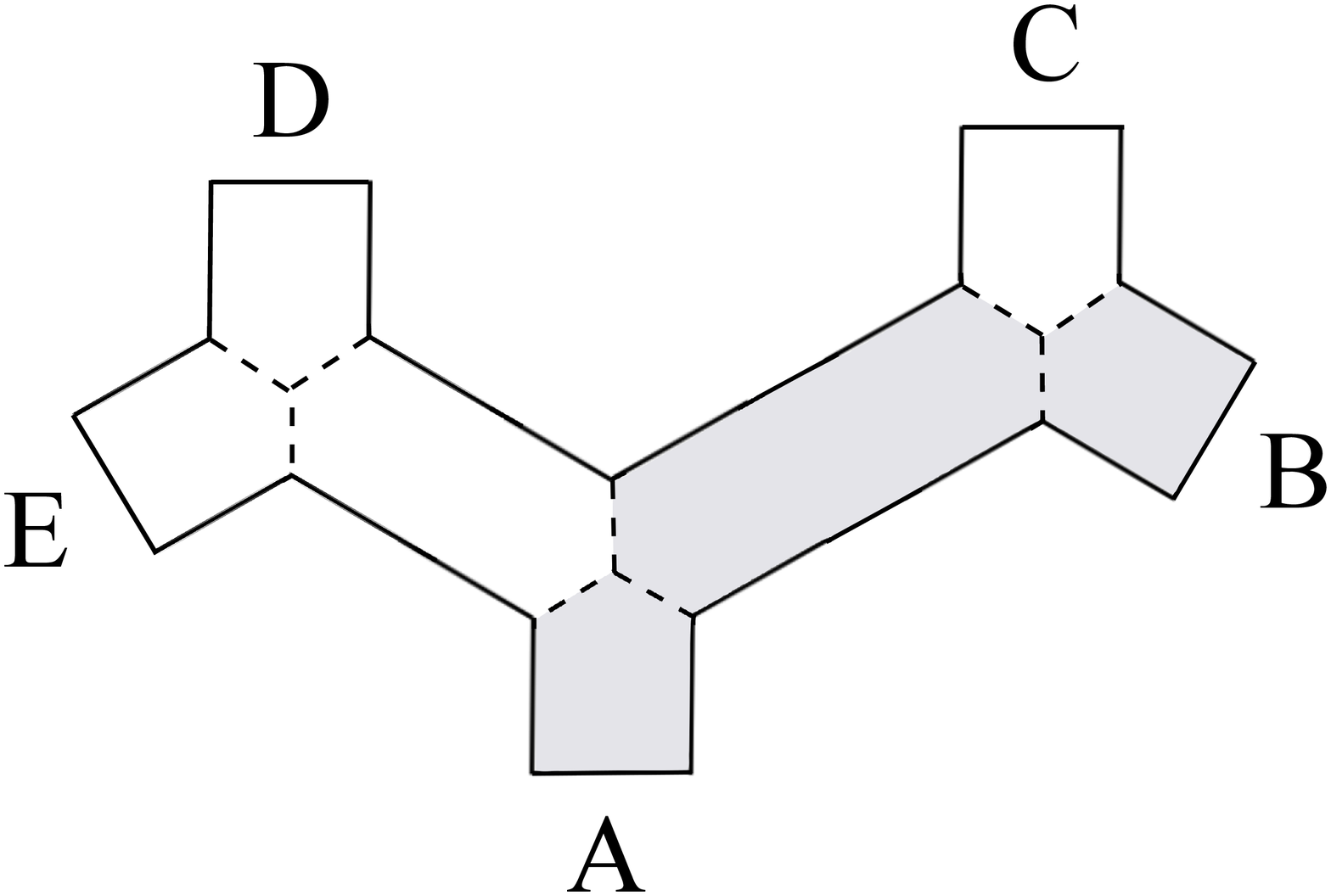} \\
(d)
 \end{center}
\end{minipage}
\begin{minipage}{0.33\hsize}
 \begin{center}
 \includegraphics[width=5cm]{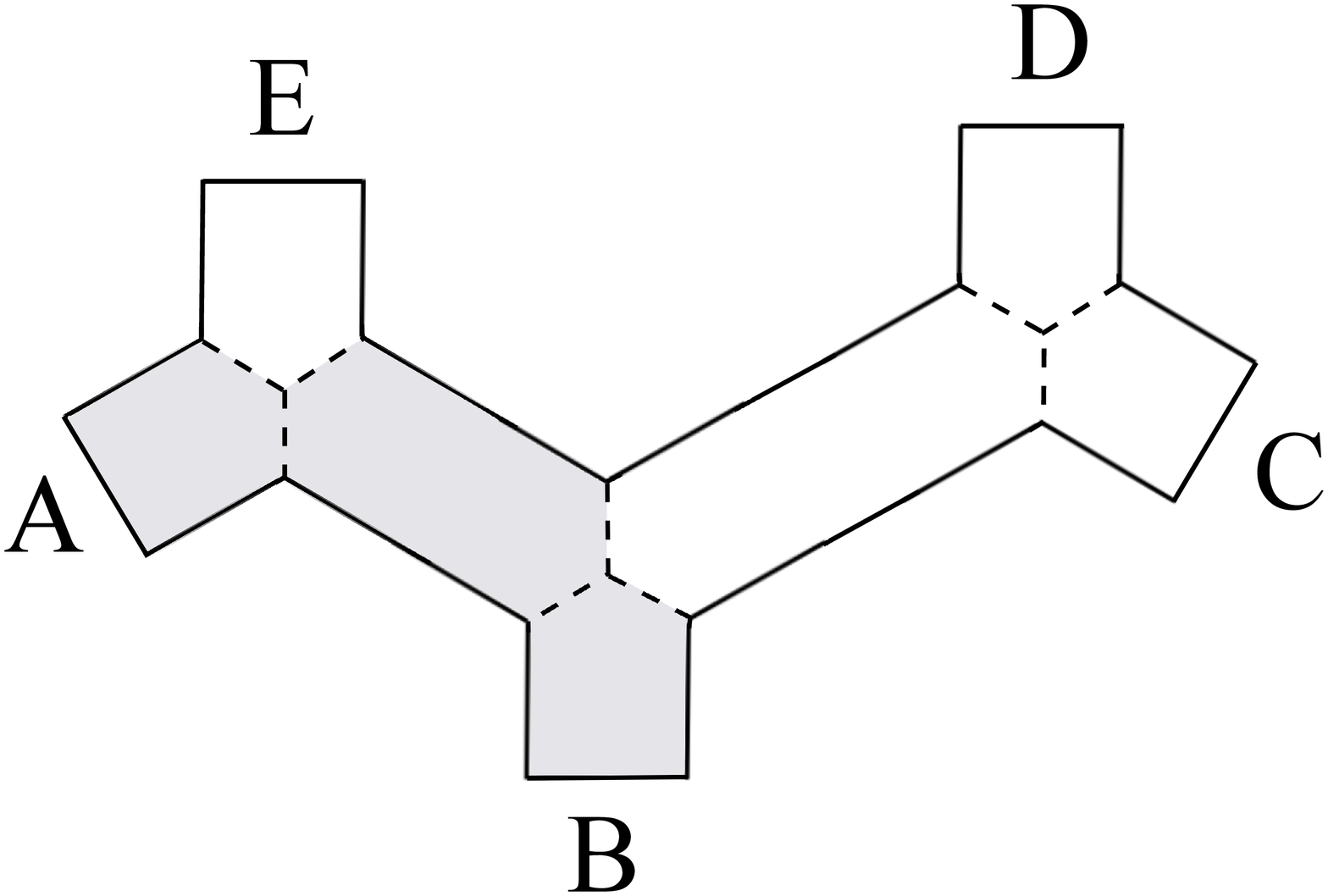} \\
(e)
 \end{center}
\end{minipage}
\vspace{2mm}
\caption{Five 2P Feynman diagrams
for two-fermion-three-boson amplitude with ordering $FFBBB$.
}\label{Feynman FFBBB 333}
\end{figure}

Now we are ready to calculate the two-fermion-three-boson amplitudes, 
for which the self-dual Feynman rules do not give the correct 
results.\cite{Michishita:Riken, Michishita:2012ku}
Let us first consider the one with ordering $FFBBB$, whose dominant part
comes from the 2P diagrams depicted in Fig.~\ref{Feynman FFBBB 333}. 
Here, Fig.~\ref{Feynman FFBBB 333}(b), in particular, includes two fermion propagators, that is, 
satisfies condition (i) 
mentioned at the end of \S\ref{four point}. The contribution from this diagram 
is in fact different from that obtained by the self-dual rules 
(even under the constraint) and improves the amplitude. 
Including this, the contribution of each 2P diagram in this case becomes
\begin{subequations}\label{cont FFBBB 333}
\begin{align}
 \mathcal{A}_{FFBBB}^{(2\textrm{P})(a)} 
=& \frac{1}{2}\int_0^\infty d^2\tau\
\langle \Big(Q\Xi_A\ \eta \Psi_B+\eta \Psi_A\ Q\Xi_B\Big)\
(\xi_{c_1} b_{c_1})\ Q\Phi_C\
b_{c_2}\ Q\Phi_D\ \eta \Phi_E\rangle_W\nonumber\\
&+\frac{1}{8}\int_0^\infty d\tau\Bigg(
\langle \Big(Q\Xi_A\ \eta \Psi_B+\eta \Psi_A\ Q\Xi_B\Big)
\nonumber\\
&\hspace{40mm} \times
(\xi_c b_c)\ \Phi_C\ \Big(Q\Phi_D\ \eta \Phi_E+\eta \Phi_D\ Q\Phi_E\Big)\rangle_W
\nonumber\\
&\hspace{20mm}
-2\langle \Big(Q\Xi_A\ \eta \Psi_B+\eta \Psi_A\ Q\Xi_B\Big)\ (\xi_c b_c)\ Q\Phi_C\ 
\eta(\Phi_D\ \Phi_E)\rangle_W
\nonumber\\
&\hspace{20mm}
-\langle \Big(Q\Phi_D\ \eta \Phi_E+\eta \Phi_D\ Q\Phi_E\Big)
\nonumber\\
&\hspace{40mm} \times
(\xi_c b_c)\ \Big(Q\Xi_A\ \eta \Psi_B+\eta \Psi_A\ Q\Xi_B\Big)\ \Phi_C\rangle_W
\nonumber\\
&\hspace{20mm}
-2\langle \eta(\Phi_D\ \Phi_E)\ (\xi_c b_c)\ \Big(Q\Xi_A\ \eta \Psi_B
+\eta \Psi_A\ Q\Xi_B\Big)\ Q\Phi_C\rangle_W\Bigg),\label{cont FFBBB 333a}\\
 \mathcal{A}_{FFBBB}^{(2\textrm{P})(b)} 
=&\frac{1}{2}\int_0^\infty d^2\tau\ \Bigg(
\langle \eta\Psi_B\ Q\Phi_C\ (\xi_{c_1} b_{c_1})\ 
Q\Phi_D\ b_{c_2}\ \eta\Phi_E\ Q\Xi_A\rangle_W
\nonumber\\
&\hspace{20mm}
+\langle Q\Xi_B\ Q\Phi_C\ (\xi_{c_1} b_{c_1})\ 
Q\Phi_D\ b_{c_2}\ \eta \Phi_E\ \eta\Psi_A\rangle_W
\Bigg)\nonumber\\
&-\frac{1}{2}\int_0^\infty d\tau\Bigg(
\langle Q\Xi_B\ \Phi_C\ (\xi_c b_c)\ 
\eta \Phi_D\ Q\Phi_E\ \eta \Psi_A\rangle_W
\nonumber\\
&\hspace{20mm}
-\langle Q\Xi_B\ Q\Phi_C\ (\xi_c b_c)\ 
\eta(\Phi_D\ \Phi_E)\ \eta\Psi_A\rangle_W
\nonumber\\
&\hspace{20mm}
- \langle Q\Phi_E\ \eta \Psi_A\ (\xi_c b_c)\ 
Q\Xi_B\ \eta (\Phi_C\ \Phi_D)\rangle_W
\nonumber\\
&\hspace{20mm}
+\langle \eta \Phi_E\ Q\Xi_A\ (\xi_c b_c)\ 
\eta \Psi_B\ Q\Phi_C\ \Phi_D\rangle_W
\nonumber\\
&\hspace{20mm}
+\langle \eta\Phi_E\ \eta\Psi_A\ (\xi_c b_c)\
Q\Xi_B\ Q\Phi_C\ \Phi_D\rangle_W
\nonumber\\
&\hspace{20mm}
-\langle \Phi_E\ \eta\Psi_A\ (\xi_c b_c)\
Q\Xi_B\ Q\Phi_C\ \eta\Phi_D\rangle_W
\Bigg),\label{cont FFBBB 333b}\\
 \mathcal{A}_{FFBBB}^{(2\textrm{P})(c)} 
=&\frac{1}{2}\int_0^\infty d^2\tau\ 
\langle Q\Phi_C\ Q\Phi_D\
(\xi_{c_1} b_{c_1})\ \eta \Phi_E\ b_{c_2}\ 
\Big(Q\Xi_A\ \eta \Psi_B+\eta \Psi_A\ Q\Xi_B\Big)\rangle_W
\nonumber\\
&+\frac{1}{8}\int_0^\infty d\tau\Bigg(
2\langle \Phi_C\ Q\Phi_D\ (\xi_c b_c)\ \eta \Phi_E\ 
\Big(Q\Xi_A\ \eta \Psi_B+\eta \Psi_A\ Q\Xi_B\Big)\rangle_W\nonumber\\
&\hspace{20mm}
+\langle \Big(Q\Phi_C\ \eta \Phi_D-\eta \Phi_C\ Q\Phi_D\Big)\nonumber\\ 
&\hspace{40mm} \times
(\xi_c b_c)\ \Phi_E\ \Big(Q\Xi_A\ \eta \Psi_B+\eta \Psi_A\ Q\Xi_B\Big)\rangle_W
\nonumber\\
&\hspace{20mm}
-2\langle Q\Phi_C\ \eta\Phi_D\ (\xi_c b_c)\
\Phi_E\ \Big(Q\Xi_A\ \eta\Psi_B+\eta\Psi_A\ Q\Xi_B\Big)\rangle_W
\nonumber\\
&\hspace{20mm}
-2\langle Q\Phi_C\ \Phi_D\ (\xi_c b_c)\
\eta\Phi_E\ \Big(Q\Xi_A\ \eta\Psi_B+\eta\Psi_A\ Q\Xi_B\Big)\rangle_W
\nonumber\\
&\hspace{20mm}
-2\langle \Big(Q\Xi_A\ \eta \Psi_B+\eta \Psi_A\ Q\Xi_B\Big)\ (\xi_c b_c)\ 
\Phi_C\ Q\Phi_D\ \eta \Phi_E\rangle_W\nonumber\\
&\hspace{20mm}
- \langle \Big(Q\Xi_A\ \eta \Psi_B+\eta \Psi_A\ Q\Xi_B\Big)\nonumber\\
&\hspace{40mm} \times
 (\xi_c b_c)\ \Big(Q\Phi_C\ \eta \Phi_D-\eta \Phi_C\ Q\Phi_D\Big)\ \Phi_E\rangle_W
\nonumber\\
&\hspace{20mm}
+2\langle \Big(Q\Xi_A\ \eta\Psi_B+\eta\Psi_A\ Q\Xi_B\Big)\ (\xi_c b_c)\
Q\Phi_C\ \eta(\Phi_D\ \Phi_E)\rangle_W
\Bigg),\label{cont FFBBB 333c}\\
 \mathcal{A}_{FFBBB}^{(2\textrm{P})(d)} 
=&\frac{1}{2}\int_0^\infty d^2\tau\
\langle Q\Phi_D\ \eta \Phi_E\ 
(\xi_{c_1} b_{c_1})\ \Big(Q\Xi_A\ b_{c_2}\ \eta \Psi_B
+\eta \Psi_A\ b_{c_2}\ Q\Xi_B\Big)\ Q\Phi_C\rangle_W\nonumber\\
&+\frac{1}{4}\int_0^\infty d\tau\Bigg(
\langle \Big(Q\Phi_D\ \eta \Phi_E+\eta \Phi_D\ Q\Phi_E\Big)\ (\xi_c b_c)\ 
\eta \Psi_A\ Q\Xi_B\ \Phi_C\rangle_W
\nonumber\\
&\hspace{20mm}
+\langle \eta(\Phi_D\ \Phi_E)\ (\xi_c b_c)\ 
\Big(Q\Xi_A\ \eta \Psi_B+\eta \Psi_A\ Q\Xi_B\Big)\ Q\Phi_C\rangle_W
\nonumber\\
&\hspace{20mm}
+\langle Q\Xi_B\ \Phi_C\ (\xi_c b_c)\
\Big(Q\Phi_D\ \eta \Phi_E+\eta \Phi_D\ Q\Phi_E\Big)\ \eta \Psi_A\rangle_W
\nonumber\\
&\hspace{20mm}
-\langle Q\Xi_B\ Q\Phi_C\ (\xi_c b_c)\ \eta(\Phi_D\ \Phi_E)\ \eta\Psi_A\rangle_W
\nonumber\\
&\hspace{20mm}
-\langle \eta\Psi_B\ Q\Phi_C\ (\xi_c b_c)\ \eta(\Phi_D\ \Phi_E)\ Q\Xi_A\rangle_W
\Bigg),\label{cont FFBBB 333d}\\
 \mathcal{A}_{FFBBB}^{(2\textrm{P})(e)} 
=&\frac{1}{2}\int_0^\infty d^2\tau\ \Bigg(
\langle \eta \Phi_E\ Q\Xi_A\ (\xi_{c_1} b_{c_1})\ 
\eta \Psi_B\ b_{c_2}\ Q\Phi_C\ Q\Phi_D\rangle_W
\nonumber\\
&\hspace{20mm}
+\langle \eta \Phi_E\ \eta\Psi_A\ (\xi_{c_1} b_{c_1})\ 
Q\Xi_B\ b_{c_2}\ Q\Phi_C\ Q\Phi_D\rangle_W
\Bigg)\nonumber\\
&+\frac{1}{4}\int_0^\infty d\tau\Bigg(
-\langle \Phi_E\ \eta \Psi_A\ (\xi_c b_c)\ Q\Xi_B\ 
\Big(Q\Phi_C\ \eta\Phi_D+\eta \Phi_C\ Q\Phi_D\Big)\rangle_W
\nonumber\\
&\hspace{20mm} 
+\langle \eta \Phi_E\ Q\Xi_A\ (\xi_c b_c)\ \eta\Psi_B\ 
\Big(Q\Phi_C\ \Phi_D-\Phi_C\ Q\Phi_D\Big)\rangle_W
\nonumber\\
&\hspace{20mm}
+\langle \eta \Phi_E\ \eta\Psi_A\ (\xi_c b_c)\ Q\Xi_B\ 
\Big(Q\Phi_C\ \Phi_D-\Phi_C\ Q\Phi_D\Big)\rangle_W
\nonumber\\
&\hspace{20mm}
+\langle \Big(Q\Phi_C\ \eta\Phi_D+\eta \Phi_C\ Q\Phi_D\Big)\ 
(\xi_c b_c)\ \Phi_E\ \eta \Psi_A\ Q\Xi_B\rangle_W
\nonumber\\
&\hspace{20mm}
+\langle \Big(Q\Phi_C\ \Phi_D-\Phi_C\ Q\Phi_D\Big) 
\nonumber\\
&\hspace{40mm} \times
(\xi_c b_c)\ \eta \Phi_E\ 
(Q\Xi_A\ \eta \Psi_B+\eta \Psi_A\ Q\Xi_B)\rangle_W
\Bigg),\label{cont FFBBB 333e}
\end{align}
\end{subequations}
after some calculations.
We moved the $Q$ and $\eta$ so as to act on the external states,
and then deformed the expression so that the external bosons
have single forms, $Q\Phi_C$, $Q\Phi_D$ and $\eta\Phi_E$.
The extra terms with less propagator
appear as the result. 

\begin{figure}[htbp]
\begin{minipage}{0.125\hsize}
 \mbox{}
\end{minipage}
\begin{minipage}{0.25\hsize}
 \begin{center}
 \includegraphics[width=3cm]{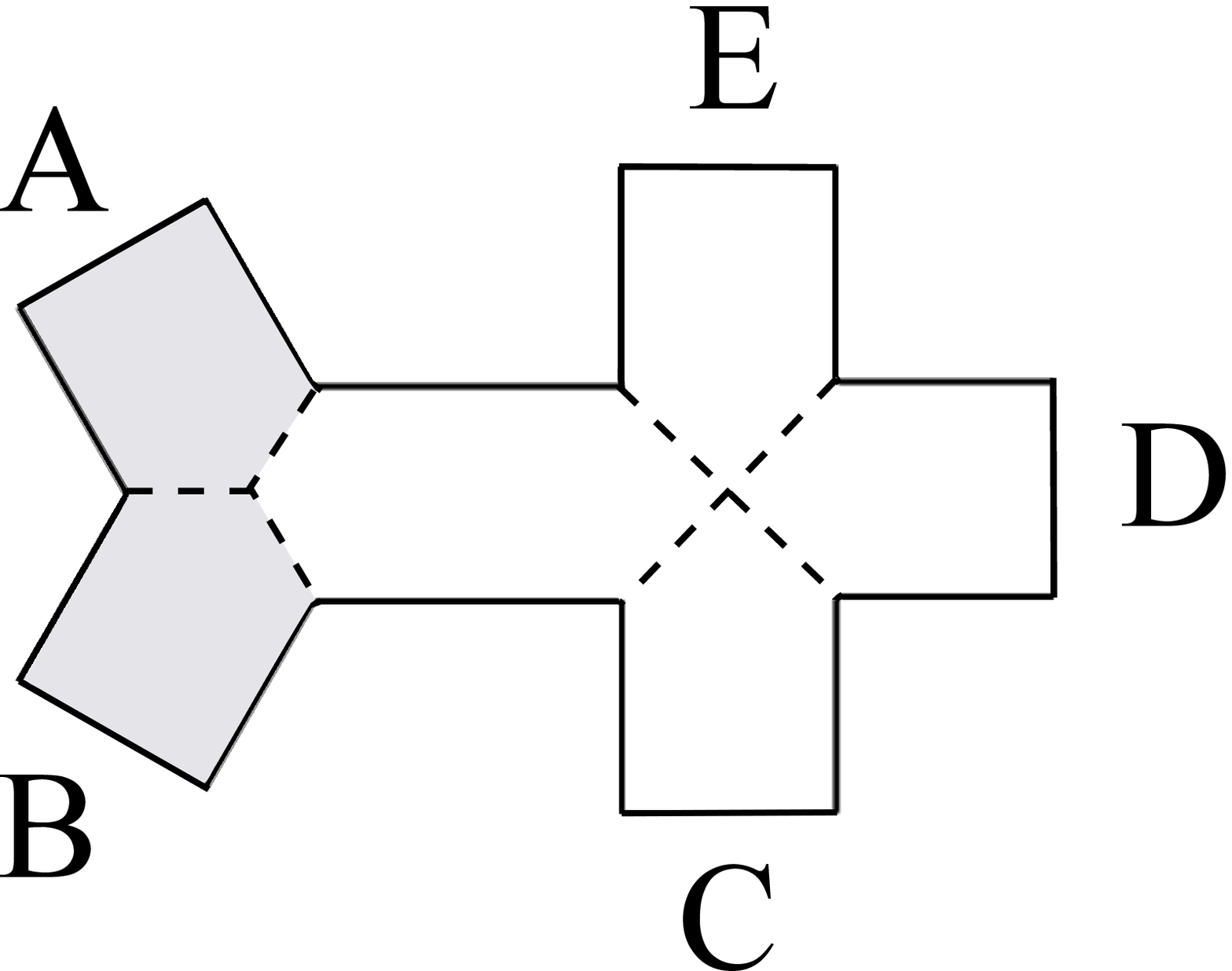}\\
(a)
 \end{center}
\end{minipage}
\begin{minipage}{0.25\hsize}
 \begin{center}
 \includegraphics[width=3cm]{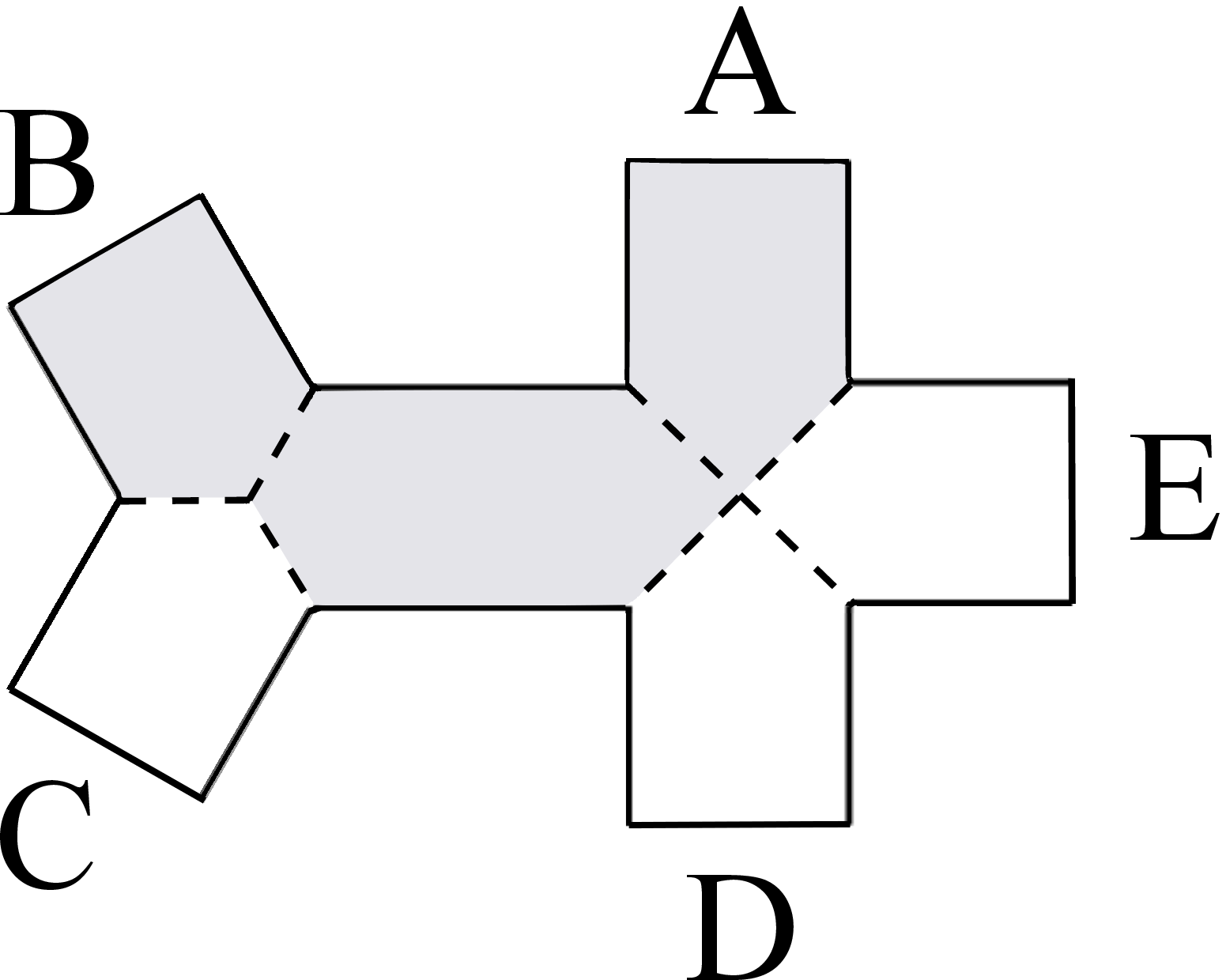}\\
(b)
 \end{center}
\end{minipage}
\begin{minipage}{0.25\hsize}
 \begin{center}
 \includegraphics[width=3cm]{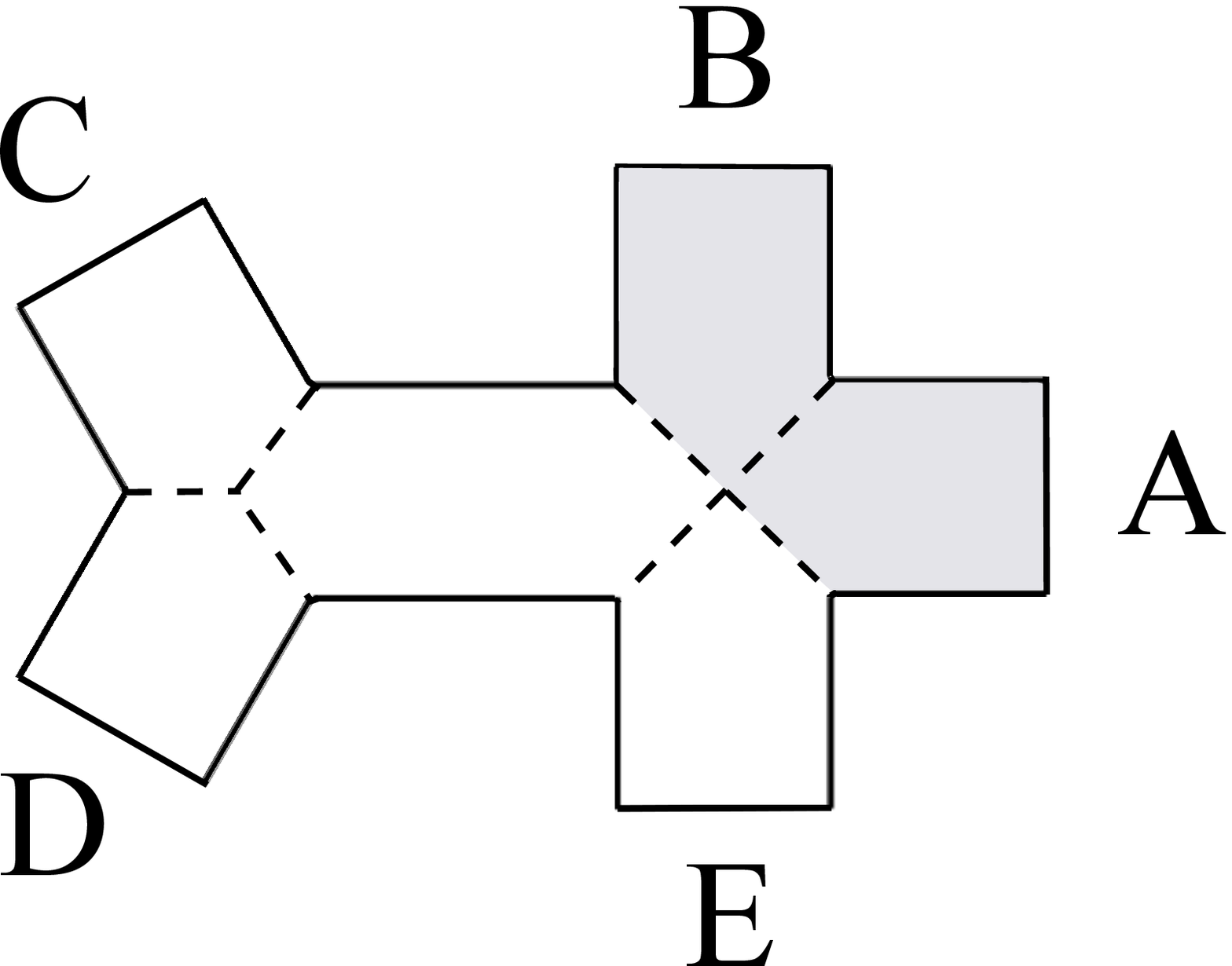} \\
(c)
 \end{center}
\end{minipage}
\begin{minipage}{0.125\hsize}
 \mbox{}
\end{minipage}
\begin{minipage}{0.23\hsize}
\mbox{}
\end{minipage}
\begin{minipage}{0.25\hsize}
 \begin{center}
 \includegraphics[width=3cm]{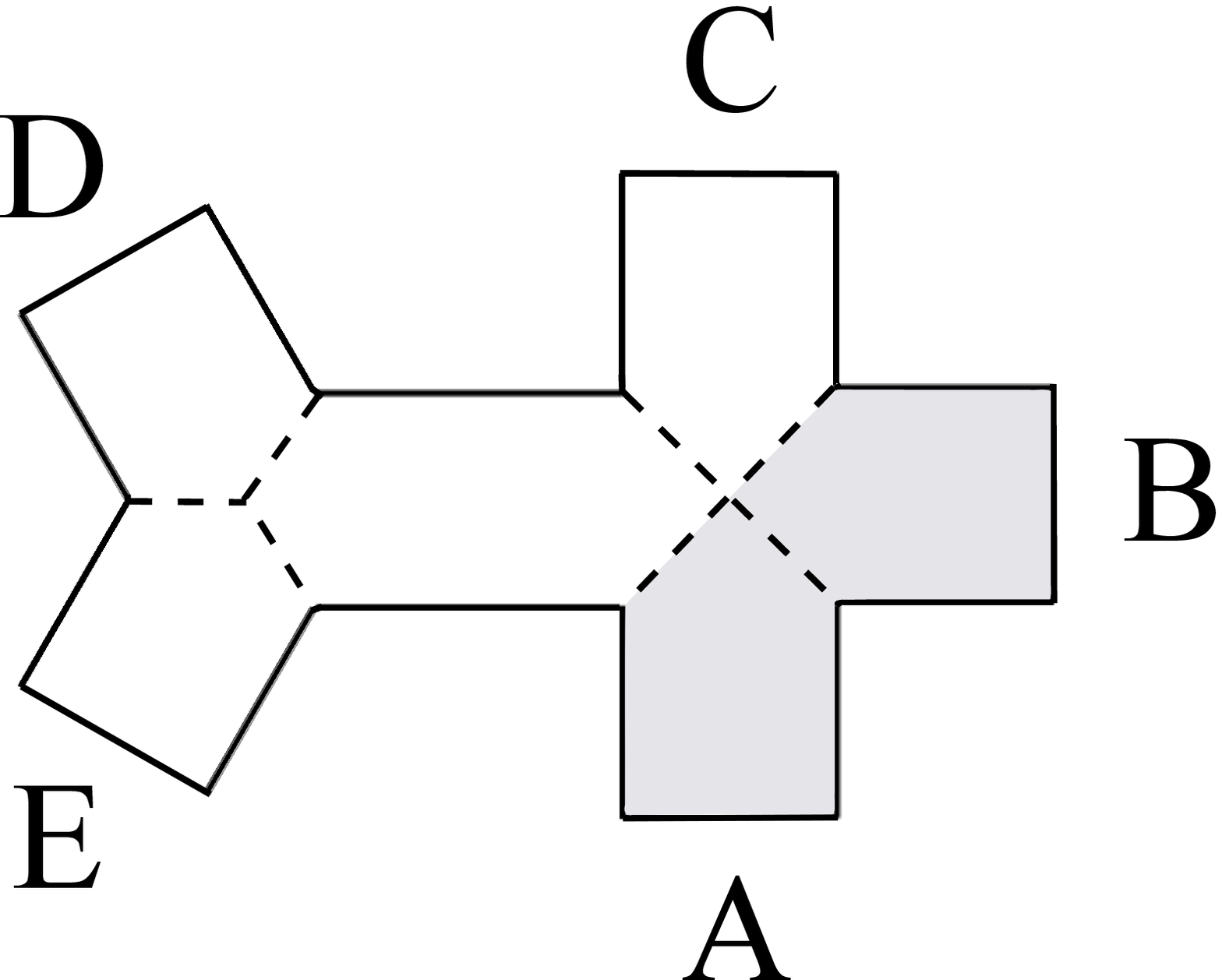} \\
(d)
 \end{center}
\end{minipage}
\begin{minipage}{0.25\hsize}
 \begin{center}
 \includegraphics[width=3cm]{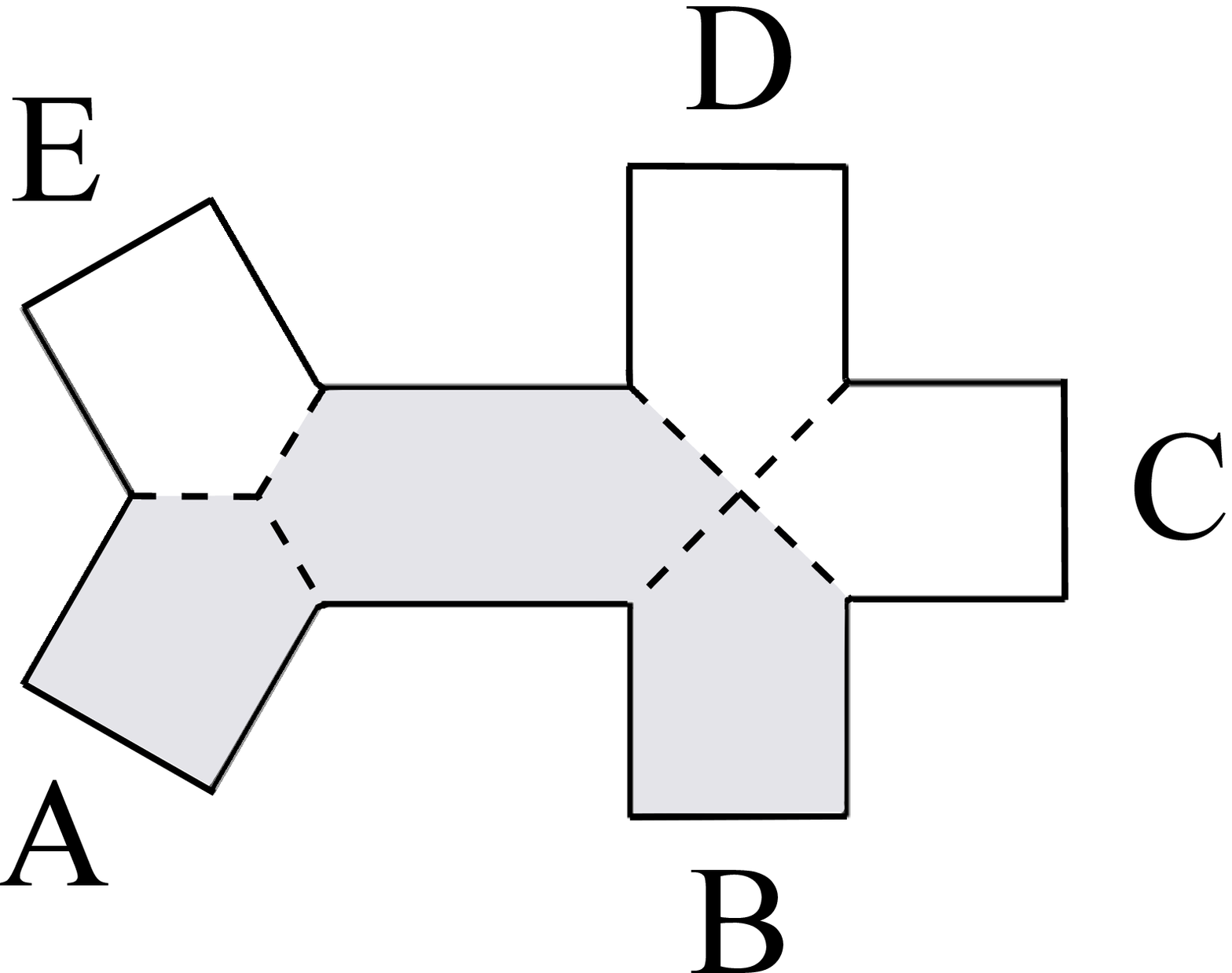} \\
(e)
 \end{center}
\end{minipage}
\vspace{2mm}
\caption{Five 1P Feynman diagrams for two-fermion-three-boson amplitude
with ordering $FFBBB$. 
}\label{Feynman FFBBB 34}
 \end{figure}

There are five different channels also in the 1P diagrams
as in Fig.~\ref{Feynman FFBBB 34}.
The contribution from each diagram is similarly evaluated as
\begin{subequations}\label{cont FFBBB 34}
\begin{align}
 \mathcal{A}_{FFBBB}^{(1\textrm{P})(a)} =&\frac{1}{24}\int_0^\infty d\tau\Bigg(
\langle \Big(Q\Xi_A\ \eta\Psi_B+\eta\Psi_A\ Q\Xi_B\Big)\ (\xi_c b_c)\ \Phi_C\ 
\Big(Q\Phi_D\ \eta\Phi_E-\eta\Phi_D\ Q\Phi_E\Big)\rangle_W
\nonumber\\
&\hspace{20mm}
-2\langle \Big(Q\Xi_A\ \eta\Psi_B+\eta\Psi_A\ Q\Xi_B\Big)\ (\xi_c b_c)\
\Big(Q\Phi_C\ \Phi_D\ \eta\Phi_E-\eta\Phi_C\ \Phi_D\ Q\Phi_E\Big)\rangle_W
\nonumber\\
&\hspace{20mm}
+\langle \Big(Q\Xi_A\ \eta\Psi_B+\eta\Psi_A\ Q\Xi_B\Big)\ (\xi_c b_c)\ 
\Big(Q\Phi_C\ \eta\Phi_D-\eta\Phi_C\ Q\Phi_D\Big)\ \Phi_E\rangle_W\Bigg),
\label{cont FFBBB 34a}\\
 \mathcal{A}_{FFBBB}^{(1\textrm{P})(b)} =&\frac{1}{4}\int_0^\infty d\tau\ \Bigg(
\langle \eta\Psi_B\ \eta\Phi_C\ (\xi_c b_c)\ Q(\Phi_D\ \Phi_E)\ Q\Xi_A\rangle_W
\nonumber\\
&\hspace{15mm}
-\langle Q\Xi_B\ \eta\Phi_C\ (\xi_c b_c)\ Q(\Phi_D\ \Phi_E)\ \eta\Psi_A\rangle_W
\nonumber\\
&\hspace{15mm}
-\langle \eta\Psi_B\ \Phi_C\ (\xi_c b_c)\  
\Big(Q\Phi_D\ \eta\Phi_E-\eta\Phi_D\ Q\Phi_E\Big)\ Q\Xi_A\rangle_W\Bigg)
\nonumber\\
&\hspace{5mm}
-\frac{1}{4}\langle Q\Xi_A\ \eta\Psi_B\ \Phi_C\ \Phi_D\ \Phi_E\rangle_W,
\label{cont FFBBB 34b}\\
\mathcal{A}_{FFBBB}^{(1\textrm{P})(c)} =& \frac{1}{8}\int_0^\infty d\tau\
\langle \Big(Q\Phi_C\ \eta\Phi_D+\eta\Phi_C\ Q\Phi_D\Big)\ (\xi_c b_c)\
\Phi_E\ \Big(Q\Xi_A\ \eta\Psi_B-\eta\Psi_A\ Q\Xi_B\Big)\rangle_W,
\label{cont FFBBB 34c}\\
\mathcal{A}_{FFBBB}^{(1\textrm{P})(d)} =& \frac{1}{8}\int_0^\infty d\tau\
\langle \Big(Q\Phi_D\ \eta\Phi_E+\eta\Phi_D\ Q\Phi_E\Big)\ (\xi_c b_c)\
\Big(Q\Xi_A\ \eta\Psi_B-\eta\Psi_A\ Q\Xi_B\Big)\ \Phi_C\rangle_W,
\label{cont FFBBB 34d}\\
 \mathcal{A}_{FFBBB}^{(1\textrm{P})(e)} =& \frac{1}{4}\int_0^\infty d\tau\ \Bigg(
\langle \Phi_E\ \eta\Psi_A\ (\xi_c b_c)\ Q\Xi_B\ 
\Big(Q\Phi_C\ \eta\Phi_D-\eta\Phi_C\ Q\Phi_D\Big)\rangle_W
\nonumber\\
&\hspace{15mm}
+\langle \eta\Phi_E\ \Big(Q\Xi_A\ (\xi_c b_c)\ \eta\Psi_B-
\eta\Psi_A\ (\xi_c b_c)\ Q\Xi_B\Big)\ Q(\Phi_C\ \Phi_D)\rangle_W
\Bigg)
\nonumber\\
&\hspace{5mm}
-\frac{1}{4}\langle\eta\Psi_A\ Q\Xi_B\ \Phi_C\ \Phi_D\ \Phi_E\rangle_W.
\label{cont FFBBB 34e}
\end{align}
\end{subequations}
Figures~\ref{Feynman FFBBB 34}(b) and (e) satisfy condition (ii), 
and their contributions (\ref{cont FFBBB 34b}) and 
(\ref{cont FFBBB 34e}) are different from those obtained by the self-dual rules.
These contributions (\ref{cont FFBBB 34}) almost cancel 
the extra terms in (\ref{cont FFBBB 333}), and give
\begin{equation}
 \sum_{i=a}^e\left(\mathcal{A}_{FFBBB}^{(2P)(i)}\right)\Big|_{\textrm{extra}}
+\sum_{i=a}^e \mathcal{A}_{FFBBB}^{(1P)(i)}=-\frac{1}{12}\langle \Big(
Q\Xi_A\ \eta\Psi_B+\eta\Psi_A\ Q\Xi_B\Big)\ \Phi_C\ \Phi_D\ \Phi_E\rangle_W.
\label{cont FFBBB extra}
\end{equation}
The nonzero result again comes from the boundary of the proper time integration
through the relation (\ref{collapse}).
\begin{figure}[htbp]
\begin{minipage}{1.0\hsize}
 \begin{center}
 \includegraphics[width=3cm]{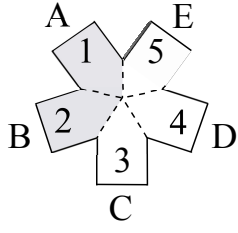}\\
 \end{center}
\end{minipage}
\vspace{2mm}
\caption{The Feynman diagram with no propagator
for two-fermion-three-boson amplitude
with ordering $FFBBB$. 
}\label{Feynman FFBBB 5}
\end{figure}

We can further draw a diagram with no propagator (NP)
using the five-string vertex as in Fig.~\ref{Feynman FFBBB 5}.
The contribution of this diagram,
\begin{equation}
 \mathcal{A}_{FFBBB}^{(\textrm{NP})} = \frac{1}{12}\langle
\Big(Q\Xi_A\ \eta \Psi_B+\eta \Psi_A\ Q\Xi_B\Big)\ \Phi_C\ \Phi_D\ \Phi_E\rangle_W,
\end{equation}
exactly cancels (\ref{cont FFBBB extra}).
The amplitude finally becomes
\begin{align}
 \mathcal{A}_{FFBBB} =&
\sum_{i=a}^e \mathcal{A}_{FFBBB}^{(2\textrm{P})(i)}
+\sum_{i=a}^e \mathcal{A}_{FFBBB}^{(1\textrm{P})(i)}
+\mathcal{A}_{FFBBB}^{(\textrm{NP})}
\nonumber\\
=&\int_0^\infty d^2\tau\ 
\Bigg(
\langle \eta \Psi_A\ \eta \Psi_B\ (\xi_{c_1} b_{c_1})\ Q\Phi_C\ b_{c_2}\
Q\Phi_D\ \eta \Phi_E\rangle_W
\nonumber\\
&\hspace{15mm}
+\langle \eta \Psi_B\ Q\Phi_C\ (\xi_{c_1} b_{c_1})\ 
Q\Phi_D\ b_{c_2}\ \eta \Phi_E\ \eta \Psi_A\rangle_W
\nonumber\\
&\hspace{15mm}
+\langle Q\Phi_C\ Q\Phi_D\ (\xi_{c_1} b_{c_1})\ \eta \Phi_E\ b_{c_2}\
\eta \Psi_A\ \eta \Psi_B\rangle_W
\nonumber\\
&\hspace{15mm}
+\langle Q\Phi_D\ \eta \Phi_E\ (\xi_{c_1} b_{c_1})\
\eta \Psi_A\ b_{c_2}\ \eta \Psi_B\ Q\Phi_C\rangle_W
\nonumber\\
&\hspace{15mm}
+\langle \eta \Phi_E\ \eta \Psi_A\ (\xi_{c_1} b_{c_1})\ \eta \Psi_B\ b_{c_2}\
Q\Phi_C\ Q\Phi_D\rangle_W
\Bigg),\label{amp FFBBB}
\end{align}
after eliminating $\Xi$ from the external fermions.
This final expression is again equal to that obtained in the CSFT,
so equivalent to the correct amplitude.

\subsubsection{Two-fermion-three-boson amplitude with ordering $FBFBB$}

\begin{figure}[htbp]
\begin{minipage}{0.33\hsize}
 \begin{center}
 \includegraphics[width=5cm]{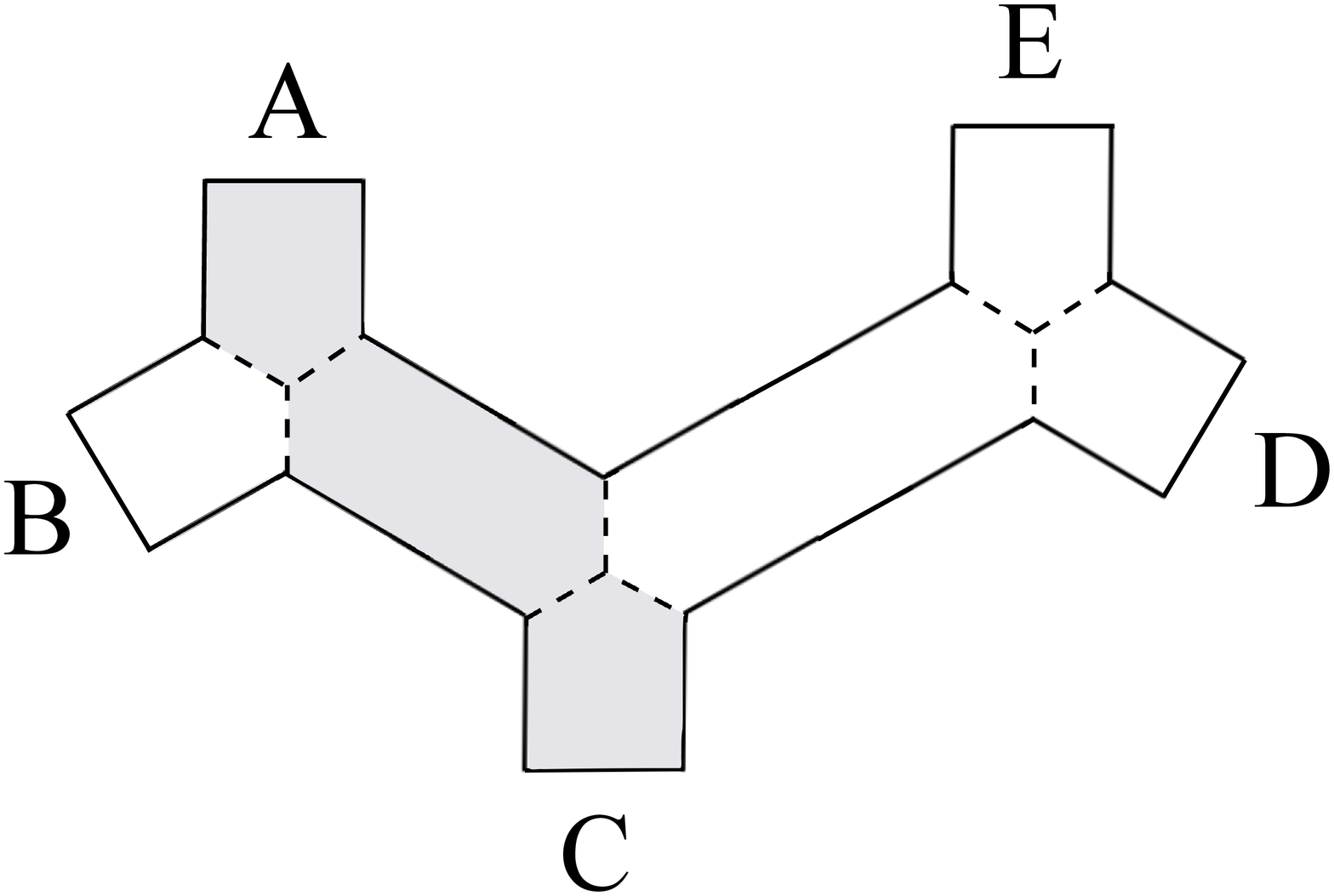}\\
(a)
 \end{center}
\end{minipage}
\begin{minipage}{0.33\hsize}
 \begin{center}
 \includegraphics[width=5cm]{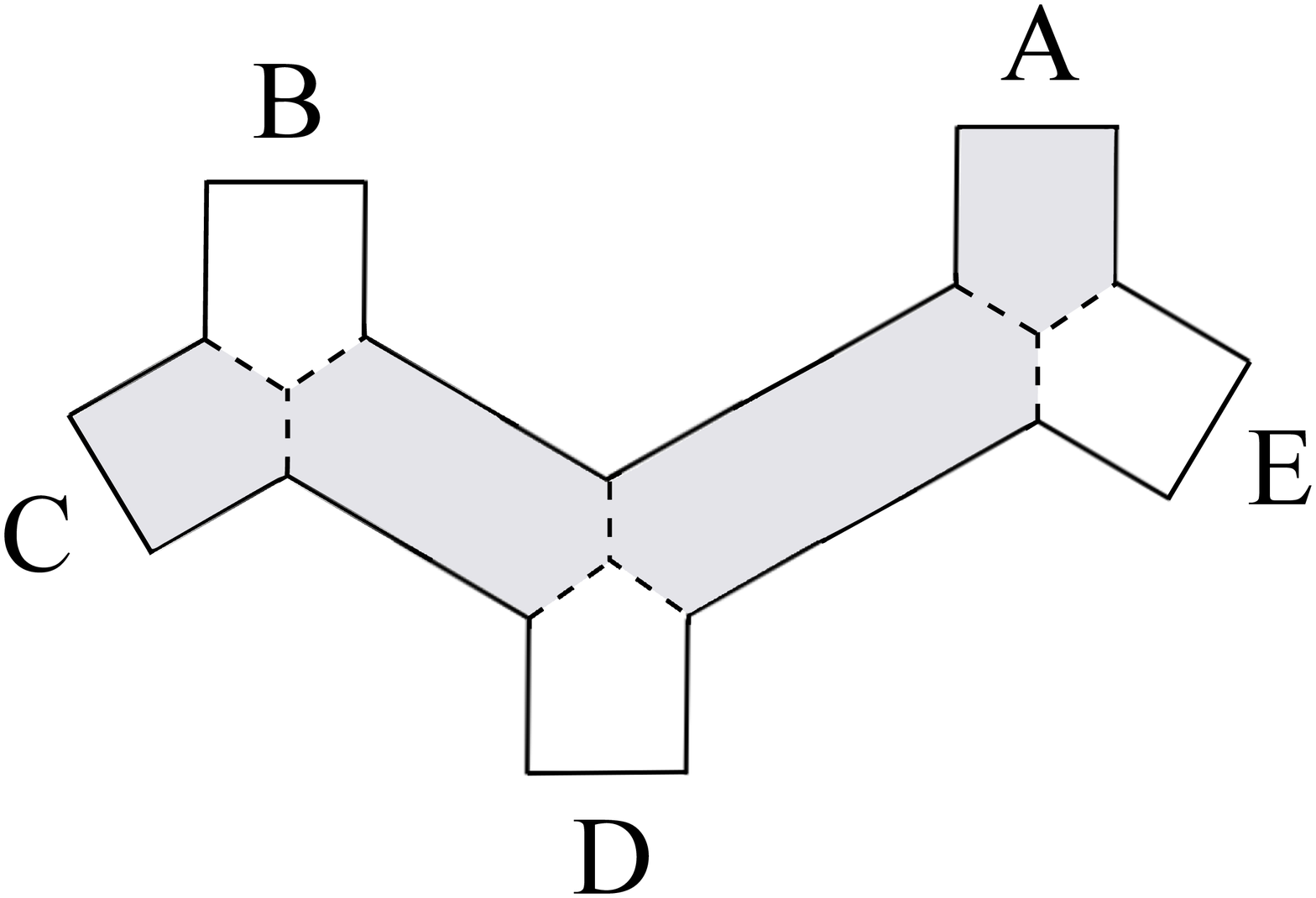}\\
(b)
 \end{center}
\end{minipage}
\begin{minipage}{0.33\hsize}
 \begin{center}
 \includegraphics[width=5cm]{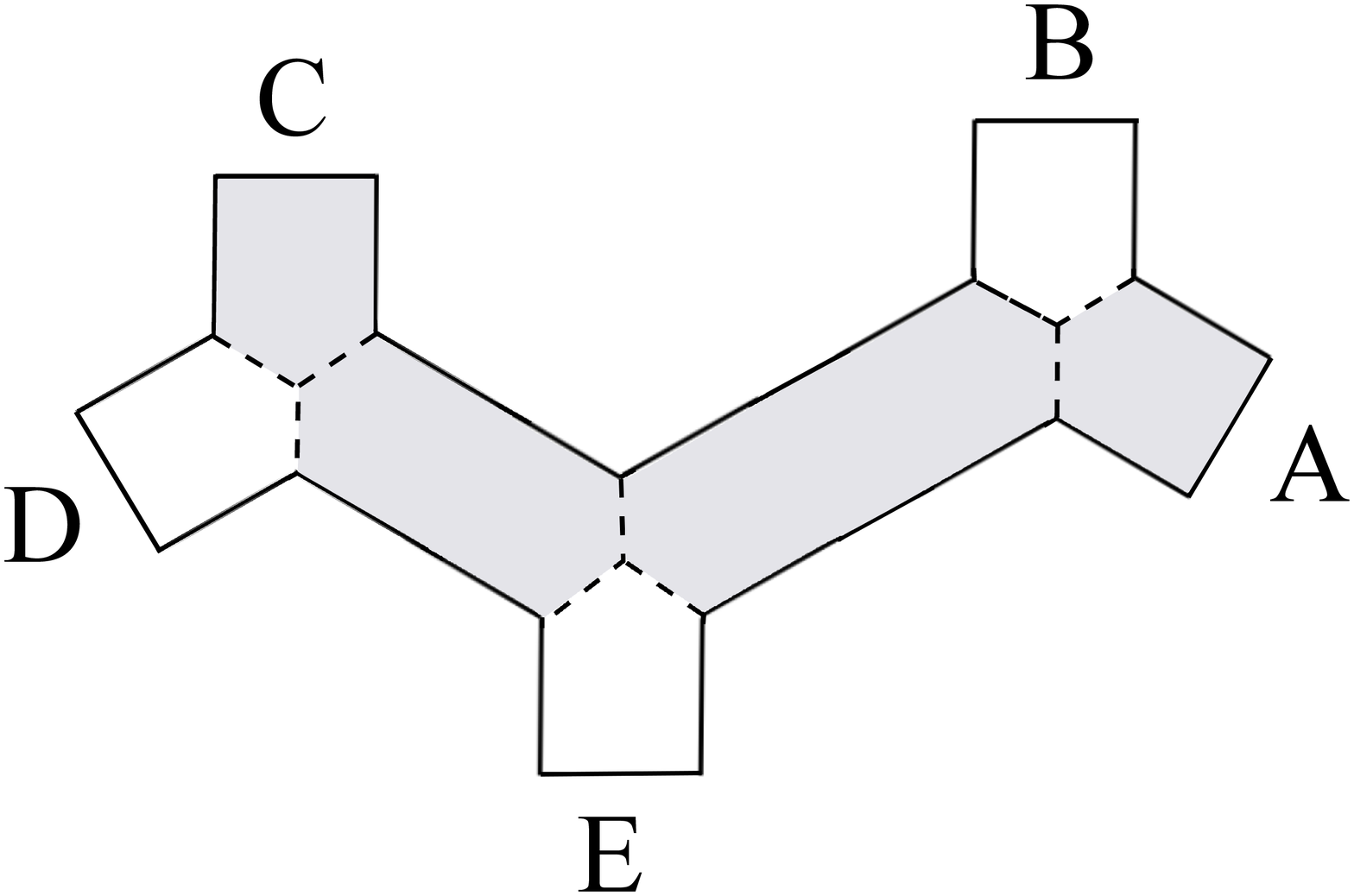} \\
(c)
 \end{center}
\end{minipage}
\begin{minipage}{0.15\hsize}
\mbox{}
\end{minipage}
\begin{minipage}{0.33\hsize}
 \begin{center}
 \includegraphics[width=5cm]{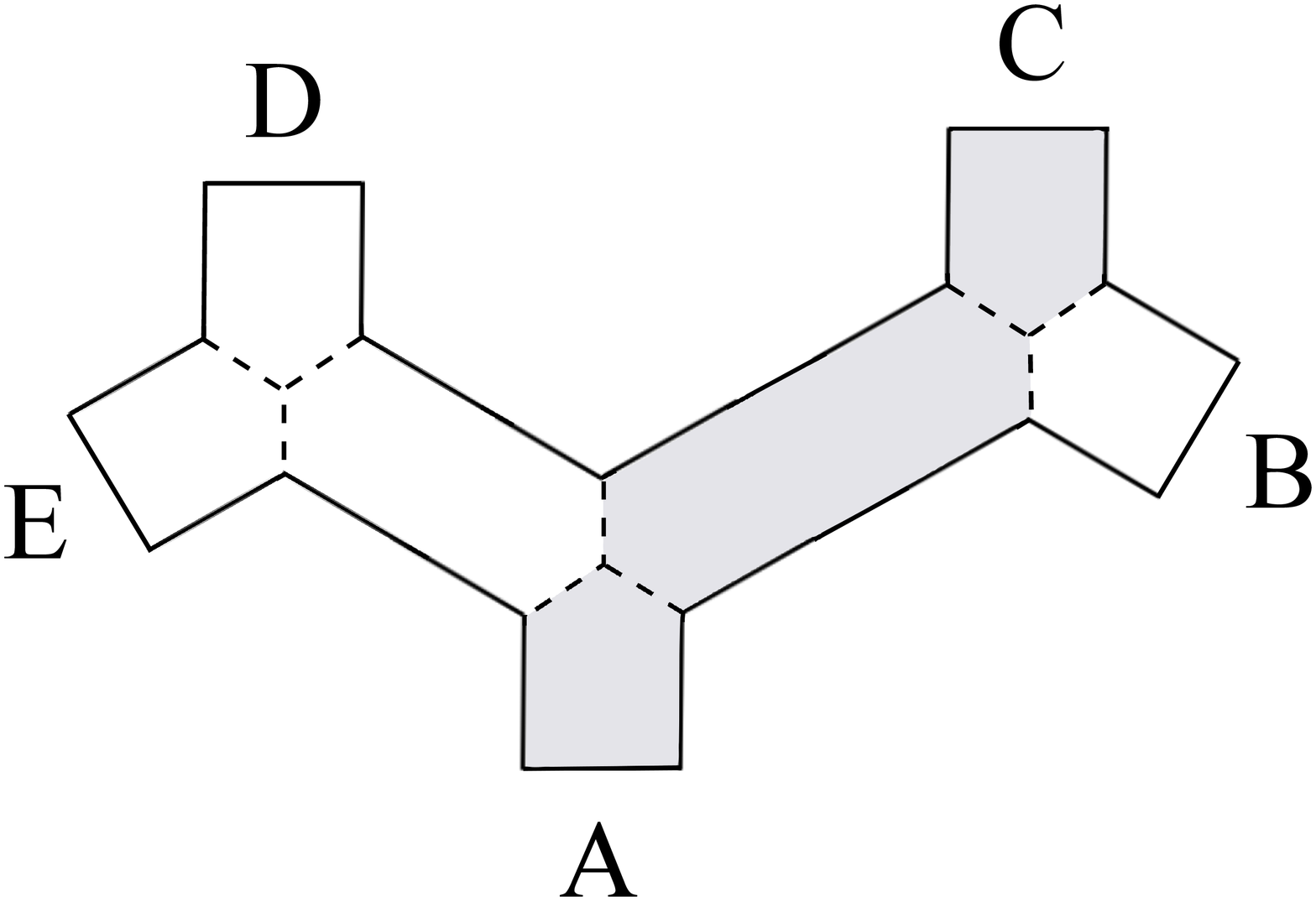} \\
(d)
 \end{center}
\end{minipage}
\begin{minipage}{0.33\hsize}
 \begin{center}
 \includegraphics[width=5cm]{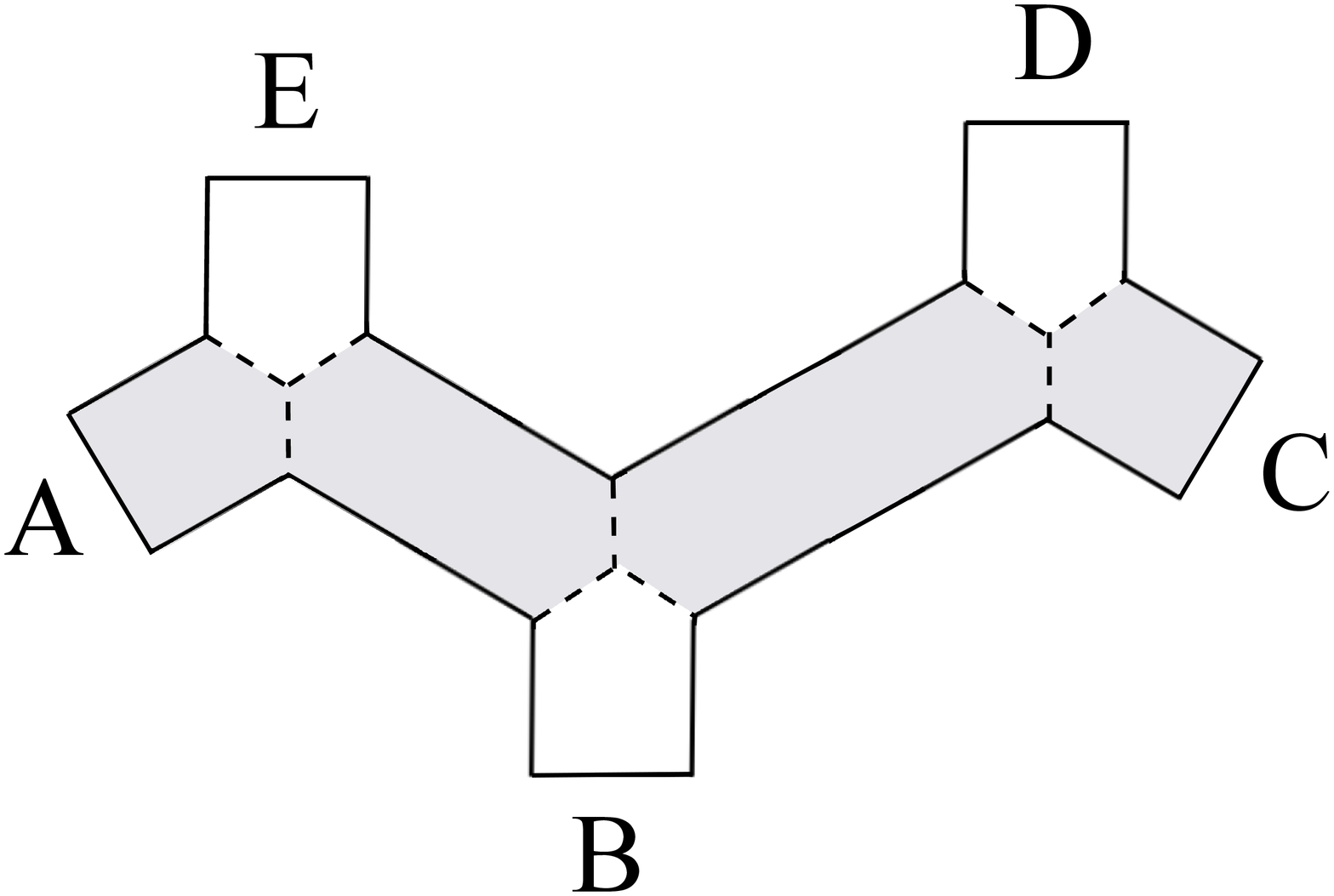} \\
(e)
 \end{center}
\end{minipage}
\vspace{2mm}
\caption{Five 2P Feynman diagrams
for two-fermion-three-boson amplitude with ordering $FBFBB$.
}\label{Feynman FBFBB 333}
 \end{figure}

\begin{figure}[htbp]
\begin{minipage}{1.0\hsize}
 \begin{center}
 \includegraphics[width=3cm]{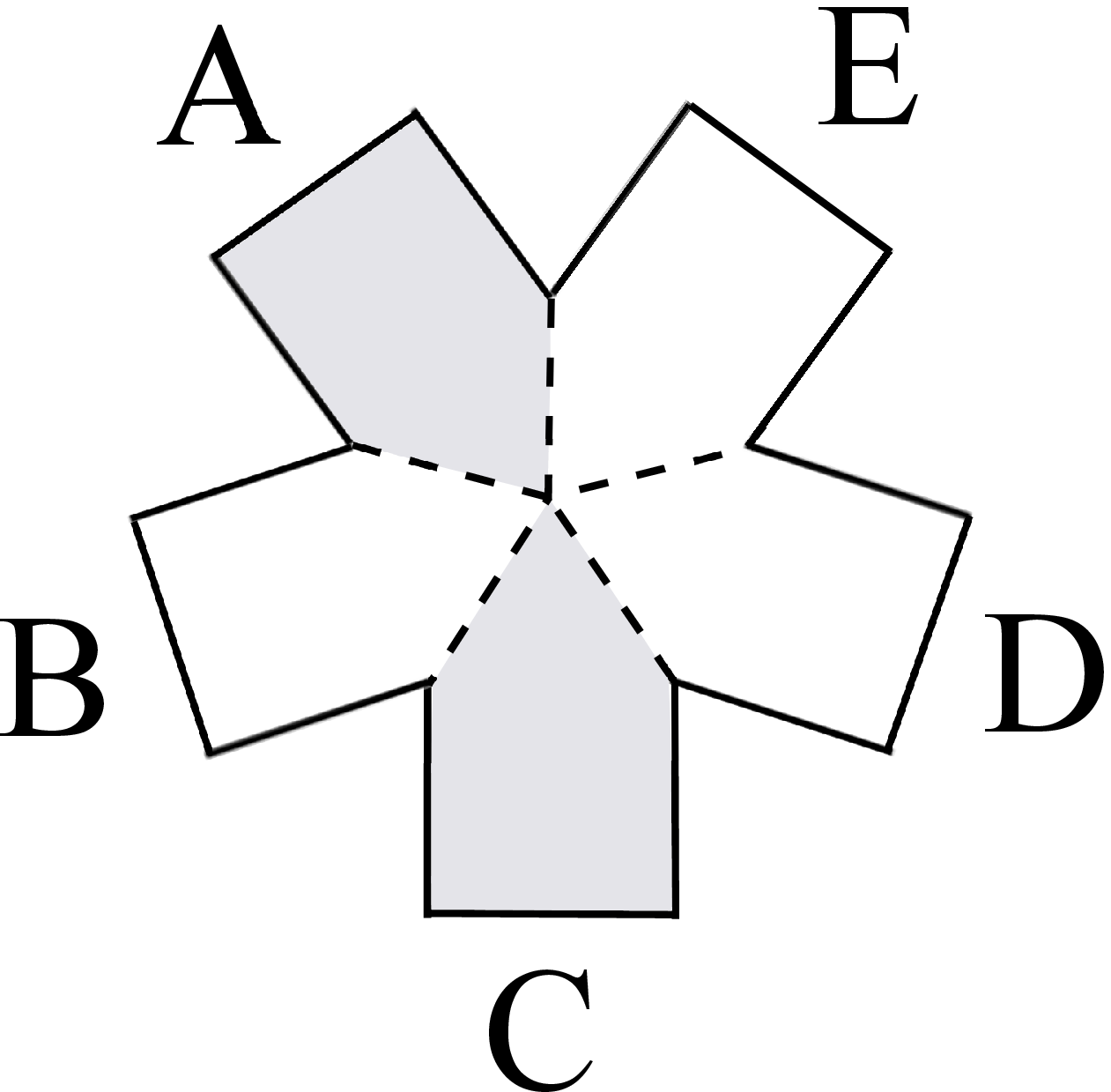}\\
 \end{center}
\end{minipage}
\vspace{2mm}
\caption{The NP Feynman diagram
for two-fermion-three-boson amplitude
with ordering $FBFBB$. 
}\label{Feynman FBFBB 5}
\end{figure}

Last is the two-fermion-three-boson amplitude
with ordering $FBFBB$, for which we can draw
the five 2P diagrams depicted in Fig.~\ref{Feynman FBFBB 333}. 
The diagrams (b), (c), and (e) satisfying condition (i) 
can contribute to the amplitude differently from the case of the self-dual rules. 
Each diagram gives the contribution,
\begin{subequations}\label{cont FBFBB 333}
\begin{align}
 \mathcal{A}_{FBFBB}^{(2\textrm{P})(a)} 
=& \frac{1}{2}\int_0^\infty d^2\tau\ 
\Bigg(
\langle Q\Xi_A\ Q\Phi_B\ (\xi_{c_1} b_{c_1})\ \eta \Psi_C\ b_{c_2}\ 
Q\Phi_D\ \eta \Phi_E\rangle_W
\nonumber\\
&\hspace{20mm}
+\langle \eta \Psi_A\ Q\Phi_B\ (\xi_{c_1} b_{c_1})\ Q\Xi_C\ b_{c_2}\ 
Q\Phi_D\ \eta \Phi_E\rangle_W
\Bigg)
\nonumber\\
&-\frac{1}{4}\int_0^\infty d\tau\Bigg(
\langle Q\Xi_A\ \Phi_B\ (\xi_c b_c)\ \eta \Psi_C\
\Big(Q\Phi_D\ \eta \Phi_E+\eta \Phi_D\ Q\Phi_E\Big)\rangle_W
\nonumber\\
&\hspace{20mm}
+\langle Q\Xi_A\ Q\Phi_B\ (\xi_c b_c)\ \eta\Psi_C\ \eta(\Phi_D\ \Phi_E)\rangle_W
\nonumber\\
&\hspace{20mm}
+\langle \eta\Psi_A\ Q\Phi_B\ (\xi_c b_c)\ Q\Xi_C\ \eta(\Phi_D\ \Phi_E)\rangle_W
\nonumber\\
&\hspace{20mm}
-\langle \Big(Q\Phi_D\ \eta \Phi_E+\eta \Phi_D\ Q\Phi_E\Big)\
(\xi_c b_c)\ Q\Xi_A\ \Phi_B\ \eta \Psi_C\rangle_W
\nonumber\\
&\hspace{20mm}
+\langle \eta(\Phi_D\ \Phi_E)\ (\xi_c b_c)\ Q\Xi_A\ Q\Phi_B\ \eta\Psi_C\rangle_W
\nonumber\\
&\hspace{20mm}
+\langle \eta(\Phi_D\ \Phi_E)\ (\xi_c b_c)\ \eta\Psi_A\ Q\Phi_B\ Q\Xi_C\rangle_W
\Bigg),
\label{cont FBFBB 333a}\\
 \mathcal{A}_{FBFBB}^{(2\textrm{P})(b)} 
=&\frac{1}{2}\int_0^\infty d^2\tau\ 
\Bigg(
\langle Q\Phi_B\ \eta \Psi_C\ (\xi_{c_1} b_{c_1})\
Q\Phi_D\ b_{c_2}\ \eta \Phi_E\ Q\Xi_A\rangle_W
\nonumber\\
&\hspace{20mm}
+\langle Q\Phi_B\ Q\Xi_C\ (\xi_{c_1} b_{c_1})\
Q\Phi_D\ b_{c_2}\ \eta\Phi_E\ \eta \Psi_A\rangle_W
\Bigg)\nonumber\\
&+\frac{1}{2}\int_0^\infty d\tau\Bigg(
\langle \Phi_B\ Q\Xi_C\ (\xi_c b_c)\
\eta \Phi_D\ Q\Phi_E\ \eta \Psi_A\rangle_W
\nonumber\\
&\hspace{20mm}
+\langle Q\Phi_B\ Q\Xi_C\ (\xi_c b_c)\ 
\eta(\Phi_D\ \Phi_E)\ \eta\Psi_A\rangle_W
\nonumber\\
&\hspace{20mm}
+\langle Q\Phi_E\ \eta \Psi_A\ (\xi_c b_c)\
\eta \Phi_B\ Q\Xi_C\ \Phi_D\rangle_W
\nonumber\\
&\hspace{20mm}
-\langle Q\Phi_E\ \eta\Psi_A\ (\xi_c b_c)\
\Phi_B\ Q\Xi_C\ \eta\Phi_D\rangle_W
\nonumber\\
&\hspace{20mm}
-\langle \eta \Phi_E\ Q\Xi_A\ (\xi_c b_c)\
Q\Phi_B\ \eta \Psi_C\ \Phi_D\rangle_W
\nonumber\\
&\hspace{20mm}
-\langle \eta\Phi_E\ \eta\Psi_A\ (\xi_c b_c)\
Q\Phi_B\ Q\Xi_C\ \Phi_D\rangle_W
\nonumber\\
&\hspace{20mm}
+\langle \Phi_E\ \eta\Psi_A\ (\xi_c b_c)\
Q\Phi_B\ Q\Xi_C\ \eta\Phi_D\rangle_W
\Bigg),
\label{cont FBFBB 333b}\\
 \mathcal{A}_{FBFBB}^{(2\textrm{P})(c)} 
=&\frac{1}{2}\int_0^\infty d^2\tau\ 
\Bigg(
\langle \eta \Psi_C\ Q\Phi_D\ (\xi_{c_1} b_{c_1})\
\eta \Phi_E\ b_{c_2}\ Q\Xi_A\ Q\Phi_B\rangle_W
\nonumber\\
&\hspace{20mm}
+\langle Q\Xi_C\ Q\Phi_D\ (\xi_{c_1} b_{c_1})\
\eta\Phi_E\ b_{c_2}\ \eta \Psi_A\ Q\Phi_B\rangle_W
\Bigg)\nonumber\\
&+\frac{1}{2}\int_0^\infty d\tau_1\ \Bigg(
\langle \eta \Psi_C\ Q\Phi_D\ (\xi_c b_c)\
\eta \Phi_E\ Q\Xi_A\ \Phi_B\rangle_W
\nonumber\\
&\hspace{20mm}
-\langle \eta \Psi_C\ Q\Phi_D\ (\xi_c b_c)\
\Phi_E\ Q\Xi_A\ \eta \Phi_B\rangle_W
\nonumber\\
&\hspace{20mm}
-\langle Q\Xi_C\ \Phi_D\ (\xi_c b_c)\
\eta\Phi_E\ \eta \Psi_A\ Q\Phi_B\rangle_W
\nonumber\\
&\hspace{20mm}
+\langle Q\Xi_A\ \Phi_B\ (\xi_c b_c)\
\eta \Psi_C\ Q\Phi_D\ \eta \Phi_E\rangle_W
\nonumber\\
&\hspace{20mm}
+\langle \eta\Psi_A\ Q\Phi_B\ (\xi_c b_c)\
Q\Xi_C\ \eta(\Phi_D\ \Phi_E)\rangle_W
\Bigg),
\label{cont FBFBB 333c}\\
 \mathcal{A}_{FBFBB}^{(2\textrm{P})(d)} 
=& \frac{1}{2}\int_0^\infty d^2\tau\ \Bigg(
\langle Q\Phi_D\ \eta \Phi_E\ (\xi_{c_1} b_{c_1})\ Q\Xi_A\ 
b_{c_2}\ Q\Phi_B\ \eta \Psi_C\rangle_W\nonumber\\
&\hspace{15mm}
+\langle Q\Phi_D\ \eta \Phi_E\
(\xi_{c_1} b_{c_1})\ \eta \Psi_A\
b_{c_2}\ Q\Phi_B\ Q\Xi_C\rangle_W
\Bigg)\nonumber\\
&+\frac{1}{4}\int_0^\infty d\tau\Bigg(
\langle \eta(\Phi_D\ \Phi_E)\ (\xi_c b_c)\
Q\Xi_A\ Q\Phi_B\ \eta\Psi_C\rangle_W
\nonumber\\
&\hspace{20mm}
+\langle \eta(\Phi_D\ \Phi_E)\ (\xi_c b_c)\
\eta\Psi_A\ Q\Phi_B\ Q\Xi_C\rangle_W
\nonumber\\
&\hspace{20mm}
-\langle \Big(Q\Phi_D\ \eta \Phi_E+\eta \Phi_D\ Q\Phi_E\Big)\ (\xi_c b_c)\
\eta \Psi_A\ \Phi_B\ Q\Xi_C\rangle_W
\nonumber\\
&\hspace{20mm}
-\langle Q\Phi_B\ \eta\Psi_C\ (\xi_c b_c)\
\eta(\Phi_D\ \Phi_E)\ Q\Xi_A\rangle_W
\nonumber\\
&\hspace{20mm}
-\langle Q\Phi_B\ Q\Xi_C\ (\xi_c b_c)\
\eta(\Phi_D\ \Phi_E)\ \eta\Psi_A\rangle_W
\nonumber\\
&\hspace{20mm}
-\langle \Phi_B\ Q\Xi_C\ (\xi_c b_c)\
\Big(Q\Phi_D\ \eta \Phi_E+\eta \Phi_D\ Q\Phi_E\Big)\ \eta \Psi_A\rangle_W
\Bigg),
\label{cont FBFBB 333d}\\
 \mathcal{A}_{FBFBB}^{(2\textrm{P})(e)} 
=& \frac{1}{2}\int_0^\infty d^2\tau\ 
\Bigg(
\langle \eta \Phi_E\ Q\Xi_A\ (\xi_{c_1} b_{c_1})\ Q\Phi_B\ b_{c_2}\ 
\eta \Psi_C\ Q\Phi_D\rangle_W
\nonumber\\
&\hspace{20mm}
+\langle \eta\Phi_E\ \eta \Psi_A\ (\xi_{c_1} b_{c_1})\ Q\Phi_B\ b_{c_2}\ 
Q\Xi_C\ Q\Phi_D\rangle_W\Bigg)
\nonumber\\
&+\frac{1}{2}\int_0^\infty d\tau\Bigg(
\langle \eta \Phi_E\ Q\Xi_A\ (\xi_c b_c)\ \Phi_B\ \eta \Psi_C\ Q\Phi_D\rangle_W
\nonumber\\
&\hspace{20mm}
+\langle \eta\Phi_E\ \eta\Psi_A\ (\xi_c b_c)\
Q\Phi_B\ Q\Xi_C\ \Phi_D\rangle_W
\nonumber\\
&\hspace{20mm}
-\langle \Phi_E\ \eta\Psi_A\ (\xi_c b_c)\
Q\Phi_B\ Q\Xi_C\ \eta\Phi_D\rangle_W
\nonumber\\
&\hspace{20mm}
-\langle Q\Xi_C\ \eta \Phi_D\ (\xi_c b_c)\ Q\Phi_E\ \eta \Psi_A\ \Phi_B\rangle_W
\nonumber\\
&\hspace{20mm}
+\langle Q\Xi_C\ \eta\Phi_D\ (\xi_c b_c)\
\Phi_E\ \eta\Psi_A\ Q\Phi_B\rangle_W
\nonumber\\
&\hspace{20mm}
+\langle Q\Xi_C\ \Phi_D\ (\xi_c b_c)\
\eta\Phi_E\ \eta\Psi_A\ Q\Phi_B\rangle_W
\Bigg),
\label{cont FBFBB 333e}
\end{align}
\end{subequations}
to the amplitude, where the dominant parts are deformed so that
the external bosons have single forms, $Q\Phi_B$, $Q\Phi_D$ and $\eta\Phi_E$.
\begin{figure}[htbp]
\begin{minipage}{0.125\hsize}
 \mbox{}
\end{minipage}
\begin{minipage}{0.25\hsize}
 \begin{center}
 \includegraphics[width=3cm]{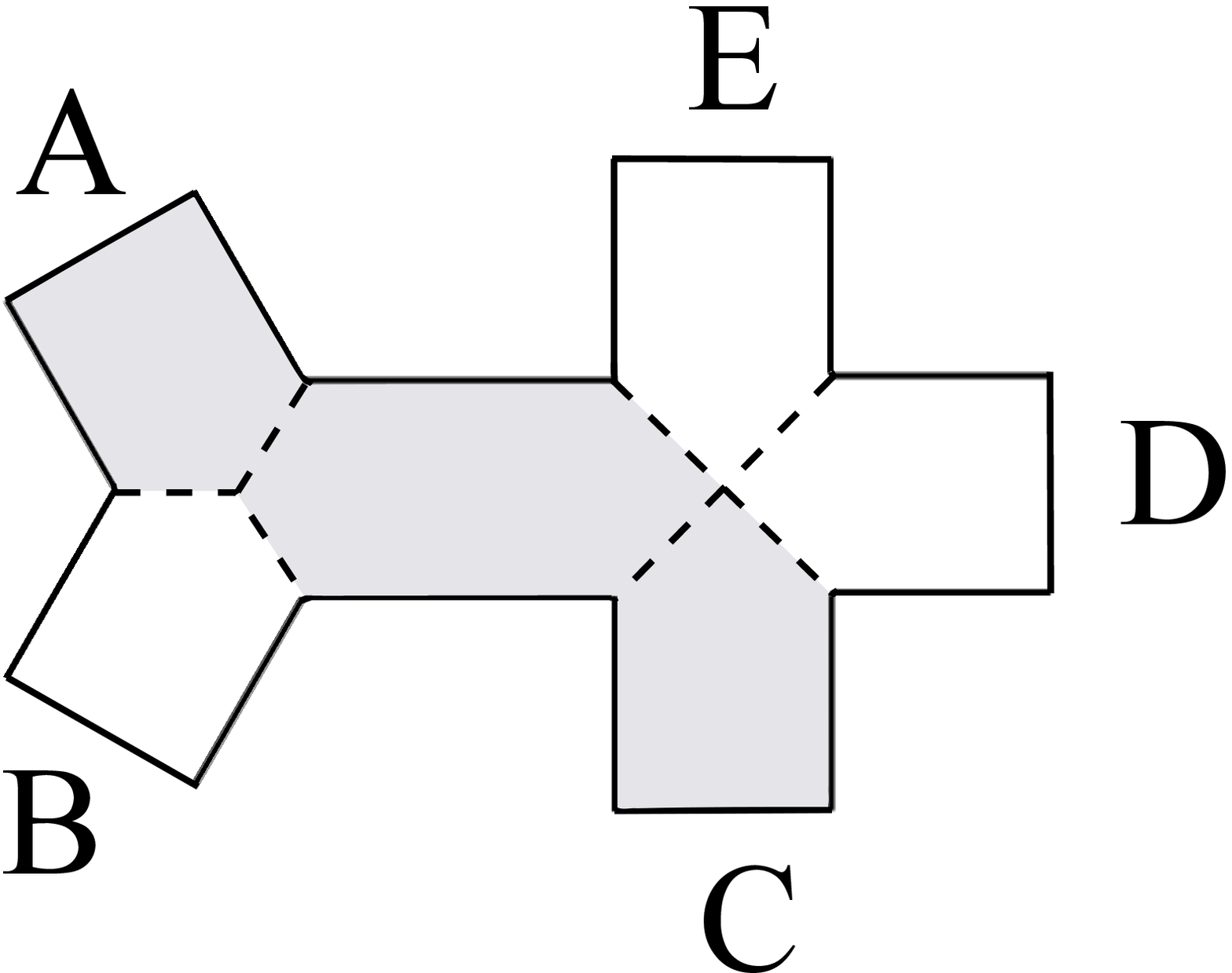}\\
(a)
 \end{center}
\end{minipage}
\begin{minipage}{0.25\hsize}
 \begin{center}
 \includegraphics[width=3cm]{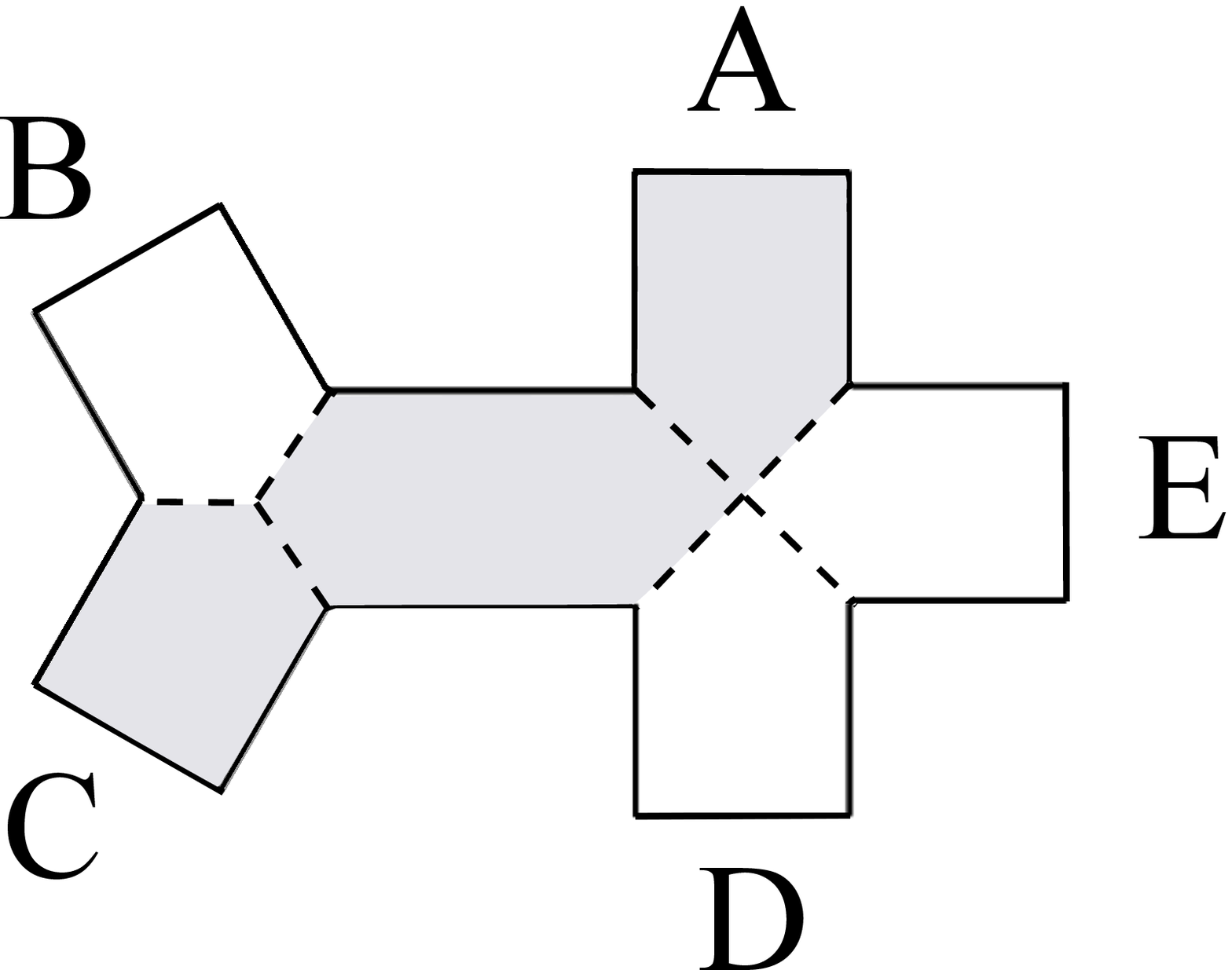}\\
(b)
 \end{center}
\end{minipage}
\begin{minipage}{0.25\hsize}
 \begin{center}
 \includegraphics[width=3cm]{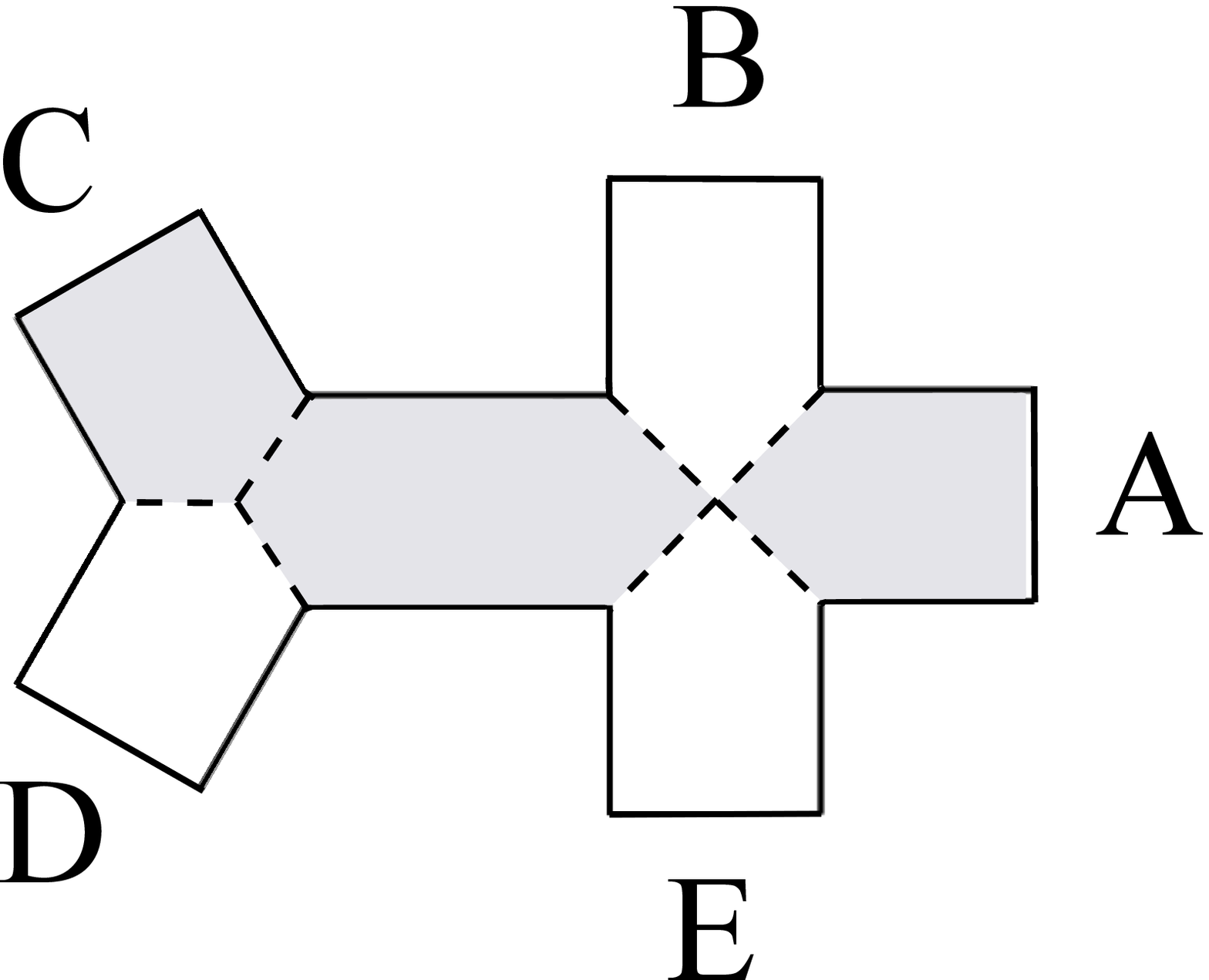} \\
(c)
 \end{center}
\end{minipage}
\begin{minipage}{0.125\hsize}
 \mbox{}
\end{minipage}
\begin{minipage}{0.23\hsize}
\mbox{}
\end{minipage}
\begin{minipage}{0.25\hsize}
 \begin{center}
 \includegraphics[width=3cm]{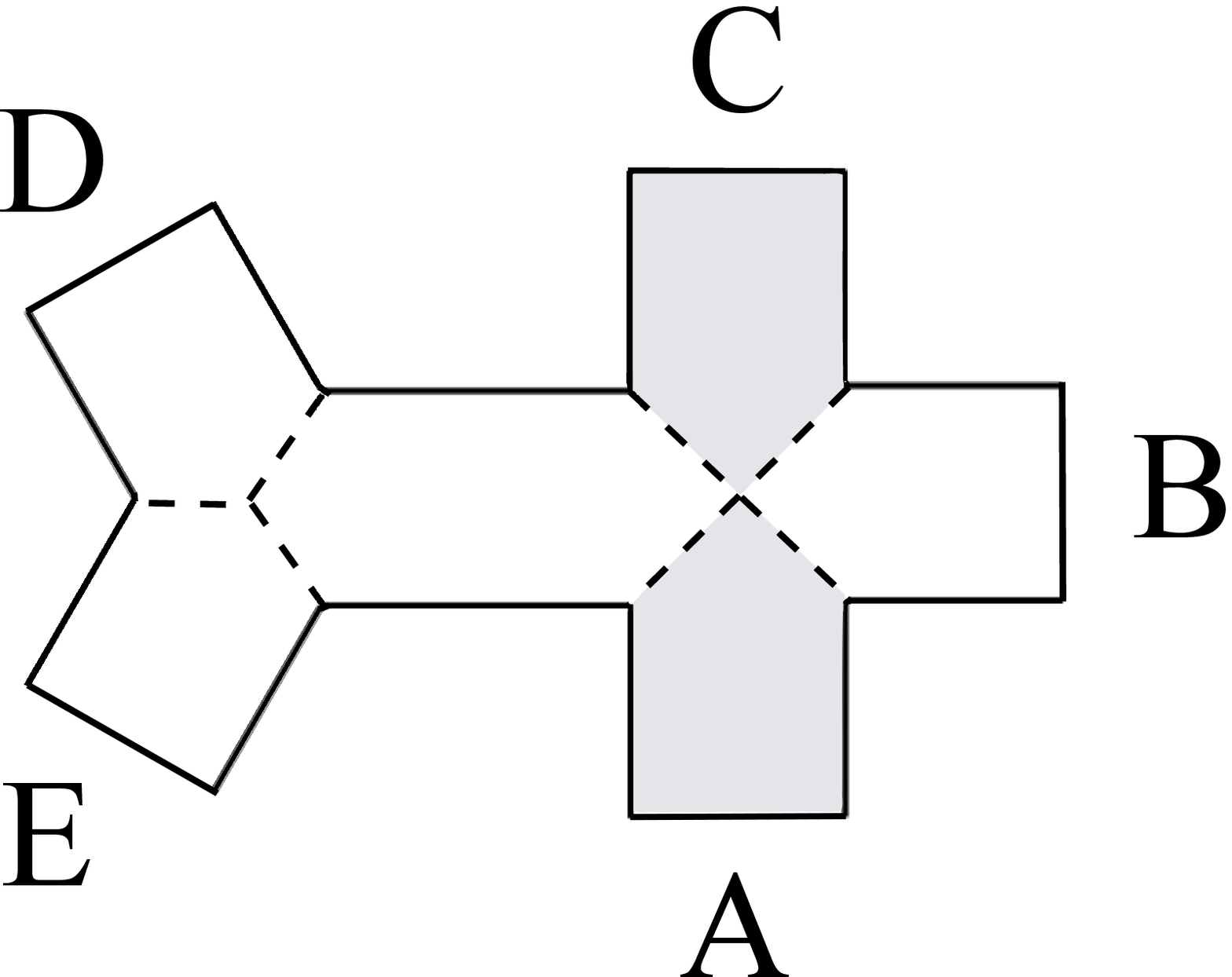} \\
(d)
 \end{center}
\end{minipage}
\begin{minipage}{0.25\hsize}
 \begin{center}
 \includegraphics[width=3cm]{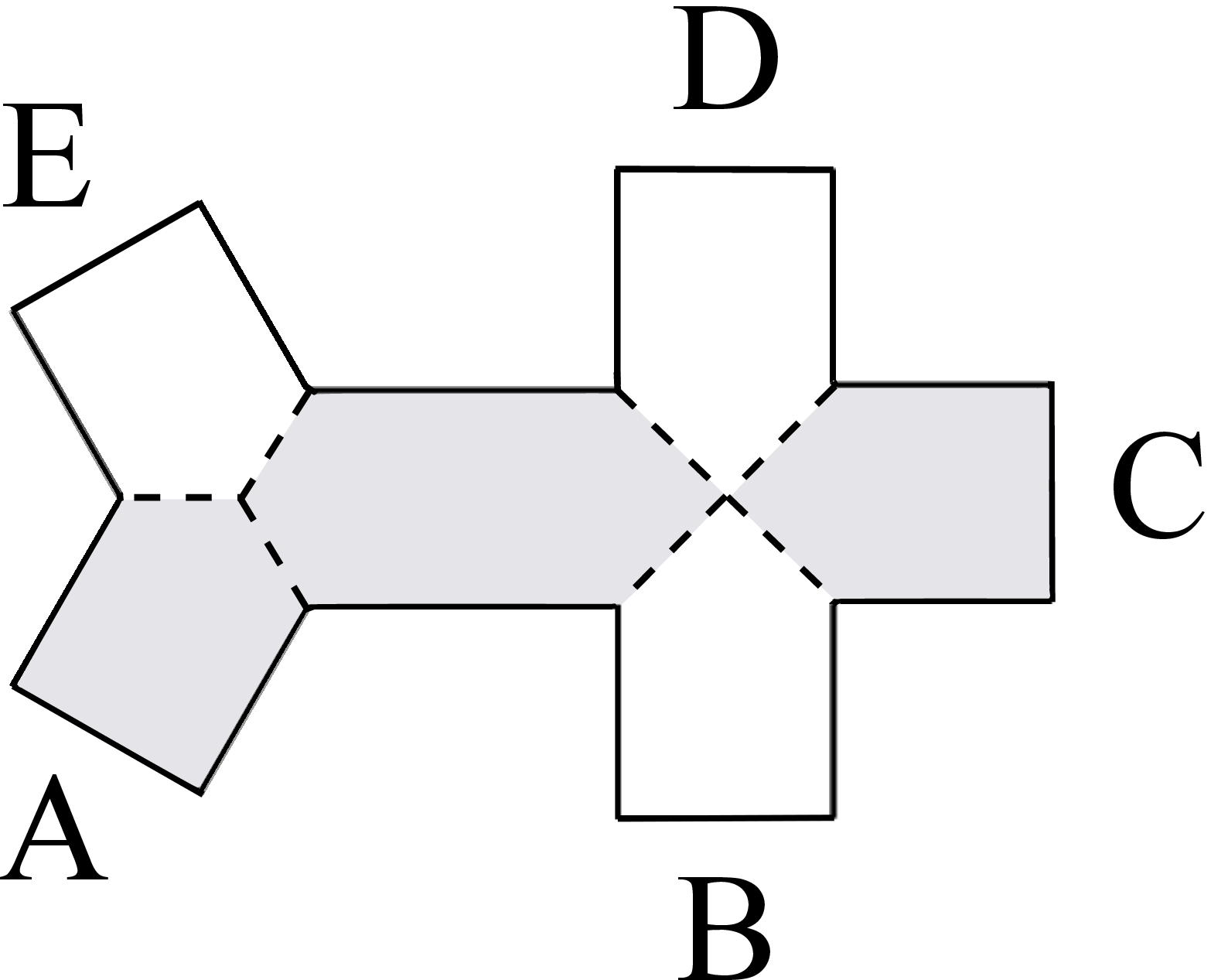} \\
(e)
 \end{center}
\end{minipage}
\vspace{2mm}
\caption{Five 1P Feynman diagrams
for two-fermion-three-boson amplitude
with ordering $FBFBB$. 
}\label{Feynman FBFBB 34}
 \end{figure}
The contributions with less propagator almost cancel with those of 
the five 1P diagrams in Fig.~\ref{Feynman FBFBB 34}: 
\begin{subequations}\label{cont FBFBB34}
\begin{align}
 \mathcal{A}_{FBFBB}^{(1\textrm{P})(a)}
=&\frac{1}{4}\int_0^\infty d\tau\Bigg(
\langle Q\Xi_A\ \eta \Phi_B\ (\xi_c b_c)\ \eta \Psi_C\ Q(\Phi_D\ \Phi_E)\rangle_W\nonumber\\
&\hspace{15mm}
-\langle \eta \Psi_A\ Q\Phi_B\ (\xi_c b_c)\ Q\Xi_C\ \eta (\Phi_D\ \Phi_E)\rangle_W
\Bigg),
\label{cont FBFBB 34a}\\
 \mathcal{A}_{FBFBB}^{(1\textrm{P})(b)} 
=&\frac{1}{4}\int_0^\infty d\tau\
\Bigg(
\langle Q\Phi_B\ \eta \Psi_C\ (\xi_c b_c)\
\eta (\Phi_D\ \Phi_E)\ Q\Xi_A\rangle_W
\nonumber\\
&\hspace{20mm}
-\langle \eta \Phi_B\ Q\Xi_C\ (\xi_c b_c)\
Q(\Phi_D\ \Phi_E)\ \eta \Psi_A\rangle_W
\Bigg),
\label{cont FBFBB 34b}\\
 \mathcal{A}_{FBFBB}^{(1\textrm{P})(c)} 
=&\frac{1}{2}\int_0^\infty d\tau\Bigg(
\langle Q\Xi_C\ \eta \Phi_D\ (\xi_c b_c)\
Q\Phi_E\ \eta \Psi_A\ \Phi_B\rangle_W
\nonumber\\
&\hspace{15mm}
-\langle Q\Xi_C\ \eta \Phi_D\ (\xi_c b_c)\
\Phi_E\ \eta \Psi_A\ Q\Phi_B\rangle_W
\nonumber\\
&\hspace{15mm}
-\langle \eta \Psi_C\ Q\Phi_D\ (\xi_c b_c)\
\eta \Phi_E\ Q\Xi_A\ \Phi_B\rangle_W
\nonumber\\
&\hspace{15mm}
+\langle \eta \Psi_C\ Q\Phi_D\ (\xi_c b_c)\
\Phi_E\ Q\Xi_A\ \eta \Phi_B\rangle_W
\Bigg),
\label{cont FBFBB 34c}\\
 \mathcal{A}_{FBFBB}^{(1\textrm{P})(d)} 
=&-\frac{1}{4}\int_0^\infty d\tau\
\Bigg(
\langle \Big(Q\Phi_D\ \eta \Phi_E+\eta \Phi_D\ Q\Phi_E\Big)\ 
(\xi_c b_c)\ Q\Xi_A\ \Phi_B\ \eta \Psi_C\rangle_W\nonumber\\
&\hspace{20mm}
-\langle \Big(Q\Phi_D\ \eta \Phi_E+\eta \Phi_D\ Q\Phi_E\Big)\ 
(\xi_c b_c)\ \eta \Psi_A\ \Phi_B\ Q\Xi_C\rangle_W
\Bigg),
\label{cont FBFBB 34d}\\
 \mathcal{A}_{FBFBB}^{(1\textrm{P})(e)} 
=&\frac{1}{2}\int_0^\infty d\tau\Bigg(
\langle \eta \Phi_E\ Q\Xi_A\ (\xi_c b_c)\
Q\Phi_B\ \eta \Psi_C\ \Phi_D\rangle_W\nonumber\\
&\hspace{15mm}
-\langle \eta \Phi_E\ Q\Xi_A\ (\xi_c b_c)\
\Phi_B\ \eta \Psi_C\ Q\Phi_D\rangle_W\nonumber\\
&\hspace{15mm}
-\langle Q\Phi_E\ \eta \Psi_A\ (\xi_c b_c)\
\eta \Phi_B\ Q\Xi_C\ \Phi_D\rangle_W\nonumber\\
&\hspace{15mm}
+\langle Q\Phi_E\ \eta\Psi_A\ (\xi_c b_c)\
\Phi_B\ Q\Xi_C\ \eta \Phi_D\rangle_W
\Bigg),
\label{cont FBFBB 34e}
\end{align}
\end{subequations}
where all the diagrams except for (d) satisfy condition (ii).
The slight remnant,
\begin{equation}
\sum_{i=a}^e\left(\mathcal{A}_{FBFBB}^{(2\textrm{P})(i)}\right)\Big|_{\textrm{extra}}
+ \sum_{i=a}^e\mathcal{A}_{FBFBB}^{(1\textrm{P})(i)}=
 \frac{1}{4}\Bigg(
\langle Q\Xi_A\ \Phi_B\ \eta \Psi_C\ \Phi_D\ \Phi_E\rangle_W
+\langle \eta \Psi_A\ \Phi_B\ Q\Xi_C\ \Phi_D\ \Phi_E \rangle_W
\Bigg),
\end{equation}
is cancelled with the contribution, 
\begin{align}
\mathcal{A}_{FBFBB}^{(\textrm{NP})} 
=& -\frac{1}{4}\Bigg(
\langle Q\Xi_A\ \Phi_B\ \eta \Psi_C\ \Phi_D\ \Phi_E\rangle_W
+\langle \eta \Psi_A\ \Phi_B\ Q\Xi_C\ \Phi_D\ \Phi_E \rangle_W
\Bigg),
\end{align}
coming from the no-propagator diagram in Fig.~\ref{Feynman FBFBB 5}.
As a consequence, the total amplitude with ordering $FBFBB$
becomes
\begin{align}
 \mathcal{A}_{FBFBB} =& \int_0^\infty d^2\tau\
\Bigg(
\langle \eta \Psi_A\ Q\Phi_B\ (\xi_{c_1} b_{c_1})\ \eta \Psi_C\ b_{c_2}\ 
Q\Phi_D\ \eta \Phi_E\rangle_W
\nonumber\\
&\hspace{15mm}
+\langle Q\Phi_B\ \eta \Psi_C\ (\xi_{c_1} b_{c_1})\
Q\Phi_D\ b_{c_2}\ \eta \Phi_E\ \eta \Psi_A\rangle_W
\nonumber\\
&\hspace{15mm}
+\langle \eta \Psi_C\ Q\Phi_D\ (\xi_{c_1} b_{c_1})\ \eta \Phi_E\
b_{c_2}\ \eta \Psi_A\ Q\Phi_B\rangle_W
\nonumber\\
&\hspace{15mm}
+\langle Q\Phi_D\ \eta \Phi_E\
(\xi_{c_1} b_{c_1})\ \eta \Psi_A\ b_{c_2}\ Q\Phi_B\ \eta \Psi_C\rangle_W
\nonumber\\
&\hspace{15mm}
+\langle \eta \Phi_E\ \eta \Psi_A\ (\xi_{c_1} b_{c_1})\
Q\Phi_B\ b_{c_2}\ \eta \Psi_C\ Q\Phi_D\rangle_W
\Bigg),
\end{align}
after eliminating the $\Xi$.
This is again in agreement with that obtained in the CSFT,
and therefore gives the well-known on-shell amplitude.

\section{Conclusion and discussion}\label{conclusion}

In this paper, we examined the symmetries of the pseudo-action,
the action supplemented by the constraint, in the WZW-type open 
superstring field theory. It was found that the pseudo-action is 
invariant under the additional symmetries provided we impose 
the constraint after the transformation. Then we proposed a prescription
for the new Feynman rules in the R sector so as to respect these symmetries. 
According to these new Feynman rules, we explicitly calculated
the on-shell four- and five-point amplitudes with the external fermions 
at the tree level. It was shown that the new rules correctly reproduce 
the well-known amplitudes in the first quantized formulation.

An important remaining problem is to clarify 
whether the new Feynman rules proposed in this paper
reproduce all the on-shell amplitudes at the tree level.
The additional symmetries 
should play an important role in solving this problem.
In order to extend the Feynman rules to those applicable
beyond the tree level, we have to fix the gauge symmetries more properly
using the Batalin-Vilkovisky method.\cite{Kroyter:2012ni,Torii:2012nj}
We may have to add a prescription, such as multiplying
each fermion loop by $1/2$, to solve the problem that may be caused by
the duplication of the off-shell fermion.
It is also worthwhile studying the off-shell amplitudes and comparing them to 
those obtained  by the rules proposed recently.\cite{Sen:2014pia}

Another interesting task is to apply similar considerations to the heterotic string 
field theory, which was also constructed based on the WZW-type 
formulation.\cite{Okawa:2004ii, Berkovits:2004xh,Kunitomo:2013mqa,Kunitomo:2014hba} 
In particular, the pseudo-action for the R sector was similarly constructed 
at some lower order in the fermion expansion.
We proposed the self-dual Feynman rules
and showed that they reproduce the on-shell four-point amplitudes.\cite{Kunitomo:2013mqa} 
We can similarly investigate, using the fermion expansion, the gauge symmetries of 
the pseudo-action and propose new Feynman rules.
It is interesting to calculate the on-shell five-point amplitudes 
by the two sets of Feynman rules, 
and confirm which rules reproduce the expected results.

\section*{Acknowledgments}

This work was initiated at the workshop on 
\lq\lq String Field Theory and Related Aspects VI\rq\rq\ 
held at SISSA in Trieste, Italy.
The author would like to thank the organizers, particularly Loriano Bonora, 
for their hospitality and providing a stimulating atmosphere.



\end{document}